%% file: peaks.v2.tex
\documentclass[onecolumn,showpacs,amsmath,amssymb,superscriptaddress,footinbib]{revtex4}

\usepackage{amsfonts}
\usepackage{amsmath}
\usepackage{aas_macros}
\usepackage{graphicx}
\usepackage{natbib}
\usepackage{color}

\input{defs.tex}

\newcommand{\beq}{\begin{equation}}
\newcommand{\eeq}{\end{equation}}
\newcommand{\beqa}{\begin{eqnarray}}
\newcommand{\eeqa}{\end{eqnarray}}

\begin{document}


\title[Density peaks, scale-dependent bias and the baryonic acoustic feature] 
      {Modeling scale-dependent bias on the baryonic acoustic scale with the statistics of peaks of Gaussian random fields}

\author{Vincent Desjacques} \email{dvince@physik.uzh.ch}
\affiliation{Institute for Theoretical Physics, University of Zurich, 8057 Zurich, Switzerland}
\author{Martin Crocce}
\affiliation{Institut de Ci\`encies de l'Espai, IEEC-CSIC, Campus UAB, Facultat de Ci\`encies, Barcelona 08193, Spain}
\author{Roman Scoccimarro}
\affiliation{Center for Cosmology and Particle Physics, Department of Physics, New York University, NY 10003, USA}
\author{Ravi K. Sheth}
\affiliation{Center for Particle Cosmology, University of Pennsylvania, 209 S 33rd Street, 
  Philadelphia, PA 19104, USA}


\begin{abstract}

Models of galaxy and halo clustering commonly assume that the tracers
can be treated as a continuous field locally biased with respect to 
the underlying mass distribution. In the peak model pioneered by 
Bardeen et al. (1986), 
one considers instead density maxima of the initial, Gaussian mass 
density field as an approximation to the formation site of virialized
objects. In this paper, the peak model is extended in two ways to 
improve its predictive accuracy. Firstly, we derive the two-point 
correlation function of initial density  peaks up to second order and 
demonstrate that a peak-background split approach can be applied to
obtain the $k$-independent and $k$-dependent peak bias factors at all
orders. Secondly, we explore the gravitational evolution of the peak
correlation function within the Zel'dovich approximation. We show that
the local (Lagrangian) bias approach emerges as a special case of the 
peak model, in which all bias parameters are scale-independent and 
there is no statistical velocity bias. We apply our formulae to study
how the Lagrangian peak biasing, the diffusion due to large scale flows 
and the mode-coupling due to nonlocal interactions affect the 
scale dependence of bias from small separations up to the baryon 
acoustic oscillation (BAO) scale. For 2$\sigma$ density peaks collapsing 
at $z=0.3$, our model predicts a $\sim$5\% residual scale-dependent bias 
around the acoustic scale that arises mostly from first-order Lagrangian 
peak biasing (as opposed to second-order gravity mode-coupling). We also 
search for a scale dependence of bias in the large scale auto-correlation 
of massive halos extracted from a very large N-body simulation provided
by the {\small MICE} collaboration. For halos with mass 
$M\gtrsim 10^{14}\hmsun$, our measurements demonstrate a scale-dependent 
bias across the BAO feature which is very well reproduced by a prediction 
based on the peak model. 

\end{abstract}

\maketitle

\setcounter{footnote}{0}

\section{Introduction}
\label{sec:intro}

A considerable amount of effort has already been invested in measuring 
the large scale distribution of galaxies, especially the galaxy two-point 
correlation function and power spectrum, to constrain viable cosmological 
models \citep[e.g.,][]{1990Natur.348..705E,1995MNRAS.276L..59B,
1999MNRAS.305..527T,2001MNRAS.327.1297P,2004ApJ...606..702T,
2005MNRAS.362..505C,2005ApJ...633..560E,2006PhRvD..74l3507T,
2006A&A...459..375H,2007MNRAS.381.1053P,2009MNRAS.400.1643S,
2009MNRAS.393.1183C,2009ApJ...696L..93M,2010MNRAS.404...60R}. 
The amplitude, shape and baryon acoustic feature in these two-point 
statistics encode a wealth of cosmological information 
\citep{1996ApJ...471...30H,1998ApJ...504L..57E,
2001ApJ...557L...7C,2003PhRvD..68f3004H,2003ApJ...594..665B,
2003PhRvD..68h3504L,2004ApJ...615..573M,2005MNRAS.357..429A,
2005MNRAS.363.1329B,2005ApJ...631....1G,2006MNRAS.366..884D,
2006ApJ...644..663Z,2006MNRAS.365..255B,2008PhRvD..77l3540P,
2009MNRAS.399.1663G,2009ApJ...693.1404S,2009ApJ...698..967Y,
2010ApJ...710.1444K}.
Ongoing and planned galaxy surveys of the high redshift Universe will 
furnish measurements of the underlying mass distribution with unprecedent 
precision and statistics. 
Alongside this great observational effort, interpreting this vast amount 
of data will require a much better understanding of the relation between 
the surveyed galaxies and the mass fluctuations they are thought to trace.  

Essentially all models of galaxy clustering assume that galaxies are 
biased tracers of the mass density fluctuation field. Although this bias 
is expected to be nonlinear, scale-dependent and stochastic
\citep{1996MNRAS.282..347M,1999MNRAS.304..767S}, the simpler, 
linear, scale-independent, deterministic model has proved to be an 
extremely useful first order approximation \citep{1984ApJ...284L...9K,
1986ApJ...304...15B,1989MNRAS.237.1127C}. However, in order to predict
corrections beyond linear order to the galaxy two-point correlation, or 
even the leading order contribution to higher order statistics such as 
the galaxy three-point correlation or bispectrum, one must address the 
complications which arise from nonlinearity, scale dependence and 
stochasticity.  For example, if the bias relation is established in 
coordinate space, then nonlinear biasing will produce scale dependence 
and stochasticity in Fourier space, and vice-versa 
\citep[e.g.,][]{1999ApJ...525..543M}.  
This randomness will add to other sources of stochasticity which may 
arise, for example, from the fact that the formation of galaxies and 
halos depends on quantities other than the mass density field (e.g., 
the surrounding tidal field). Moreover, the bias may be established in 
the initial conditions (Lagrangian bias) or, alternatively, at the 
present time (Eulerian bias). In the former case, the bias between 
the tracers and the mass will be affected by the subsequent, nonlinear 
gravitational evolution. This will introduce additional nonlinearity, 
scale dependence and stochasticity. 
Furthermore, if the velocities of the tracers differ from those of 
the mass elements, then this will complicate the application of the 
continuity equation to describe the redshift evolution of bias.

Current analytic approaches to galaxy and dark matter halo clustering 
take into account some  of these complications. In most models, the 
fundamental quantity is the overdensity of tracers 
$\delta_{\rm g,h}(R,\vx)$ within a sphere of radius $R$ centered at 
position $\vx$.  
It is commonly assumed that $\delta_{\rm g,h}$ is solely a function of 
the local mass overdensity $\delta_{\rm m}$ 
\citep{1993ApJ...413..447F,1988ApJ...333...21S}
\citep[see also][]{2009JCAP...08..020M}, whose Taylor 
expansion coefficients are the galaxy or halo bias parameters $b_N$
\citep{1997MNRAS.284..189M,2001ApJ...546...20S,2010MNRAS.402..589M}.
If this bias is established at a different time than the epoch at which 
the tracers are observed, then this local bias scheme is combined with 
some (Eulerian or Lagrangian) perturbative treatment of gravitational 
instability 
\citep[see][for a review of perturbation theory]{2002PhR...367....1B} 
to predict the galaxy or halo power spectrum, bispectrum etc. 
\citep[e.g.,][]{1998MNRAS.301..797H,1998MNRAS.297..692C,
1998MNRAS.298.1097P,2000ApJ...537...37T,2000ApJ...530...36B,
2001ApJ...546...20S,2006PhRvD..74j3512M,2007PhRvD..75f3512S,
2007PhRvD..76h3004S,2007PASJ...59...93N,2008PhRvD..78h3519M,
2009ApJ...691..569J,2009PhRvD..80h3528S,2010arXiv1009.1131V}. 
This formalism can be extended to include stochasticity by 
formulating the biasing in terms of the conditional probability 
distribution $P(\delta_{\rm g,h}|\delta_{\rm m})$ of $\delta_{\rm g,h}$ 
at a given $\delta_{\rm m}$ \citep[e.g.,][]{1998ApJ...504..601P,
1999ApJ...520...24D,1999ApJ...525..543M,2000ApJ...537...37T,
2000ApJ...542..559T}. One of the main caveats with such local biasing 
schemes is that galaxies (or halos) are treated as though they define 
a continuous field smoothed on some scale $R$, whereas they are, in 
fact, discrete objects.

The peaks approach to galaxy and dark matter halo clustering is interesting 
because it exhibits all of the complications mentioned above while also 
accounting for the discrete nature of the tracers (after all, peaks 
define a point process).  In this model, the fundamental quantity is 
the set of positions which are local maxima of the density field (from 
which a peak overabundance $\dpk(R,\vx)$ in spheres of radius $R$ could 
in principle be derived). Since the evolved density field is highly 
nonlinear, the peak constraint is generally applied to the initial 
(Lagrangian) Gaussian density field, with the assumption that the most 
prominent peaks should be in one-to-one correspondence with luminous 
galaxies or massive halos in the low redshift Universe 
\citep[see, e.g.,][for numerical studies of this association]
{1988ApJ...327..507F,1993MNRAS.265..689K,2002MNRAS.332..339P}. 
Peak abundances, profiles and correlation functions in real and redshift 
space have been studied in the literature
~\citep{1985MNRAS.217..805P,1985ApJ...297...16H,1986ApJ...304...15B,
1986PhRvL..56.1878O,1987MNRAS.225..777C,1989MNRAS.238..319C,
1989MNRAS.238..293L,1990MNRAS.243..133P,1995MNRAS.272..447R,
1998ApJ...499..548M,2008PhRvD..78j3503D,2010PhRvD..81b3526D}. Some of 
these results have been 
used to interpret the abundance and clustering of rich clusters 
\citep{1985ApJ...297..365K,1988lsmu.book..419B,1989ApJ...345....3C,
1997MNRAS.284..189M,1998ApJ...509..494C,2001NYASA.927....1S}, constrain 
the power spectrum of mass fluctuations \citep{1998ApJ...495..554C, 
2010MNRAS.401.1989D}, and study evolution bias \citep{2008MNRAS.385L..78P} 
and assembly bias \citep{2008ApJ...687...12D}.

On asymptotically large scales, peaks are linearly biased tracers of the 
mass density field, and this bias is scale independent
\citep{1984ApJ...284L...9K,1984ApJ...285L...1P,1986ApJ...304...15B,
1997MNRAS.284..189M}.
However, these conclusions are based on a configuration space argument
known as the peak background split -- which establishes a relation 
between the sensitivity of the peak bias factors and the peak abundances
on the peak height -- whereas a Fourier space analysis suggests that the 
linear bias factor of peaks is the sum of two terms, one of which is 
$k$-dependent \citep{1999ApJ...525..543M,2008PhRvD..78j3503D}. In 
configuration space, this leads to scale dependence of the bias and 
stochasticity. The $k$-dependence of the linear peak bias arises from 
the peak constraint, i.e. the fact that one must specify not only the 
value of the mass density field but also its first two derivatives to
define a peak. Therefore, this is a model in which the bias depends on 
quantities other than the local density. Moreover, as mentioned above,
the peak biasing is applied to the initial Gaussian density field so 
that the late time peak bias is modified by nonlinear evolution and 
associated stochasticity. In this regards, peaks exhibit a nontrivial 
velocity bias \citep{2010PhRvD..81b3526D}, which further complicates 
the nonlinear evolution.  

In the peak model, both the constant and the $k$-dependent piece of the 
linear bias factor depend on peak height. As shown in 
\citep{2010PhRvD..81b3526D}, the scale-independent contribution can be 
derived from the peak background split argument. In the first half of 
this paper, we demonstrate that the Fourier space approach also 
predicts constant and $k$-dependent contributions to the second and 
higher order peak bias factors. We then show that the scale-independent 
parts of all these nonlinear peak bias factors can be also derived from 
the peak background split argument, thus generalizing the result of 
\citep{2010PhRvD..81b3526D}.
We go on to show how the peak background split approach can be used 
to determine the scale-dependent part of the peak bias factors, first 
at linear order, and then for all nonlinear orders as well.  
This is particularly interesting because it illustrates how the peak 
background split argument should be implemented if the abundance of 
the biased tracers (in this case, peaks) depends on quantities other 
than the local mass overdensity (in this case, the first and second 
derivatives of the mass density field).  

As recognized in \cite{2008PhRvD..78j3503D}, the $k$-dependence of the 
{\it first order} peak bias strongly amplifies the contrast of the
baryon acoustic oscillation 
\citep[or BAO. see][and references therein]{2009arXiv0910.5224B} 
in the correlation of initial density maxima.  However, this calculation 
was performed for peaks identified in the initial conditions, so there 
was no clear connection with the clustering of dark matter halos and 
galaxies. This is also true of all the results presented in the first 
half of this paper. To remedy this problem, we show in the second half 
how the effects of the (nonlinear, nonlocal) gravitational evolution of 
density peaks can be incorporated in the peak model. This allows us to 
ascertain the extent to which the initial scale dependence of bias 
across the BAO survives at late times. Our analysis incorporates two 
main complications that are usually ignored in local bias schemes. 
Namely, peak biasing depends on more than just the value of the local 
density, and peaks exhibit a velocity bias which (in addition to merging)
complicates analyses based on the continuity equation. Finally, we show 
that taking into account these effects is of more than academic interest:  
Our peaks model provides a very good description of the scale dependence 
of the bias of massive halos in numerical simulations -- halos that are 
expected to host the luminous red galaxies (LRGs) which are often 
targeted in BAO experiments. 

The paper is organized as follows. Section \S\ref{sec:defs} briefly 
reviews known results and introduce some useful definitions. Section
\S\ref{sec:xipk2nd} focuses on the correlation of initial density 
peaks of a Gaussian random field. It is shown that the scale-dependent 
and scale-independent parts of the peak bias parameters can be derived 
from a peak-background split argument. Section \S\ref{sec:evol} 
considers the gravitational evolution of the peak correlation function 
in the Zel'dovich approximation. It is shown that, in addition to 
gravity mode-coupling, the Lagrangian peak biasing can generate a 
significant scale-dependent bias across the baryonic acoustic feature 
at the collapse epoch. Measurements of bias at BAO scales from the 
clustering of massive halos are also presented and compared with the
model. Section \S\ref{sec:conclusion} summarizes our results. Technical 
details of the calculation can be found in Appendix \S\ref{app:xpk2L} 
and \ref{app:xpkevol}.

\section{Definitions, notations and known results}
\label{sec:defs}

We begin by introducing some definitions and reviewing known results
about the clustering of density peaks  in Gaussian random
fields. Next, we derive the peak correlation at second  order. This
result will serve as input to the calculation of the evolved
correlation of density peaks.

\subsection{Spectral moments}
\label{sub:moments}

The statistical  properties of density peaks depend not only on the
underlying density field, but also on its first and second
derivatives.  We are, therefore, interested in the linear (Gaussian)
density field $\delta$ and its first and second derivatives,
$\partial_i\delta$ and $\partial_i\partial_j\delta$. In this regard,
it is convenient to introduce the normalized  variables
$\nu=\delta/\sigma_0$, $\eta_i\equiv\partial_i\delta/\sigma_1$ and
$\zeta_{ij}\equiv \partial_i\partial_j\delta/\sigma_2$, where the
$\sigma_n$ are the spectral moments of the matter power spectrum,
\begin{equation}
\sigma_n^2(R_S,z_0) \equiv \frac{1}{2\pi^2}\int_0^\infty\!\!dk\,k^{2(n+1)}\,
  P_\delta(k,z_0) W(k R_S)^2\;.
 \label{eq:mspec}
\end{equation} 
$P_\delta(k,z_0)$ denotes the dimensionless power spectrum of the
linear density field at redshift $z_0$, and $W$ is a spherically
symmetric smoothing kernel of length $R_S$ introduced to ensure
convergence of all spectral moments. A Gaussian filter will be adopted
throughout this paper.  We will use the notation $P_{\delta_S}(k,z_0)$
to denote  $P_\delta(k,z_0)W(k R_S)^2$.  The ratio $\sigma_0/\sigma_1$
is proportional to the typical separation between zero-crossings of
the density field~\citep{1986ApJ...304...15B}.   For subsequent use,
we also define the spectral parameters
\begin{equation}
 \gamma_n(R_S)=\frac{\sigma_n^2}{\sigma_{n-1}\sigma_{n+1}}
 \label{eq:gammas}
\end{equation}
which reflect the range over which $k^{2(n-1)}P_{\delta_S}(k,z_0)$ is
large. We will also work with the  scaled velocities  ${\rm v}_i\equiv
{\rm v}_{{\rm p}i}/(aHf)$ and with the curvature
$u=-\partial^2\delta/\sigma_2$. Here, ${\rm v}_{{\rm p}i}(\vx)$ is the
$i$-th component of the (proper) peculiar velocity, $H\equiv d\ln a/dt$, 
$f\equiv d\ln D/d\ln a$ is  the logarithmic derivative of the linear 
theory growth rate $D(z_0)$ and $\partial^2=\partial^i\partial_i$ is 
the Laplacian.  Note that ${\rm v}_i$ has dimensions of length.

The analogous quantities to $\sigma_n^2$ at non-zero separation are 
defined as follows:
\begin{equation}
 \xi_\ell^{(n)}(R_S,r,z_0)= \frac{1}{2\pi^2}\int_0^\infty\!\! dk\,
  k^{2(n+1)} P_{\delta_S}(k,z_0)\; j_\ell(kr)\;,
 \label{xielln}
\end{equation}
where $j_\ell(x)$ are spherical Bessel functions. As $\ell$ gets
larger, these harmonic transforms become increasingly sensitive to
small scale power.  The auto- and cross-correlations of the fields
$v_i(\vx)$, $\eta_i(\vx)$, $\nu(\vx)$, $u(\vx)$ and $\zeta_{ij}(\vx)$
can generally be decomposed  into components with definite
transformation properties under rotations.  \cite{2008PhRvD..78j3503D}
gives explicit expressions for the isotropic and homogeneous linear
density field.

\subsection{Peak biasing and 2-point correlation function at the first order}
\label{sub:bias}

Although density peaks form a well-behaved point process, the
large scale asympotics of the two-point correlation $\xpk(r,z)$ and
line of sight mean streaming $v_{12}(\vr,z)\cdot\rvh$ of peaks of
height $\nu$ and curvature $u$ identified on a scale $R_S$ in the
initial Gaussian density field linearly extrapolated at redshift $z_0$
can be thought of as arising from the continuous, deterministic bias 
relation \citep{2008PhRvD..78j3503D,2010PhRvD..81b3526D}
\begin{align}
\dpk(\nu,u,R_S,\vx) &= b_\nu \delta_S(\vx,z_0) 
-b_\zeta \partial^2\delta_S(\vx,z_0), \\
\vvpk(R_S,\vx,z_0) &= \vv_S(\vx,z_0)-\frac{\sigma_0^2}{\sigma_1^2}\,
\partial\delta_S(\vx,z_0)\;,
\label{eq:pkbiasing}
\end{align}
which is nonlocal owing to the smoothing of the mass distribution. Here, 
$\dpk$ and $\vvpk$ are the average peak overdensity and velocity,
$\delta_S$ and $\vv_S$ are the mass density and velocity smoothed at 
scale $R_S$ (so as to retain only the large scale, coherent motion of the 
peak), $\partial^2$ is the Laplacian, and the bias parameters $b_\nu$ and 
$b_\zeta$ are
\begin{equation}
b_\nu(\nu,u,R_S,z_0) \equiv
\frac{1}{\sigma_0(R_S,z_0)}\left(\frac{\nu-\gamma_1 u}{1-\gamma_1^2}\right),
\qquad
b_\zeta(\nu,u,R_S,z_0) \equiv 
\frac{1}{\sigma_2(R_S,z_0)}\left(\frac{u-\gamma_1\nu}{1-\gamma_1^2}\right)\;.
\label{eq:biasvu}
\end{equation}
The bias coefficient $b_\nu$ is dimensionless, whereas $b_\zeta$ has
units of (length)$^2$. In fact, $b_\nu$ is precisely the amplification
factor found by \cite{1986ApJ...304...15B} who neglected derivatives
of the density correlation function (i.e. their analysis assumes
$b_\zeta\equiv 0$). Unlike $\vvpk$, $\dpk$ does not depend on
$z_0$ (as expected) because the redshift dependence of $b_\nu$,
$b_\zeta$ cancels the factor $D(z_0)$ coming from $\delta_S(\vx,z_0)$.
Note also that, if $b_\nu>0$ the effective peak density $\dpk(\vx)$ 
can be less than -1 in deep voids. However, this is not a problem because
$\dpk(\vx)$ is not an observable quantity (this is {\it not} a 
count-in-cell density).

In what follows, we will focus on the clustering of initial density
peaks of significance $\nu$, for which the first order bias parameters
are
\begin{align}
\label{eq:1stbias}
\bv(\nu,R_S,z_0) &\equiv
\frac{1}{\sigma_0(R_S,z_0)}\left(\frac{\nu-\gamma_1\bar{u}}{1-\gamma_1^2}
\right) \\
\bz(\nu,R_S,z_0) &\equiv 
\frac{1}{\sigma_2(R_S,z_0)}\left(\frac{\bar{u}-\gamma_1\nu}{1-\gamma_1^2}
\right)\;.
\end{align}
Here, the overline denotes the averaging over the peak curvature, so
that $\bar u\equiv {\bar u}(\nu,R_S)$ is the mean curvature of peaks of 
height $\nu$ on filtering scale $R_S$. It is convenient to define the 
quantity $\bspk$ as the Fourier space multiplication by 
\begin{equation}
\bspk(k,z_0) = b_\nu + b_\zeta k^2 \;,
\label{eq:pkbiask}
\end{equation}
where we have omitted the explicit dependence on $\nu$, $u$ and $R_S$
for brevity.  Although $\bspk(k,z_0) $  has the same functional form as 
Eq.~(57) of \cite{1999ApJ...525..543M}, this author approximated density 
peaks by density extrema. Therefore, our coefficients agree with his 
expressions only in the limit $\nu\gg 1$, in which $b_\nu\to\nu/\sigma_0$
and $b_\zeta\to 0$. As we will see shortly, the 
product of $\bspk(k,z_0)$ factors can be used to define spatial bias 
parameters at all orders. For peaks of significance $\nu$, the 
first order biasing is equivalent to the Fourier space multiplication 
by $\bias{I}(k,z_0)\equiv \bb_{\rm spk}(k,z_0)$, i.e.
\begin{equation}
\bias{I}(k,z_0)\equiv \bias{10}+\bias{01}k^2\qquad \mbox{where}\qquad
\bias{10}\equiv\bv,~~\bias{01}\equiv\bz\;.
\label{eq:bias1st}
\end{equation}
We emphasize that this result is exact: there are no higher powers such 
as $k^4$, etc. In $\bias{ij}$, $i$ and $j$ count the number of factors 
of $b_\nu$ and $b_\zeta$, respectively \citep[our notation should not 
be confounded with that of][]{2010PhRvD..81f3530G}. In 
Sec.\S\ref{sub:pksplit}, we will demonstrate that the $\bias{i0}$ are 
the bias parameters in the local bias model. Eq.(\ref{eq:bias1st}) 
defines the first order bias for peaks of height $\nu$. Notice that, in 
real space, $\bias{I}(k,z_0)=\bias{10}-\bias{01}\partial^2$ is a 
differential operator acting on fields and correlation functions. Hence, 
the first order average peak overabundance can also be rewritten
$\dpk(\nu,\vx,z_0)=(\bias{I}\delta_S)(\vx,z_0)$.

Using the peak bias~(\ref{eq:pkbiask}),  it is straightforward to show
that the real space cross- and auto-power spectrum are
\begin{align}
 \ppd^{(1)}(\nu,R_S,k,z_0) &= \bias{I}(k,z_0)\,P_\delta(k,z_0)\,W(k R_S)
 \label{eq:ppd}\\
 \ppk^{(1)}(\nu,R_S,k) &= \bias{I}^2(k,z_0)\,P_\delta(k,z_0)\,W^2(k R_S)\;.
 \label{eq:ppk}
\end{align}
The corresponding relations for the correlation functions are
\begin{align}
\xpd^{(1)}(\nu,R_S,r,z_0) &= \bigl(\bias{I}\xi_0^{(0)}\bigr)=
\bias{10}\,\xi_0^{(0)}\!(R_S,r,z_0)+\bias{01}\,
\xi_0^{(1)}\!(R_S,r,z_0)\,
\label{eq:xpd}\\
\xpk^{(1)}(\nu,R_S,r) &= \bigl(\bias{I}^2\xi_0^{(0)}\bigr)=
\bias{10}^2\,\xi_0^{(0)}\!(R_S,r,z_0) + 2\bias{10}\bias{01}\,
\xi_0^{(1)}\!(R_S,r,z_0)+\bias{01}^2\,\xi_0^{(2)}\!(R_S,r,z_0)\;.
\label{eq:xpk}
\end{align} 
Note that the cross-correlations with the linear density field
$\delta(\vx,z_0)$ depend explicitly on $z_0$.  As shown in
\cite{2008PhRvD..78j3503D,2010PhRvD..81b3526D}, these expressions
agree with those obtained from a rather lengthy derivation based on
the peak constraint, which involves joint probability distributions of
the density field and its derivatives.  It is worth noticing that,
while expressions (\ref{eq:ppk}) and (\ref{eq:xpk}) are only valid at
leading order, the cross-correlation functions (\ref{eq:ppd}) and
(\ref{eq:xpd}) are exact to all orders.

We emphasize that the biasing (\ref{eq:pkbiasing})  is a mean bias
relation that does not contain any information about
stochasticity. Due to the  discrete nature of density peaks however,
one can expect that the average peak overabundance $\dpk(\vx)$ 
in a cell centered at $\vx$ generally be a random function of the 
underlying matter density (and its derivatives) in some neighborhood 
of that point. In fact, while the bias is deterministic  in Fourier 
space, it is generally stochastic and scale-dependent in configuration 
space \citep{2010PhRvD..81b3526D}.

\subsection{Velocities}
\label{sub:velocities}

In what follows, we will be interested in the gravitational evolution
of the correlation of initial density peaks for which the velocity
field also matters. As can be seen from, e.g., the average bias 
relation (\ref{eq:pkbiasing}), peaks locally move with the dark matter
(since the gradient of the density vanishes at the position of a peak).
However, the three-dimensional velocity dispersion of  peaks,  
$\sigma^2_{\rm vpk}=\la\vvpk^2\ra$, is smaller than the mass velocity
dispersion $\sigma_{-1}^2$ \citep{1986ApJ...304...15B,1987AcPhH..62..263S},
\begin{equation}
 \sigma^2_{\rm vpk} = \sigma_{-1}^2\, (1 - \gamma_0^2)\;,
 \label{eq:sigmavpk}
\end{equation}
because large scale flows are more likely to be directed towards peaks 
than to be oriented randomly. As recognized in \cite{2010PhRvD..81b3526D},
the $k$-dependence of the first order peak bias $\bspk(k)$ leads
to a $k$-dependence of the peak velocity statistics even though the peaks 
move with the dark matter flows. Taking the divergence of the peak velocity
Eq.~(\ref{eq:pkbiasing}) and Fourier transforming, we find
\begin{equation}
  \theta_{\rm pk}(R_S,\vk,z_0)=\left(1 - \frac{\sigma_0^2}{\sigma_1^2}\,
  k^2\right)W(k R_S)\,\theta(\vk,z_0)\equiv \bvpk(k)\,\theta_S(\vk,z_0)\;,
  \label{eq:vpkbiask}
\end{equation}
where $\theta\equiv\nabla\cdot\vv$ is the velocity divergence. This
defines the {\it linear} velocity bias  factor $\bvpk(k)$ for peaks 
of significance $\nu$ and curvature $u$. Note that it does not depend 
on $\nu$, $u$ nor on redshift and, for the highest peaks, it remains 
scale dependent even though the spatial bias $\bias{I}(k,z_0)$ has no 
$k$-dependence. Nonetheless, for notational consistency, we define
\begin{equation}
\bias{vpk}(k)\equiv\bar{b}_{\rm vpk}(k)=\bvpk(k)
\end{equation}
as being the velocity bias of peaks of height $\nu$. 

\subsection{Smoothing scale and peak height}
\label{sub:smoothing}

To illustrate the key predictions of the peak formalism, we will present 
results for the two-point correlation of density peaks in a $\Lambda$CDM 
cosmology with $h=0.7$, $\Omega_{\rm m}=0.279$, $\Omega_{\rm b}=0.0462$, 
$n_s=0.96$ and normalization $\sigma_8=0.81$ consistent with the latest
constraints from the CMB \citep{2010arXiv1001.4538K}. The sound horizon 
at recombination is $r_0\approx 105\hmpc$.

The peak height $\nu$ and the filtering radius $R_S$ could in
principle be treated as two independent variables.  However, in order
to make as much connection with dark matter halos (and, to a lesser
extent, galaxies) as possible, we assume that density maxima with
height $\nu=\dsc(z_0)/\sigma_0(R_S)$ identified in the smoothed
density field linearly extrapolated at $z_0$ are related to dark
matter halos of mass $M_S\propto R_S^3$ collapsing at redshift $z_0$,  
where $\dsc(z_0)$ is the critical density for collapse in the spherical
model \citep{1972ApJ...176....1G,1974ApJ...187..425P} and we use a
Gaussian filter to relate smoothing scale to mass. A more realistic
treatment should include non-spherical collapse since the maxima of a
Gaussian density field are inherently triaxial
\citep{1970Afz.....6..581D,1996ApJS..103....1B,2001MNRAS.323....1S,
2008MNRAS.388..638D}.
In the background cosmology we assume, the linear critical density 
for spherical collapse at $z_0=0.3$ is  $\dsc\approx 1.681$.  The
Gaussian smoothing scale at which $\nu=1$ is $R_{S_\star}\approx
0.8\hmpc$, which corresponds to a characteristic mass scale
$M_{S_\star}\approx 6.2\times 10^{11}\hmsun$.

While there is a direct correspondence between  massive halos in the
evolved density field and the largest maxima  of the initial density
field, the extent to which galaxy-sized halos trace the initial
density maxima is unclear.   Therefore, we will only consider mass
scales $M_S$ significantly larger than the characteristic mass for
clustering, $M_{S_\star}$, for  which the peak model is expected to work
best. For sake of illustration, we will present  results for $\nu=2$
(2$\sigma$) and $\nu=3$ (3$\sigma$) density peaks. At redshift 
$z_0=0.3$, this corresponds to a filtering length $R_S = 2.9\hmpc$ 
and $R_S=5.3\hmpc$ or, equivalently, a mass scale 
$M_S = 3.0\times 10^{13}\hmsun$ and $1.6\times 10^{14}\hmsun$. To help 
set scales in the discussion which follows, the associated values of
$(\gamma_1,\sigma_0/\sigma_1,\bias{10},\bias{01})$ are
$(0.65,3.7,0.8,21.1\hhmpc)$ and $(0.68,6.2,3.5,72.4\hhmpc)$, 
respectively (note that bias factors here are Lagrangian ones). The
three-dimensional velocity dispersion of these peaks is
$\sigma_{-1}^2\,(1-\gamma_0^2)$: for our two smoothing scales, this
corresponds to $(7.75\hmpc)^2$ and $(7.03\hmpc)^2$ (recall that our
velocities are in units of  $aHf\approx 58\hkmsmpc$ at $z=0.3$, so
dispersions have dimensions of (length)$^2$).

\section{Correlation of initial density peaks at second order}
\label{sec:xipk2nd}

\subsection{The general formula}
\label{sub:xipk2nd}

Correlations of density maxima can be evaluated using the Kac-Rice
formula \citep{Kac1943,Rice1945}. In this approach, $\eta_i(\vx)$ is
Taylor-expanded around the position $\vx_{\rm pk}$ of a local
maximum. The number density of peaks of height $\nu'$ at position
$\vx$ in the smoothed density field $\delta_S$ reads as (we
drop the subscript $S$ in the right-hand side for notational
convenience)
\begin{equation}
\npk(\nu',R_S,\vx)\equiv
\frac{3^{3/2}}{R_1^3}|\det\zeta(\vx)|\,
\delta^{(3)}\!\!\left[\veta(\vx)\right]\theta\!\left[\lambda_3(\vx)\right]
\delta\!\left[\nu(\vx)-\nu'\right]
\label{eq:npkx}
\end{equation}
where
\beq
R_1 \equiv \sqrt{3}\, {\sigma_1 \over \sigma_2}
\eeq
is the characteristic radius of a peak. Note that Eq.~(\ref{eq:npkx}) 
is independent of the redshift $z_0$ for peaks at fixed $\nu$. The 
three-dimensional Dirac delta $\delta^{(3)}\!(\veta)$ ensures that all 
extrema are included. The product of the theta function $\theta(\lambda_3)$, 
where $\lambda_3$ is the lowest eigenvalue of the shear tensor $\zeta_{ij}$, 
and the Dirac delta $\delta(\nu-\nu')$ further restrict the set to density 
maxima with specific height $\nu'$.  The 2-point correlation function for 
maxima of a given significance separated by a  distance 
$r=|\vr|=|\vx_2-\vx_1|$ thus is
\begin{equation}
1+\xpk(\nu,R_S,r)=
\frac{\la\npk(\nu,R_S,\vx_1)\npk(\nu,R_S,\vx_2)\ra}
{\bnpk^2(\nu,R_S)}\;,
\label{eq:2ptkacrice}
\end{equation}
where $\bnpk(\nu,R_S)$ is the differential average number density of 
peaks of height $\nu$ on filtering scale $R_S$ 
\citep{1986ApJ...304...15B},
\begin{equation}
\bnpk(\nu,R_S)=\frac{1}{(2\pi)^2 R_1^3}\,e^{-\nu^2/2}\,
G_0^{(1)}(\gamma_1,\gamma_1\nu)\;.
\label{eq:bnpk}
\end{equation}
Note that it does not depend on $z_0$ (or, equivalently, on the 
amplitude of density fluctuations) at fixed $\nu$.  The function
$G_0^{(\alpha)}(\gamma_1,\gamma_1\nu)$ is defined in Eq.(\ref{eq:Gn}).
For the 2$\sigma$ and 3$\sigma$ density peaks considered here, the
mean abundance is $\bnpk=9.0\times 10^{-5}$ and $4.8\times
10^{-6}\hhhmpc$, respectively. While the calculation of
Eq.(\ref{eq:2ptkacrice}) at first order in the mass correlation and
its derivatives is rather straightforward \citep{2008PhRvD..78j3503D}
(this is Eq.\ref{eq:xpk}), at second order it is quite involved. The 
main steps are detailed in Appendix  \ref{app:xpk2L}. Fortunately, 
most of the terms nicely combine together, and the final result can 
be recast into the compact form
\begin{align}
\label{eq:xpkeasy}
\xpk(\nu,R_S,r) &= \bigl(\bias{I}^2\xi_0^{(0)}\bigr) +\frac{1}{2}
\bigl(\xi_0^{(0)}\bias{II}^2\xi_0^{(0)}\bigr)-\frac{3}{\sigma_1^2}
\bigl(\xi_1^{(1/2)}\bias{II}\xi_1^{(1/2)}\bigr)
-\frac{5}{\sigma_2^2}\bigl(\xi_2^{(1)}\bias{II}\xi_2^{(1)}\bigr)
\biggl(1+\frac{2}{5}\partial_\alpha\ln G_0^{(\alpha)}\!(\gamma_1,\gamma_1\nu)
\Bigl\rvert_{\alpha=1}\biggr) \nonumber \\
&\quad + \frac{5}{2\sigma_2^4}\Bigl[\bigl(\xi_0^{(2)}\bigr)^2+\frac{10}{7}
\bigl(\xi_2^{(2)}\bigr)^2+\frac{18}{7}\bigl(\xi_4^{(2)}\bigr)^2\Bigr]
\biggl(1+\frac{2}{5}\partial_\alpha\ln G_0^{(\alpha)}\!(\gamma_1,\gamma_1\nu)
\Bigl\rvert_{\alpha=1}\biggr)^2 \nonumber \\ 
&\quad +\frac{3}{2\sigma_1^4}\Bigl[\bigl(\xi_0^{(1)}\bigr)^2
+2\bigl(\xi_2^{(1)}\bigr)^2\Bigr]+\frac{3}{\sigma_1^2\sigma_2^2}
\Bigl[3\bigl(\xi_3^{(3/2)}\bigr)^2+2\bigl(\xi_1^{(3/2)}\bigr)^2\Bigr]\;.
\end{align}
In the right-hand side of Eq.(\ref{eq:xpkeasy}), all the correlations
are function of $R_S$, $r$ and $z_0$. More precisely, the first line
contains terms involving first  and second order peak bias  parameters
$\bias{I}$ and $\bias{II}$, the second line has a $\nu$-dependence 
through the function 
$1+(2/5)\partial_\alpha\ln G_0^{\alpha}(\gamma_1,\gamma_1\nu)|_{\alpha=1}$
(which is displayed in Fig.{\ref{fig:lng0}}), and the  last two terms
depend on the separation $r$ (and $R_S$) only. Note that this expression 
exhibits not only terms quadratic in bias parameters but, unlike standard
local bias (Eulerian or Lagrangian), also terms linear in them. These
terms involve derivatives $\xi_{j\ne 0}^{(n)}$ of the linear mass 
correlation $\xi_0^{(0)}$ that vanish at zero lag. They arise because 
the peak correlation depends also on the statistical properties of 
$\eta_i$ and $\zeta_{ij}$.

\begin{figure*}
\center
\resizebox{0.45\textwidth}{!}{\includegraphics{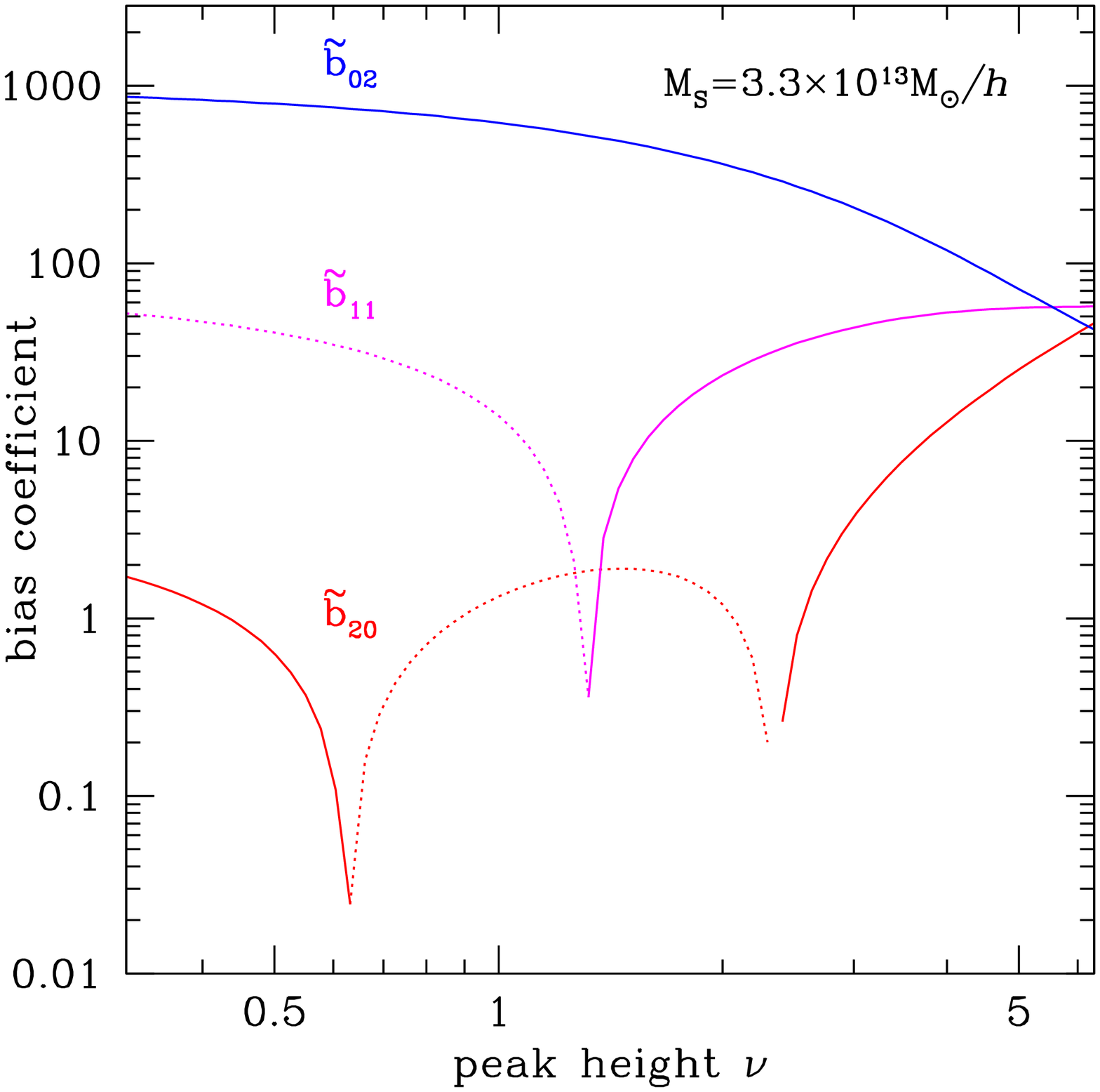}}
\resizebox{0.45\textwidth}{!}{\includegraphics{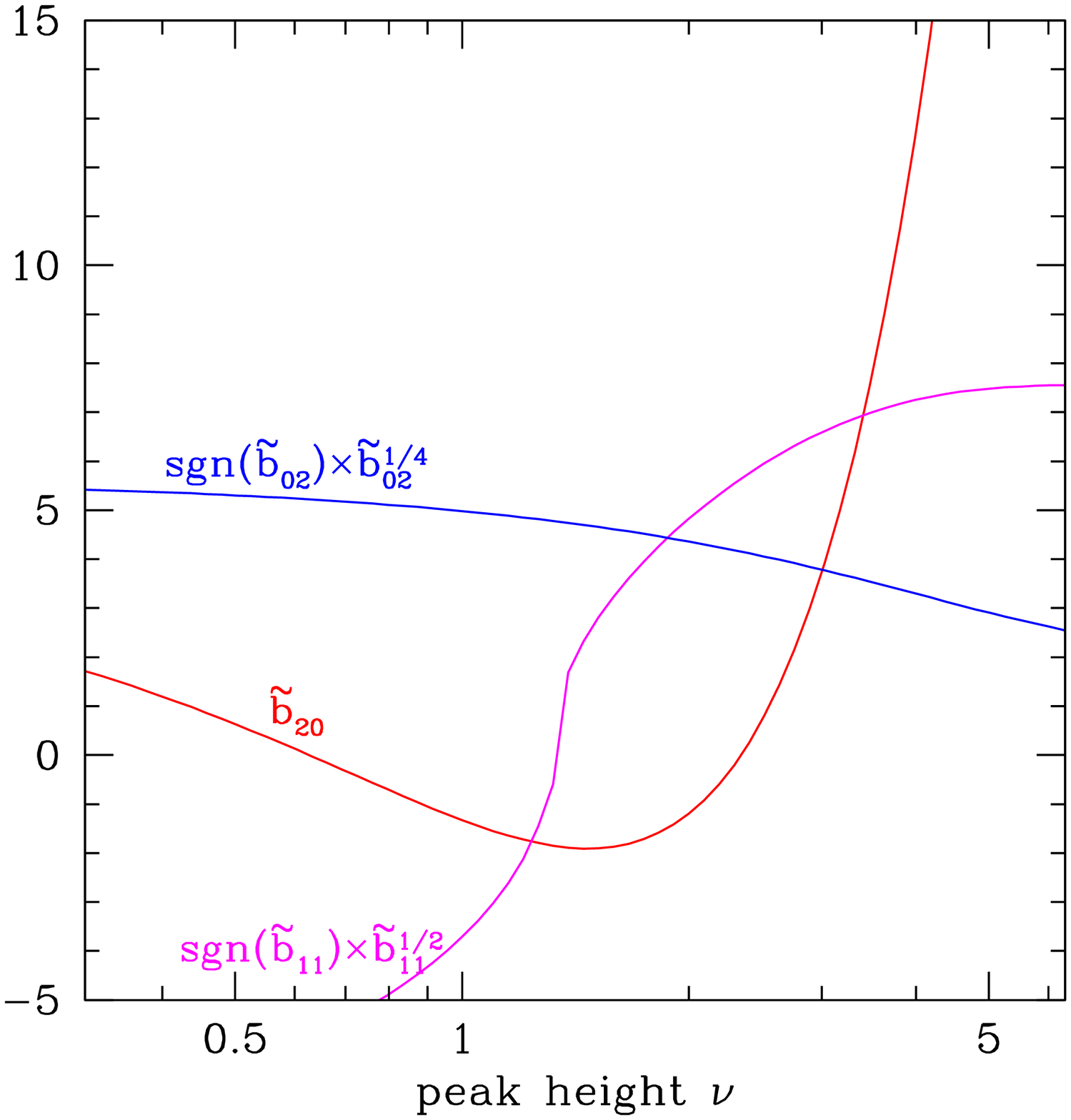}}
\caption{{\it Left panel}~: Lagrangian bias coefficients characterizing 
the second order peak bias $\bias{II}(q_1,q_2)$, Eqs. (\ref{eq:bvv}) 
-- (\ref{eq:bzz}), as a function of peak height for a filtering radius
$R_S=2.9\hmpc$ or, equivalently, a mass scale $M_S=3\times 10^{13}\hmsun$. 
The shape parameter is $\gamma_1\approx 0.65$. 
For the $2\sigma$ peaks considered in subsequent illustrations, 
$\bias{20}$ is negative, $\bias{20}\approx -1.2$. 
{\it Right panel}~: The second  and fourth root $\bias{11}^{1/2}$ 
and $\bias{02}^{1/4}$ define a characteristic scale below which 
the scale dependence of $\bias{II}$ is large. In the limit $\nu\to\infty$, 
$\bias{02}$ becomes negative and converges towards 
$-\sigma_2^{-2}(1-\gamma_1^2)^{-1}$, whereas $\bias{11}$ asymptotes 
to the constant $(\gamma_1/\sigma_1)^2(1-\gamma_1^2)^{-1}$. Note that, in 
contrast to $\bias{11}^{1/2}$ and $\bias{02}^{1/4}$ that 
have units of length, $\bias{20}$ is dimensionless.}
\label{fig:bias}
\end{figure*}

In analogy with $\bias{I}(k,z_0)$, the action of the second order peak 
bias $\bias{II}$ is defined as the Fourier space  multiplication by
\begin{equation}
\label{eq:bnII}
\bias{II}(q_1,q_2,z_0) \equiv \ov{\bspk(q_1,z_0)\bspk(q_2,z_0)}
-\left(1-\gamma_1^2\right)^{-1}\biggl[\frac{1}{\sigma_0^2}
+\frac{(q_1 q_2)^2}{\sigma_2^2}
-\frac{\gamma_1^2}{\sigma_1^2}\bigl(q_1^2+q_2^2\bigr)\biggr]
\equiv \bias{20}+\bias{11}\bigl(q_1^2+q_2^2\bigr)+\bias{02}\,q_1^2 q_2^2\;,
\end{equation}
where $q_1$ and $q_2$ are wavemodes and the coefficients $\bias{20}$, 
$\bias{11}$ and $\bias{02}$ describing the peak bias at 
second order are
\begin{align}
\bias{20}(\nu,R_S,z_0) &\equiv
\bb_{\nu\nu}-\frac{1}{\sigma_0^2\left(1-\gamma_1^2\right)}=
\frac{1}{\sigma_0^2}\left[
\frac{\nu^2-2\gamma_1\nu\bar{u}+\gamma_1^2\ov{u^2}}
{\left(1-\gamma_1^2\right)^2}
-\frac{1}{\left(1-\gamma_1^2\right)}\right] \label{eq:bvv} \\
\bias{11}(\nu,R_S,z_0) &\equiv
\bb_{\nu\zeta}+\frac{\gamma_1^2}{\sigma_1^2\left(1-\gamma_1^2\right)}=
\frac{1}{\sigma_0\sigma_2}\left[\frac{\left(1+\gamma_1^2\right)
\nu\bar{u}-\gamma_1\bigl[\nu^2+\ov{u^2}\bigr]}
{\left(1-\gamma_1^2\right)^2}+\frac{\gamma_1}
{\left(1-\gamma_1^2\right)}\right] \label{eq:bvz} \\
\bias{02}(\nu,R_S,z_0) &\equiv
\bb_{\zeta\zeta}-\frac{1}{\sigma_2^2\left(1-\gamma_1^2\right)}=
\frac{1}{\sigma_2^2}\left[\frac{\ov{u^2}-2\gamma_1\nu\bar{u}
+\gamma_1^2\nu^2}{\left(1-\gamma_1^2\right)^2}
-\frac{1}{\left(1-\gamma_1^2\right)}\right]\;. \label{eq:bzz}
\end{align}
See Fig.~\ref{fig:bias} for the relative size of these contributions 
as a function of peak height. As shown in Appendix \S\ref{app:xpk2L}, 
$\bb_{\nu\nu}$, $\bb_{\nu\zeta}$ and $\bb_{\zeta\zeta}$ arise upon 
averaging the products $b_\nu b_\nu$, $b_\nu b_\zeta$ and 
$b_\zeta b_\zeta$ over the peak curvature. In this respect, 
$\overline{u^n}\equiv G_n^{(1)}(\gamma_1,\gamma_1\nu)/ 
G_0^{(1)}(\gamma_1,\gamma_1\nu)$ is
the $n$th moment of the peak curvature at a given significance
$\nu$. For sake of completeness, $\bias{II}^2$ acts on the functions
$\xi_{\ell_1}^{(n_1)}(r)$ and $\xi_{\ell_2}^{(n_2)}(r)$ according to
\begin{equation}
\bigl(\xi_{\ell_1}^{(n_1)}\bias{II}^2\xi_{\ell_2}^{(n_2)}\bigr)
\equiv\frac{1}{4\pi^4}\int_0^\infty\!\!dq_1\int_0^\infty\!\!dq_2\,
q_1^{2(n_1+1)}q_2^{2(n_2+1)}\bias{II}^2(q_1,q_2,z_0)P_{\delta_S}(q_1,z_0)
P_{\delta_S}(q_2,z_0)j_{\ell_1}(q_1 r) j_{\ell_2}(q_2 r)\;.
\end{equation} 
When $\ell_1=\ell_2=0$, the real space counterpart of
$\bias{II}(q_1,q_2,z_0)$ is readily obtained by making the replacement 
$q^2\to -\partial^2$ (which reflects the fact that
$\xi_0^{(n)}(R_S,r,z_0)$ is a solution to the Helmholtz  equation
$(\partial^2+k^2)\xi(r)=0$).   Eq.(\ref{eq:xpkeasy}) is the main result
of this Section. We note that \cite{1995MNRAS.272..447R} also
computed second order corrections to the peak correlation 
$\xpk(\nu,R_S,r)$ for which, however, they did not provide any explicit 
expression.

Before illustrating the impact of the second order terms on the 
correlation of initial density maxima, we remark that, although 
the calculation of the peak correlation at third order is very involved, 
the contribution proportional to $(\xi_0^{(0)})^3$ can be derived 
relatively easily. We find
\begin{equation}
\frac{1}{6}\left[\bb_{\nu\nu\nu}-\frac{3\,\bv}{\sigma_0^2
\left(1-\gamma_1^2\right)}\right]^2\bigl(\xi_0^{(0)}\bigr)^3\equiv 
\frac{1}{6}\bias{30}^2 \bigl(\xi_0^{(0)}\bigr)^3\;,
\end{equation}
where the third order coefficient $\bb_{\nu\nu\nu}$ is defined as
\begin{equation}
\label{eq:bvvv}
\bb_{\nu\nu\nu}(\nu,R_S,z_0) =
\frac{\nu^3-3\gamma_1\nu^2\bar{u}+3\gamma_1^2\nu\ov{u^2}-\gamma_1^3\ov{u^3}}
{\sigma_0^3\left(1-\gamma_1^2\right)^3}\;.
\end{equation}
Thus, up to third order, the peak correlation may be cast into the form
\begin{equation}
\label{eq:localbpk}
 \xpk(\nu,R_S,r)\approx \bias{10}^2 \xi_0^{(0)}
 +\frac{1}{2}\bias{20}^2\bigl(\xi_0^{(0)}\bigr)^2
 +\frac{1}{6}\bias{30}^2\bigl(\xi_0^{(0)}\bigr)^3
+\mbox{additional terms}\;,
\end{equation}
where the missing terms, while of the same order as the ones we display,
have a more complicated structure (see Eq.(\ref{eq:xpkeasy}) for second
order contributions).
In the limit $\nu\gg 1$, the scale-independent pieces $\bias{10}$,
$\bias{20}$ and $\bias{30}$ to the bias asymptote to the values 
$\bias{10}\to\nu/\sigma_0$, $\bias{20}\to(\nu/\sigma_0)^2$
and $\bias{30}\to (\nu/\sigma_0)^3$ obtained in the high level 
excursion set approximation \citep{1984ApJ...284L...9K}. We will see 
shortly that these bias factors are indeed equal to the peak-background 
split biases derived from the average peak abundance Eq.(\ref{eq:bnpk}).

In Fig.\ref{fig:bias}, the second order Lagrangian biases $\bias{20}$, 
$\bias{11}$ and $\bias{02}$ are shown as a function of the 
peak height for the mass scale $M_S=3.3\times 10^{13}\hmsun$ (left panel). 
The second  and fourth root $\bias{11}^{1/2}$ and 
$\bias{02}^{1/4}$, respectively, define a characteristic comoving 
scale below which the corresponding scale-dependent terms 
$\bias{11}q^2$ and $\bias{02}q^4$ are large. In the limit 
$\nu\to\infty$, the scale-independent piece $\bias{20}$ increasingly 
dominates whereas $\bias{11}$ and $\bias{02}$ tend towards 
the constant value $(\gamma_1/\sigma_1)^2(1-\gamma_1^2)^{-1}$ and
$-\sigma_2^{-2}(1-\gamma_1^2)^{-1}$. For a realistic threshold height
$\nu<4$ however, the scale-dependent contributions cannot be neglected
since $|\bias{11}|$ and/or $|\bias{02}|$ are typically much
larger than $|\bias{20}|$. Although the exact value of the second order 
biases somewhat changes with the mass scale $M_S$, their overall behavior 
varies little over the range $M_S\sim 10^{12}-10^{14}\hmsun$ as $\gamma_1$ 
weakly depends on $R_S$. Therefore, our conclusions hold regardless the 
exact amount of smoothing. It should also be noted that, if one wishes 
to associate these bias factors to halos of mass 
$M_S=3.3\times 10^{13}\hmsun$, then the variation with $\nu$ is in fact 
a variation with redshift. 

\begin{figure*}
\center
\resizebox{0.45\textwidth}{!}{\includegraphics{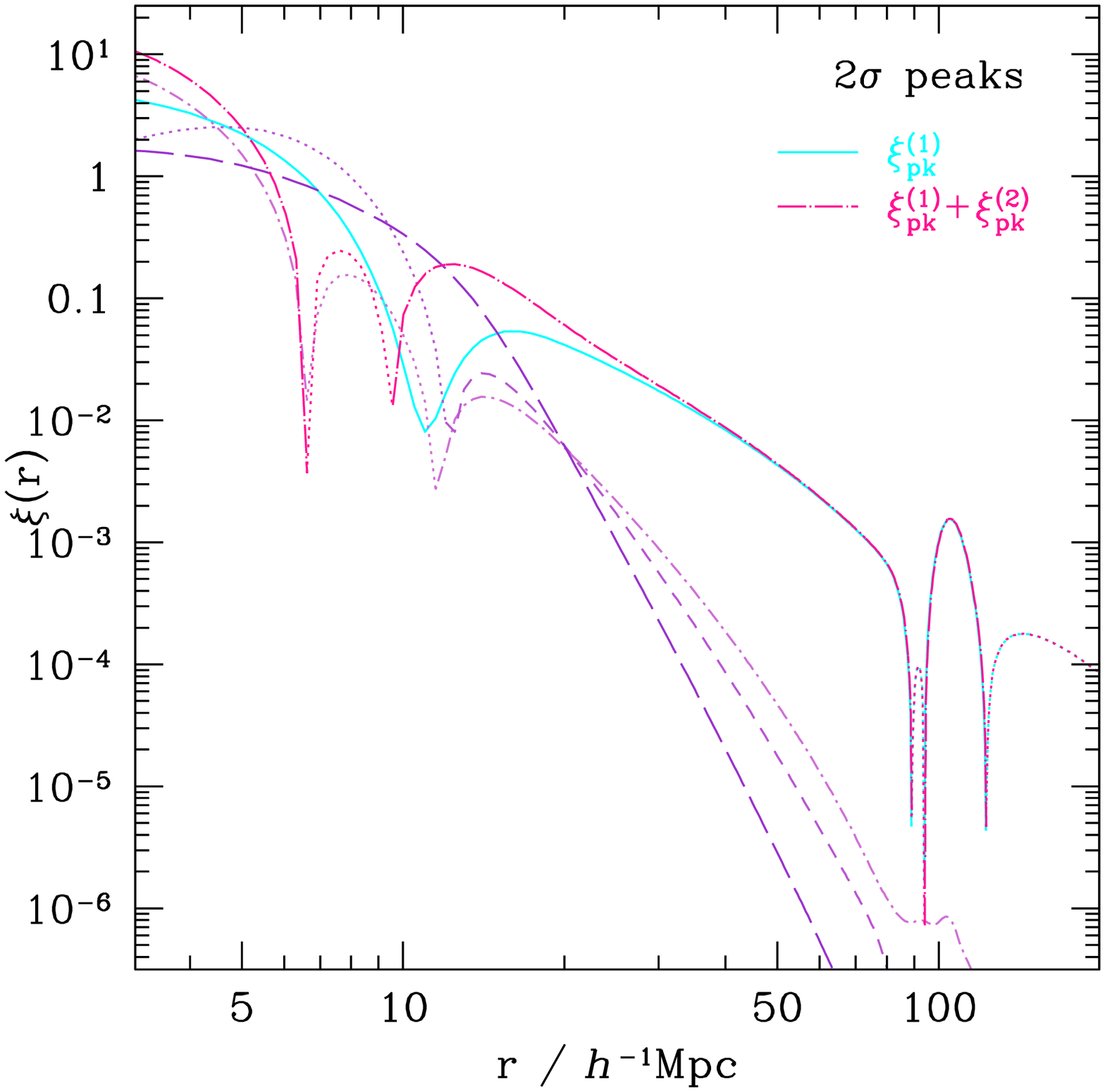}}
\resizebox{0.45\textwidth}{!}{\includegraphics{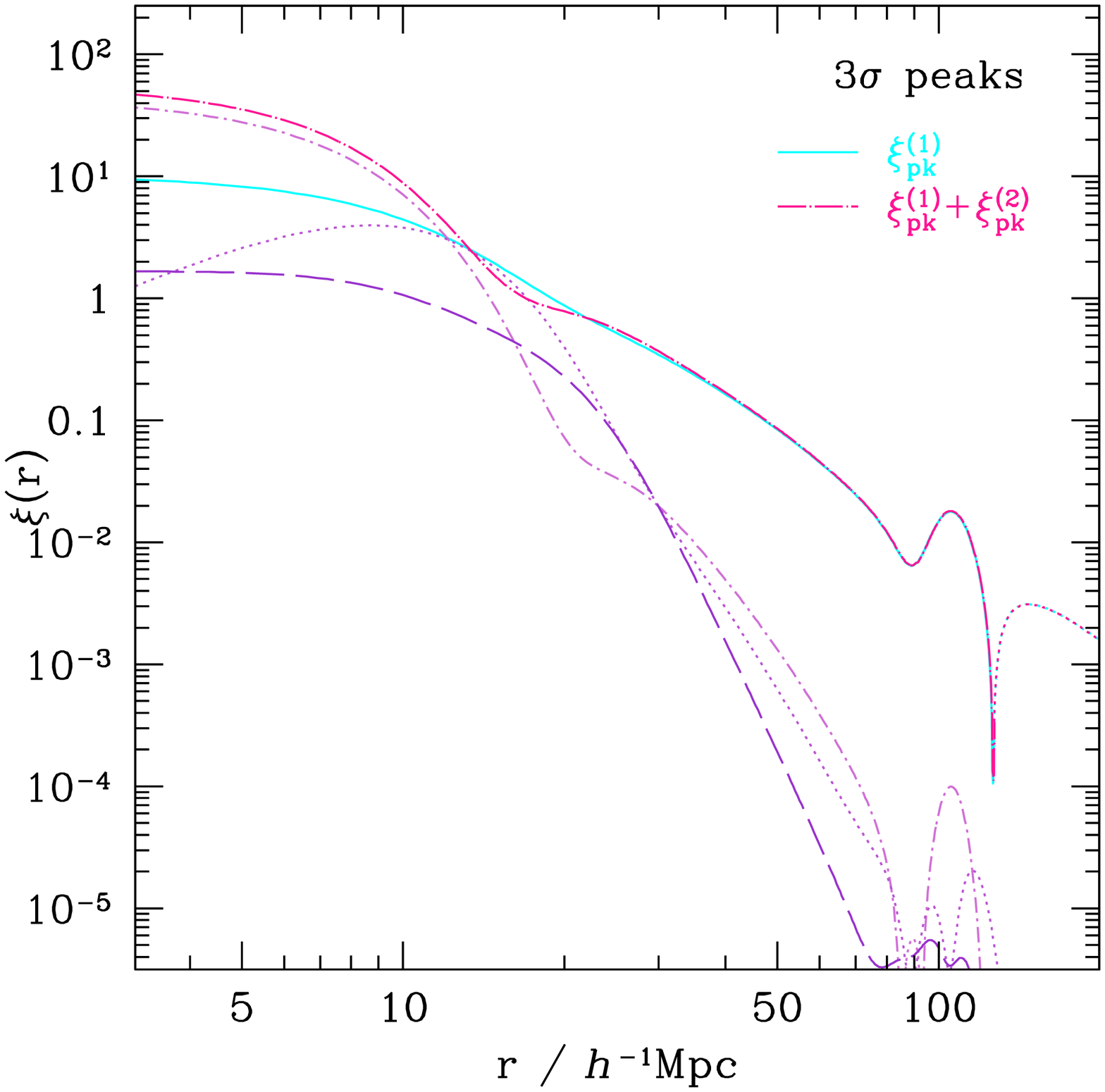}}
\caption{The correlation of initial density peaks at the second order
is shown as the long dashed-dotted (magenta) curve for 2$\sigma$ (left
panel) and 3$\sigma$ (right panel) density peaks collapsing at
redshift $z_0=0.3$ according to the spherical collapse prescription.
For the Gaussian filter used in this paper, this corresponds to a mass 
scale $M_S=3\times 10^{13}$ and $2\times 10^{14}\hmsun$, respectively.  
The individual contributions appearing in Eq.~(\ref{eq:xpkeasy}) are 
shown separately.  Namely, the solid (cyan) curve is the first order 
contribution $\bias{I}^2\xi_0^{(0)}$, whereas the second order term 
quadratic in $\bias{II}$,  linear in $\bias{II}$ and independent
of $\bias{II}$ are shown as the short dashed-dotted, short-dashed and 
long-dashed curve, respectively. A dotted line indicates  negative 
values.}
\label{fig:xpk}
\end{figure*}

The correlation $\xpk(\nu,R_S,r)$ is shown in Fig.\ref{fig:xpk} for 
the 2$\sigma$ and 3$\sigma$ initial density peaks collapsing at 
redshift $z_0=0.3$. The solid (green) curve represents the first order 
term $\bias{I}^2\xi_0^{(0)}$ (Eq.(\ref{eq:xpk})) while the 
long-dashed-dotted curve is the full second order correlation 
(Eq.(\ref{eq:xpkeasy}). We have also plotted the second order 
contributions quadratic in $\bias{II}$, linear in $\bias{II}$ and 
independent of $\bias{II}$ separately. They are shown as the 
short-dashed-dotted, short-dashed and long-dashed curve, respectively. 
Notice that $(\bias{20},\bias{11},\bias{02})=(-1.2,23,363)$ and 
$(7.8,285,3927)$ for the low and high threshold, respectively. For the 
3$\sigma$ peaks, the term linear in $\bias{II}$ is negative over the 
range of distances considered and, thus, appears as a dotted line. In 
fact, the piece linear in $\bias{II}$ is the only negative contribution 
at small separations but it vanishes at zero lag. Since the true peak 
correlation rapidly converges to -1 for $r<R_1$ (as shown by 
\cite{1989MNRAS.238..293L}, 
the small scale behavior of $\xpk$ is dominated by an exponential term 
$\exp(-R_1^2/r^2)$), small scale exclusion should manifest itself in 
higher-order terms, but it is beyond the scope of this paper to calculate 
them. We can also observe that the correlation of 2$\sigma$ peaks is 
negative on scales $r\sim 5-10\hmpc$. However, this is likely an artifact 
of truncating the expansion at second order in the correlations 
$\xi_\ell^{(n)}$.

\begin{figure*}
\center
\resizebox{0.45\textwidth}{!}{\includegraphics{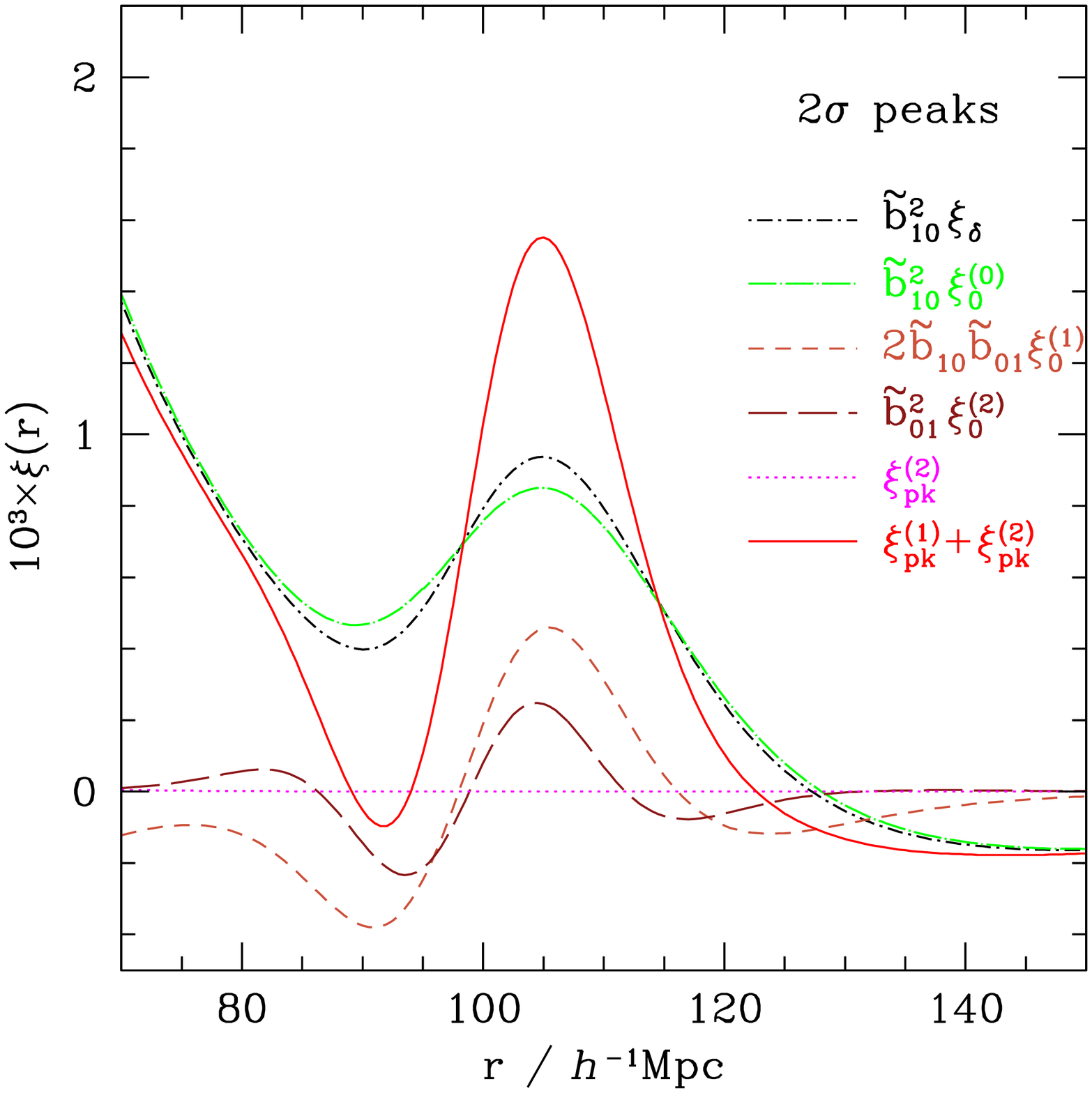}}
\resizebox{0.45\textwidth}{!}{\includegraphics{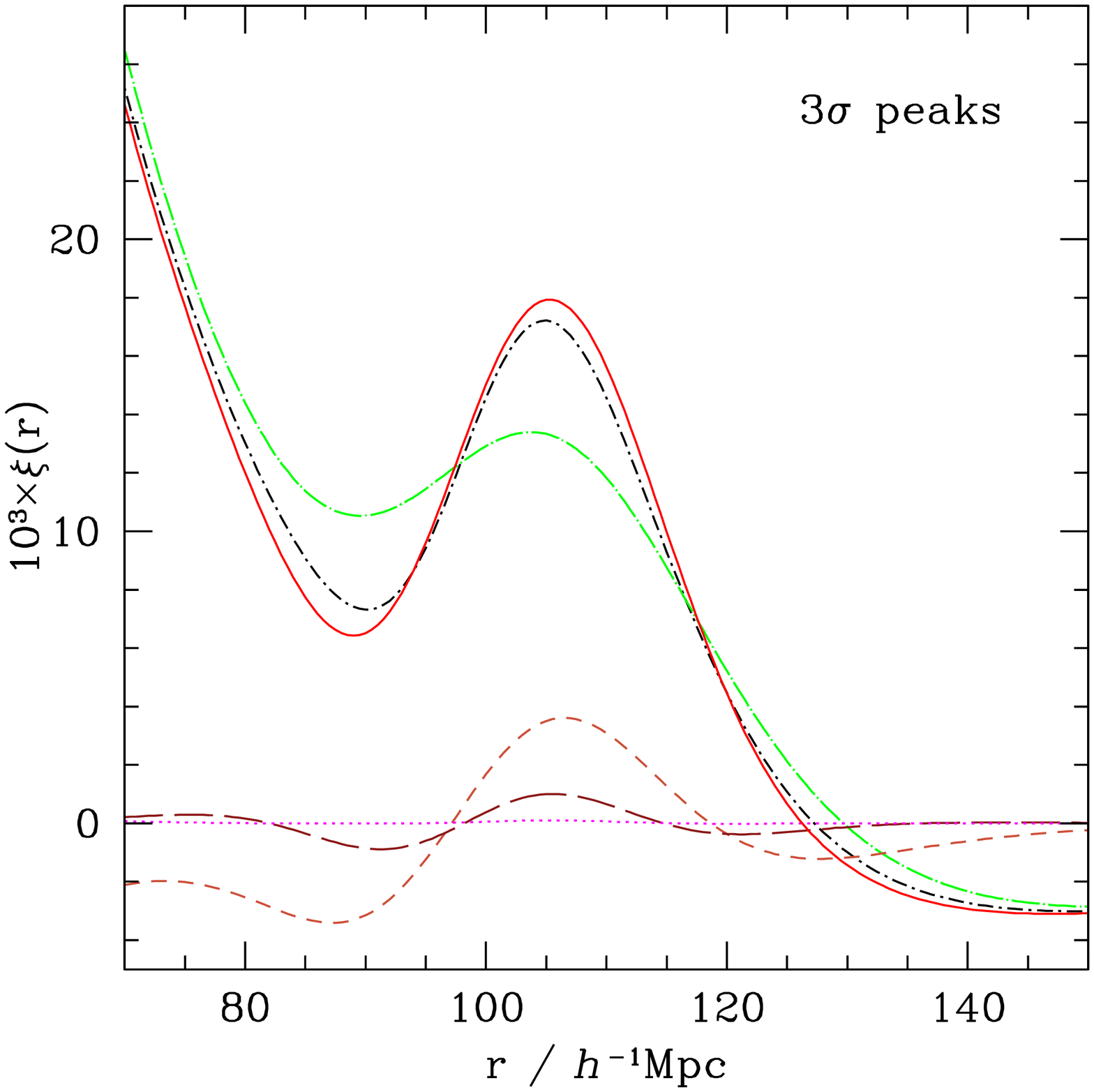}}
\caption{A comparison between the initial {\it unsmoothed} density 
correlation $\xi(r,z_0)$ (black, dotted-dashed)  and the initial peak 
correlation $\xpk(\nu,R_S,r)$ (red, solid) around the BAO. To obtain 
the peak correlation, the density field was smoothed with a Gaussian 
filter on mass scale $M_S=3\times 10^{13}$ (left panel) and 
$2\times 10^{14}\hmsun$ (right panel).  The dotted-long dashed, 
short-dashed and long-dashed curves represent the individual contributions
$\bias{10}^2\xi_0^{(0)}$, $2\bias{10}\bias{01}\xi_0^{(1)}$ and 
$\bias{01}^2\xi_0^{(2)}$ to the first order peak correlation 
(Eq.\ref{eq:xpk}). A nonzero $\bias{01}$ restores, and even amplifies 
the acoustic peak otherwise smeared out upon filtering the mass density 
field. The dotted curve indicates the second order correction to the peak 
correlation. Results are shown for the CDM transfer function considered 
in this paper.}
\label{fig:bao}
\end{figure*}

Figure \ref{fig:bao} focuses on the baryon acoustic oscillation. At
separation $r>80\hmpc$, second order corrections are negligibly small, so 
that $\xpk(\nu,R_S,r)$ is given by Eq.(\ref{eq:xpk}) with an accuracy
better than 1\%. For comparison, we also plot the first order correlation 
$\bias{10}^2\xi_\delta$ arising in a local biasing scheme with same value 
of $\bias{10}$. It is important to note that $\xi_\delta$ is the correlation
of the {\it unsmoothed}, linear mass density field (in practice we use 
$R_S=0.1\hmpc$). It is quite remarkable how the ``sombrero''-shaped terms
$2\bias{10}\bias{01}\xi_0^{(1)}$and $\bias{01}^2\xi_0^{(2)}$ restore 
the contrast of the baryonic feature otherwise smeared out by the large 
filtering (Recall that $R_S=2.9$ and 5.3$\hmpc$ for the 2$\sigma$ and 
3$\sigma$ peaks, respectively). The BAO in the peak correlations is even
sharpened relative to the BAO in $\xi_\delta$. A thorough discussion of 
this effect can be found in \cite{2008PhRvD..78j3503D}. In \S\ref{sec:evol}, 
we will see that, although most of the initial enhancement of the BAO 
contrast is smeared out by the gravitational motions of the peaks, some of 
it survives at the epoch of collapse. 

\subsection{A peak-background split derivation of the peak bias factors}
\label{sub:pksplit}

The peak-background split \citep{1986ApJ...304...15B,1989MNRAS.237.1127C,
1996MNRAS.282..347M,1999MNRAS.308..119S} is a heuristic argument that
furnishes another way to derive the large scale bias of density peaks.
This approach is quite different from ours because it is based on number
counts in configuration space and, thus, does not make reference to the
bias in Fourier space.

There are two ways in which the peak-background split is implemented.  
In the first 
\citep{1986ApJ...304...15B,1989MNRAS.237.1127C,1997MNRAS.284..189M},
the $N$-th order bias parameter $b_N(\nu)$ is related to the 
$N$-th order derivative of the differential number density 
$\bar{n}(\nu)$ of virialized objects according to
\begin{equation}
 b_N(\nu,z_0)\equiv \left(-\frac{1}{\sigma_0(z_0)}\right)^N
 \bar{n}(\nu)^{-1}\,\frac{\partial^N[\bar{n}(\nu)]}{\partial \nu^N}\;,
\label{eq:pksplit}
\end{equation}
with the important caveat that the mass function is universal (i.e. it
depends on $\nu$ solely). For density peaks, on setting 
$\bar{n}=\bnpk(\nu,R_S)$  and performing the derivatives with respect 
to $\nu$ at fixed smoothing radius $R_S$, one obtains 
\citep{1997MNRAS.284..189M}
\begin{equation}
\label{eq:bnN}
 b_N(\nu,R_S,z_0) \equiv \left(-\frac{1}{\sigma_0(R_S,z_0)}\right)^N\,
 \bnpk(\nu,R_S)^{-1}\frac{\partial^N[\bnpk(\nu,R_S)]}{\partial\nu^N}\;. 
\end{equation}
As noted by \cite{2010PhRvD..81b3526D}, the first order peak-background
split bias is
\begin{equation}
\label{eq:bnI}
b_I(\nu,R_S,z_0)=\bias{10}(\nu,R_S,z_0)\;.
\end{equation}
This means that the large scale, constant and deterministic bias factor 
returned by the peak-background split argument is exactly the same as in 
our approach, when we are on large enough scales that the $k$-dependence 
associated with the $\bias{01}k^2$ term can be ignored. It turns out
that Eq.(\ref{eq:bnI}) generalizes as follows:  higher order derivatives 
of the peak number density~(\ref{eq:bnpk}) with respect to $\nu$ 
\citep[which are reported in][]{1997MNRAS.284..189M} result in the large 
scale, $k$-independent peak bias coefficients 
\begin{equation}
b_{II}(\nu,R_S,z_0)=\bias{20}(\nu,R_S,z_0),\qquad 
b_{III}(\nu,R_S,z_0)=\bias{30}(\nu,R_S,z_0),\qquad 
\mbox{etc.}\;.
\end{equation}
However, derivatives of Eq.(\ref{eq:bnpk}) cannot produce the 
$k$-dependent bias terms like $\bias{01}$, $\bias{11}$ etc.,
which arise owing to the constraints imposed by derivatives of the mass
density field.  

Therefore, we will now consider the second implementation of the 
peak background split \citep{1996MNRAS.282..347M,1999MNRAS.308..119S}
in which the dependence of the mass function on the overdensity of the 
background is derived explicitly. The ratio of this conditional mass 
function to the universal one is then expanded in powers of the 
background density.  The bias factors are the coefficients of this 
expansion. We will demonstrate below that this is the correct approach to 
recover the scale or $k$-dependence of the peak bias parameters.

The key quantity is the average number density of peaks identified on 
scale $R_S$ as a function of the overdensity $\delta_B$ defined on 
another smoothing scale $R_B$ (we use the subscript $B$ because we are 
mainly be interested in the regime in which the scale $R_B$ of the
background satisfies $R_B \gg R_S$). This conditional peak number 
density is 
\begin{equation}
 \bnpk(\nu,R_S|\delta_B,R_B) = 
 \frac{G_0^{(0)}(\tilde\gamma_1,\tilde\gamma_1\tilde\nu)}{(2\pi)^{3/2} R_1^3}\,
   \frac{\exp[-(\nu - \epsilon\nu_B)^2/2(1 - \epsilon^2)]}
        {\sqrt{2\pi (1 - \epsilon^2)}}\;,
\label{eq:bnpk|bkgnd}
\end{equation}
where 
\begin{gather}
  \label{eq:pbsvar1}
 \nu_B \equiv \frac{\delta_B}{\sigma_{0B}}, \qquad 
 \langle\nu \nu_B\rangle \equiv \epsilon 
           =  \frac{\sigma_{0\times}^2}{\sigma_{0S}\sigma_{0B}}, \qquad
 \langle u \nu_B\rangle \equiv \gamma_1\epsilon r, \qquad 
      r \equiv \frac{\langle k^2\rangle_\times}{\langle k^2\rangle_S} 
         = \frac{\sigma_{1\times}^2/\sigma_{1S}^2}
         {\sigma_{0\times}^2/\sigma_{0S}^2}, \\
  \label{eq:pbsvar2}
 \la u|\nu,\nu_B\ra \equiv \tilde\gamma_1\tilde\nu 
       = \gamma_1\nu \biggl(\frac{1-\epsilon^2 r}{1-\epsilon^2}\biggr) - 
         \gamma_1\biggl(\frac{1-r}{1-\epsilon^2}\biggr) \epsilon\nu_B, 
         \qquad
 {\rm Var}(u|\nu,\nu_B) \equiv 1 - \tilde\gamma_1^2 \equiv 1 - \gamma_1^2 \,
  \left[1 + \epsilon^2 \frac{(1-r)^2}{1-\epsilon^2}\right],
\end{gather}
with $\sigma_{nS}\equiv \sigma_n(R_S,z_0)$, 
$\sigma_{nB}\equiv\sigma_n(R_B,z_0)$, and we have defined 
\begin{equation}
 \sigma_{n\times}^2 \equiv \frac{1}{2\pi^2}
  \int_0^\infty\!\!dk\, k^{2(n+1)}\,P_\delta(k)\,W(kR_S)W(kR_B)
\end{equation}
\citep[see equation~E5 of][]{1986ApJ...304...15B}. Here, the $\times$ 
denotes the splitting of smoothing scales, i.e. one filter is of size 
$R_S$, the other of size $R_B$. We have deliberately written $r$ as an 
average of $k^2$ to emphasize that we naively expect it to give rise 
to the $k^2$ dependence of peak bias. This will eventually be proven
correct. In addition, note that $\epsilon \nu_B 
= (\delta_B/\sigma_{0S})\,(\sigma_{0\times}^2/\sigma_{0B}^2)$.  
In what follows, it will be convenient to define 
$\Sigma^2_{\times B} \equiv \sigma_{0\times}^2/\sigma_{0B}^2$. 
When $R_B\gg R_S$, this ratio is of order unity (there is a form factor 
that depends on the shape of the smoothing filter).  

Notice that the integral of Eq.~(\ref{eq:bnpk|bkgnd}) over all $\delta_B$ 
gives the unconditional number density $\bnpk(\nu,R_S)$ of 
Eq.~(\ref{eq:bnpk}).  The peak background split expands the ratio 
$\bnpk(\nu,R_S|\delta_B,R_B)/\bnpk(\nu,R_S)$ in powers of $\delta_B$.  
This ratio is then interpreted as representing the average 
overabundance of peaks in regions which have mass overdensity 
$\delta_B$ although, strictly speaking, it is a statement about 
cells of overdensity $\delta_B$ that have a peak at their center.  
Therefore, it is {\em not} a statement about randomly placed cells, 
even though, as we discuss below, it is often treated as such.  

If we set $r\to 0$ and $\epsilon\to 0$ then the coefficient of 
the term of order $\delta_B$ gives $\bias{10}$, that of order gives 
$\delta_B^2$ gives $\bias{20}$, etc.  Note that in this limit, 
$\tilde\gamma_1\to \gamma_1$ and 
$\tilde\gamma_1\tilde\nu \to \gamma_1\nu (1 - \delta_B/\delta_S)$, 
so that expanding Eq.(\ref{eq:bnpk|bkgnd}) in powers of $\delta_B$ 
will be the same as differentiating Eq.(\ref{eq:bnpk}) with respect 
to $\delta_S$.  These derivatives result in the large scale, 
$k$-independent, peak bias factors we have been denoting as 
$\bias{10}$, $\bias{20}$ etc.  As noted above, however, these 
derivatives cannot produce the $k$-dependent bias terms like 
$\bias{01}$, $\bias{11}$, etc.  

Setting $\epsilon\to 0$ but keeping the $r$ dependence means 
that $\tilde\gamma_1\to \gamma_1$ but $\tilde\gamma_1\tilde\nu \to 
 \gamma_1\nu [1 - (1-r)(\delta_B/\delta_S)\Sigma^2_{\times B}]$.  
As a result, derivatives of $G_0^{(0)}$ with respect to $\delta_B$ 
will introduce terms which depend on $r$; these are terms which 
could {\em not} have been obtained by differentiating the 
unconditional mass function.  For example, to first order in $\delta_B$, 
\begin{equation}
 G_0^{(0)}(\tilde\gamma_1,\tilde\gamma_1\tilde\nu) \approx 
 G_0^{(0)}(\gamma_1,\gamma_1\nu)\, 
 \biggl[1 - (1-r)\,\frac{(\bar{u} - \gamma_1\nu)\gamma_1\nu}{1-\gamma_1^2}\,
       \frac{\delta_B}{\delta_S} \Sigma_{\times B}^2 \biggr]
\end{equation}
so that Eq.(\ref{eq:bnpk|bkgnd}) becomes 
\begin{equation}
 \bnpk(\nu,R_S|\delta_B,R_B) \approx \bnpk(\nu,R_S)\,
 \biggl[1 + \nu \frac{\delta_B}{\sigma_{0S}}\,\Sigma_{\times B}^2
         - (1-r)\,\frac{(\bar{u} - \gamma_1\nu)\gamma_1\nu}{1-\gamma_1^2}\,
           \frac{\delta_B}{\delta_S} \Sigma_{\times B}^2 \biggr] .
\end{equation}
Hence, in this limit, 
\begin{align}
  \label{eq:mpkdensity}
 \la\dpk|\delta_B\ra 
  &\equiv \frac{\bnpk(\nu,R_S|\delta_B,R_B)}{\bnpk(\nu,R_S)} - 1 
  \approx \frac{\delta_B}{\sigma_{0S}}\,\Sigma_{\times B}^2
        \left(\nu-\,\frac{(\gamma_1\bar{u}-\gamma_1^2\nu)}{1-\gamma_1^2}\right)
          + r\gamma_1\nu\,\frac{\delta_B}{\delta_S} \Sigma_{\times B}^2 \,
               \frac{(\bar{u}-\gamma_1\nu)}{1-\gamma_1^2}\, \nonumber\\
  &= \frac{\sigma_{0 \times}^2}{\sigma_0^2(R_B)}
            \frac{\delta_B}{\sigma_{0S}}\,
            \left(\frac{\nu-\gamma_1\bar{u}}{1-\gamma_1^2}\right)
          + \frac{\sigma_{1\times}^2}{\sigma_{0B}^2} 
            \frac{\delta_B}{\sigma_{2S}} 
            \left(\frac{\bar{u}-\gamma_1\nu}{1-\gamma_1^2}\right)\,
  = \frac{\sigma_{0\times}^2}{\sigma_{0B}^2}\, \bias{10} \, \delta_B 
          + \frac{\sigma_{1\times}^2}{\sigma_{0B}^2}\, \bias{01}\,\delta_B 
\end{align}
where $\bias{10}$ and $\bias{01}$ were defined in Eq.(\ref{eq:bias1st}).  
Therefore, the cross correlation between the average overdensity of peaks 
defined on scale $R_S$ and the mass overdensity on scale $R_B$ is
\begin{equation}
 \left\la\dpk\delta_B\right\ra(R_B) = \sigma_{0\times}^2 \, 
 \bias{10} + \sigma_{1\times}^2\, \bias{01} \qquad 
 {\rm on\ large\ scales\ } R_B\gg R_S.
 \label{bxlargeR}
\end{equation}
Note that this final expression is the Fourier transform of 
$(\bias{10} + \bias{01} k^2) P_\delta(k,z_0)W(kR_S)W(kR_B)$.  
Thus, we have shown explicitly that our implementation of the peak 
background split argument has produced the same linear, scale dependent 
bias as the Fourier space argument.  

At this point, one may be worried there is an inconsistency in the
above expansion since $\epsilon\to 0$ necessarily implies $r\to 0$. For
example, for Gaussian smoothing of a powerlaw spectrum $P(k)\propto k^n$, 
we find $r\simeq 2(R_S/R_B)^2$ and $\epsilon\simeq (2 R_S/R_B)^{3+n/2}$ 
at sufficiently large $R_B$ \citep{1986ApJ...304...15B}. For a realistic 
spectral index $-3<n<3$, $\epsilon$ and, obviously, $r$ always vanish in 
the limit $R_B\to\infty$. In fact, there is no problem here since one
can, at least formally, treat $\epsilon$ and $r$ as two independent, 
small parameters in addition to $\delta_B$. Hence, setting $r\to 0$ at 
fixed value of $\delta_B$ and $\epsilon$ for instance corresponds to 
retaining the change in the small scale density $\delta_S$ while 
neglecting the change in the curvature $u$ induced by the background 
wave $\delta_B$. 

The higher order bias factors can be derived in an analogous way. Each 
term of order $\delta_B^N$ will include terms of order $r^m$ with 
$m\le N$. Terms proportional to $r$ give $\bias{01}, \bias{11}$, 
etc.; terms proportional to $r^2$ give $\bias{02}$, $\bias{12}$, etc. 
Thus, our analysis provides a simple way of determining all the additional 
$k$-dependent higher-order bias terms which arise in the peaks model. The 
$k$-dependent polynomials associated to these bias factors are determined 
from symmetry considerations. For instance, $\bias{12}$ multiplies the 
polynomial $q_1^2 q_2^2+q_1^2 q_3^2+q_2^2 q_3^2$. As a general rule, the 
bias factor $\bias{ij}$ is associated to the monomonial symmetric function 
$m_{(1_1,\cdots,1_j)}(q_1^2,\cdots,q_{i+j}^2)$. 

Our peak-background split derivation is very interesting for the 
following reasons.  Firstly, all the analytic models of halo/galaxy 
clustering {\it assume} that the (scale-independent) bias coefficients 
multiplying  powers of the mass correlation function $\xi_0^{(0)}$ 
(Eq.~\ref{eq:localbpk}) are the peak-background split biases  
(Eq.~\ref{eq:pksplit}). In the peak model, this equivalence can be 
derived from first principles. Secondly, the peak-background split 
holds even though the mean number density $\bnpk(\nu,R_S)$ is not 
universal (due to its $R_S$ dependence). For a spherical collapse
prescription $\nu=\dsc(z_0)/\sigma_{0S}$, this implies that it is 
the abundance of virialized objects of mass $M_S$ as a function of
collapse redshift $z_0$ which is related to the bias parameters, and 
not the mass function -- the abundance as a function of mass $M_S$ -- 
at fixed redshift $z_0$.
Thirdly, when correctly implemented, the second of our prescriptions 
yields the correct scale-dependent bias factors, despite the 
non-universality which comes from the fact that the background density 
is correlated with the curvature. 
And finally, the peak-background split approximation has been shown 
to provide a good description of large scale peak bias in simulations 
\citep{1997MNRAS.284..189M,2001PASJ...53..155T} although, recently, 
deviations at the 5-10\% level have been reported 
\citep{2010MNRAS.402..589M,2010arXiv1001.3162T}.
Since our expressions reproduce this limit, we have confidence that
our approach will furnish a good  approximation at the smaller scales
where the bias parameters become scale-dependent (which we showed is 
reproduced by the peak-background split). 

Summarizing, we have shown how to determine the cross correlation 
between the peak point process and the smoothed, linear mass overdensity 
$\delta_B$ order by order.  In fact, because we have the full expression 
for the correlation between peaks and the surrounding density field, this 
cross-correlation can be computed exactly.  
The calculation simplifies by noting that the integral to be done is 
$f(u)\,g(u|\nu,\nu_B)\,g(\nu|\nu_B)\,\nu_B\, g(\nu_B)$, where $g$ is a 
Gaussian variate. This can be rewritten as 
$f(u)\,g(\nu)\,g(u|\nu)\,\nu_B\, g(\nu_B|\nu,u)$. Assuming that $\nu_B$ 
ranges over all $[-\infty,\infty]$ simplifies the integrals to compute, 
and one finds 
\begin{equation}
  \label{eq:peakmass}
 \sigma^2_{{\rm pk},\delta}(R_B) = 
   \sigma_{0\times}^2 \,\bias{10} + \sigma_{1\times}^2\,\bias{01} 
   \qquad {\rm at\ all\ scales}.
\end{equation}
This expression, which is exact to all orders, coincides with the average 
mass profile around an initial density peak 
\citep{1986ApJ...304...15B,2010PhRvD..81b3526D}. 
This average density profile has been shown to provide an accurate 
description of the cross-correlation between the Stokes $Q$ and $U$ 
polarization parameters and the temperature peaks of (two-dimensional) 
Gaussian CMB maps \citep{2010arXiv1001.4538K}.

\subsection{Relation to the local bias model}
\label{sub:localbias}

The local bias model is commonly used to model the clustering of dark 
matter halos and galaxies. In its Lagrangian formulation, this model
assumes that the tracer overdensity $\delta_{\rm h}$ of halos of mass
$M_S$ is a local deterministic function of the linear mass density 
field smoothed on scale $R_B$. This function may be written as the 
Taylor series \citep[e.g.,][]{1993ApJ...413..447F}
\begin{equation}
\delta_{\rm h}(\nu,M_S,R_B,\vx) = 
b_0 + \sum_{N=1}^\infty b_N(\nu)\frac{\delta_B^N(\vx)}{N!}\;,
\label{eq:localbias}
\end{equation}
where the dependence on $M_S$ arises through the peak-background split 
bias parameters $b_N(\nu)$ solely. The value of $b_0$ is set by requiring 
$\la\delta_{\rm h}(\nu,M_S,R_B,\vx)\ra=0$. The choice of $R_B$ is 
arbitrary as long as the constraint $R_B\gg R_S$ is satisfied. The 
large scale clustering is then independent of the smoothing on scales
 $k R_B\ll 1$ (see \cite{2007PhRvD..75f3512S,2008PhRvD..78h3519M}; see 
also \cite{2006PhRvD..74j3512M} for a discussion of smoothing in the 
context of local Eulerian biasing).
Therefore, the cross correlation between the tracers overabundance and 
the mass overdensity in spheres of radius $R_B$ is
\begin{equation}
  \label{eq:crosslocal}
 \sigma^2_{{\rm h},\delta}(R_B) 
  \equiv \left\la \delta_B(\vx)\,\delta_{\rm h}(\nu,M_S,R_B,\vx)\right\ra 
    = \sum_{N=1}^\infty b_N(\nu)\frac{\la\delta_B(\vx)\delta_B^N(\vx)\ra}{N!}
    \approx \left(b_I+ \frac{1}{2}b_{III}\,\sigma_{0B}^2\right)
    \sigma_{0B}^2 \;.
\end{equation}
A comparison with Eq.(\ref{eq:peakmass}) reveals that, even on the 
largest scales, the cross correlation between the peaks and the mass 
distributions, $\sigma_{{\rm pk},\delta_B}^2
\approx b_I\sigma_{0\times}^2$, differs from the local bias expression, 
which involves all the bias factors with odd $N$ values.  This can be 
traced to the fact that the peaks expression is for cells centered on 
peaks, and only the mass field is smoothed on scale $R_B$, whereas the 
local bias expansion is for randomly placed cells, where both the 
tracer field and the mass have been smoothed on scale $R_B$.  

For similar reasons, the auto-correlation function, at second order in 
the local bias model, is
\begin{equation}
  \label{eq:autolocal}
  \xi_{\rm h}(\nu,M_S,R_B,r) \approx b_I^2\,\xi_0^{(0)}\!(R_B,r,z_0)
  + \frac{1}{2}b_{II}^2\,\bigl[\xi_0^{(0)}\!(R_B,r,z_0)\bigr]^2\;,
\end{equation}
where we have ignored extra terms involving powers of $\sigma_{0B}^2$
which arise owing to the continuous nature of the bias relation
(\ref{eq:localbias}). 

This demonstrates that, for large separations $r\gg R_B$, the peak 
correlation function $\xpk(\nu,R_S,r)$ is consistent with that of the 
local bias model although, for peaks, the bias factors are $k$-dependent. 
It is pretty clear from Eqs.(\ref{eq:autolocal}) and (\ref{eq:localbpk}) 
that the local bias scheme is a special case of the more general peak model 
which, on large scales, approaches a local, 
deterministic, scale independent relation only in the high peak limit 
$\nu\gg 1$. Notwithstanding this, the previous analysis shows that
\begin{equation}
 \xpk(\nu\gg 1,R_S,r)\approx b_I^2\xi_0^{(0)}\!(R_S,r,z_0)
 +\frac{1}{2}\left(b_{II}^2+\frac{2 b_{II}\gamma_1^2}
   {\sigma_1^2\left(1-\gamma_1^2\right)}\partial^2\right)
 \Bigl[\xi_0^{(0)}\!(R_S,r,z_0)\Bigr]^2+\mbox{other terms}\;.
\end{equation}
on smaller scales where the $k$-dependence of peak bias matters. As can
be seen, the second order peak bias (the term in curly brackets) remains 
scale-dependent even for the highest peaks. Therefore, if the peaks model 
is correct, then we shall expect deviations from the local bias model on 
mildly nonlinear scales $k\sim 0.01-0.1\hmmpc$, even for the most prominent 
density maxima.

\section{Gravitational evolution of the peak correlation function}
\label{sec:evol}

Thus far, we have explored the scale dependence of bias in the
2-point correlation  of local maxima of the primordial density field.
However, nonlinear collapse and pairwise motions induced by 
gravitational instabilities will distort the initial correlation. 
Since the precise calculation of the dynamical evolution of 
$\xpk(\nu,R_S,r)$ is rather involved, we will assume that the initial 
density peaks are test particles which flow locally with the dark matter
according to the Zel'dovich approximation (i.e. peaks move along
straight lines). We will calculate the correlation of their positions as
a function of redshift and show that we recover a velocity damping factor
and a mode-coupling power similar to that found in (Eulerian) 
renormalized perturbation theory \citep[RPT, see][]{2006PhRvD..73f3519C}.

\subsection{The peak correlation function in the Zel'dovich approximation}
\label{sub:zaevol}

In the Lagrangian approach, the Eulerian comoving position and proper
velocity of a density peak can generally be expressed as a mapping
\begin{equation}
\vxpk(z)=\vq_{\rm pk}+\vvs(\vq_{\rm pk},z)\;,\qquad 
\vvpk(z)=a(z)\,\dot{\vvs}(\vq_{\rm pk},z)\;,
\label{eq:map} 
\end{equation}
where $\vq_{\rm pk}$ is the initial peak position, $\vvs(\vq,z)$ is
the displacement field, $a$ is the scale factor and the peak velocity 
is in standard units (i.e. not scaled by $aHf$). A dot denotes a
derivative with respect to cosmic time.  We will assume that the local 
maxima are test particles that do not interact with each other. 
Therefore, at the first order, the peak position is described by the 
Zeldovich approximation \citep{1970A&A.....5...84Z}, in which the 
displacement factorizes into a time and a spatial component,
\begin{equation}
\vvs=-D(z)\grad\Phi(\vq),~~~\dot{\vvs}=-\beta(z)\grad\Phi(\vq)\;.
\label{eq:za}
\end{equation} 
Here, $D(z)$ is the growth factor of linear mass density perturbations
and $\Phi(\vq)$ is the perturbation potential linearly extrapolated
to present time. Explicitly, $\Phi(\vq)=\phi(\vq,z)/4\pi G\rb(z)a^2
D(z)$ where $\phi(\vq,z)$ is the Newtonian gravitational potential and
$\rb(z)$ is the average matter density. Finally, $\beta(z)=HDf$, where
$H(z)$ is the Hubble constant, is proportional to the logarithmic 
derivative $f=d\ln D/d\ln a$. Note that $f(z)$ scales as 
$\Omega_m(z)^{0.6}$ for a wide range of CDM cosmologies 
\citep{1980lssu.book.....P}.

Let us now consider an ensemble of realizations of initial peak
positions.   The correlation function $\xpk(\nu,R_S,r,z)$ (we will
henceforth omit the dependence on $\nu$ and $R_S$ for brevity) is
related to the zeroth moment of  the joint probability
$P_2(\vv_1,\vv_2;\vr,z|{\rm pk})$ to have a pair of peaks  separated
by a distance $\vr$ and with normalized velocities $\vv_1$  and
$\vv_2$. Following \cite{1996ApJ...472....1B}, we write
\begin{equation}
\bnpk^2\left[1+\xpk(r,z)\right]=\int\!\!d^3\vv_1d^3\vv_2\,
P_2(\vv_1,\vv_2;\vr,z|{\rm pk})\;.
\end{equation}
As we will see shortly, even though the probability
$P_2(\vv_1,\vv_2;\vr,z|{\rm pk})$ depends upon the distance
$r=\vr\cdot\rvh$ {\it and} the unit direction vector $\rvh$, the peak
correlation depends only on $r$.  When the peak motions are governed
by the Zel'dovich approximation Eq.(\ref{eq:za}), we can easily relate
$P_2(\vv_1,\vv_2;r,z|{\rm pk})$ to the joint probability distribution
at the initial redshift $z_i\gg 1$,
\begin{equation}
\label{eq:pk2if}
P_2(\vups_1,\vups_2;r,z|{\rm pk})=P_2(\vups_1,\vups_2;\vr-\sigma_v
\Delta \vups_{12},z_i|{\rm pk})
=\int\!\!d^3\vr'\,
\delta^{(3)}\!\left(\vr'-\vr+\sigma_v\Delta\vups_{12}\right)
P_2(\vups_1,\vups_2;\vr',z_i|{\rm pk})\;,
\end{equation}
where $\sigma_v\equiv\sigma_{-1}(z)$ is the 3-dimensional rms velocity
variance of the matter, and $\Delta\vups_{12}$ is the velocity
difference  $\vups_2-\vups_1$. Note that in this equation and those that 
follow, velocities have been scaled by $aHf\sigma_{-1}$, i.e., 
$\vups\equiv\vv/\sigma_{-1}$. Eq.(\ref{eq:pk2if}) is a consequence of 
Liouville's theorem, which states that the phase space density of peaks
is conserved \citep[so, as shown in][one can easily obtain a differential 
equation describing the evolution of the $n$-particle distribution functions]
{1996ApJ...472....1B}. It is especially useful because we know
how to calculate 2-point distributions subjects to the peak constraint
in the Gaussian initial conditions,
\begin{equation}
P_2(\vups_1,\vups_2;\vr',z_i|{\rm pk})=
\int\!\!d^6\zeta_1d^6\zeta_2\,\npk(\vx_1')\npk(\vx_2')
P_2(\vw_1,\vw_2;\vr',z_i)\;.
\end{equation}
Here, $P_2$ is a joint-probability for the 13-dimensional vector 
$\vw=(\upsilon_i,\eta_i,\nu,\zeta_A)$ of variables at position $\vx_1'$ and
$\vx_2'$, and $\npk(\vx)$ is the Klimontovitch density Eq.(\ref{eq:npkx}).
Expressing the Dirac delta as the Fourier transform of a uniform
distribution, we find
\begin{align}
\label{eq:xpkz1}
\lefteqn{\bnpk^2\left[1+\xpk(r,z)\right]} \\
&\quad=\int\!\!\frac{d^3\vk}{(2\pi)^3}\int\!\!d^3\vr'\,
e^{i\vk\cdot(\vr-\vr')}
\int\!\!d^6\zeta_1d^6\zeta_2\,\npk(\vx_1')\npk(\vx_2')
\left\{\int\!\!d^3\vups_1d^3\vups_2\,P_2(\vw_1,\vw_2;\vr',z_i)\,
e^{i\sigma_v\vk\cdot\Delta\vups_{12}}\right\} \nonumber \\
&\quad=\int\!\!\frac{d^3\vk}{(2\pi)^3}\int\!\!d^3\vr'\,
e^{i\vk\cdot(\vr-\vr')}
\int\!\!d^6\zeta_1d^6\zeta_2\,\npk(\vx_1')\npk(\vx_2') 
P_2(\vy_1,\vy_2;\vr',z_i)
\left\{\int\!\!d^3\vups_1d^3\vups_2\,
P_2(\vups_1,\vups_2|\vy_1,\vy_2;\vr',z_i)\,
e^{i\sigma_v\vk\cdot\Delta\vups_{12}}\right\} \nonumber \;,
\end{align}
where the vector $\vy$ corresponds to $(\eta_i,\nu,\zeta_A)$. The second 
equality follows from Bayes' theorem. To integrate over the velocities, 
we use the identity
\begin{equation}
\label{eq:probid}
\int\!\!d^N\vy\,y_{i_1}\cdots y_{i_n}\,P(\vy)\, e^{i\vj\cdot\vy} 
= (-i)^n\frac{\partial}{\partial J_{i_1}}\cdots\frac{\partial}
{\partial J_{i_n}}
\exp\left(-\frac{1}{2}\vj^\dagger\Sigma\vj+i\vj^\dagger\Xi\right)\;,
\end{equation}
which follows from the relation 
\begin{equation}
P(\vy)=\frac{1}{(2\pi)^N}\int\!\!d^N\!\vj\,
\exp\left[-i\vj^\dagger\Bigl(\vy-\Xi\Bigr)
-\frac{1}{2}\vj^\dagger\Sigma\vj\right]\;.
\end{equation}
Here, $P(\vy)$ is a $N$-dimensional Gaussian multivariate of
covariance matrix $\Sigma=\la\vy\vy^\dagger\ra$ and centered at
$\vy=\Xi$. Eq.(\ref{eq:probid}) is a very useful relation since it
allows us to circumvent the inversion of the covariance matrix. On
inserting the above identity into Eq.(\ref{eq:xpkz1}), the peak
correlation function may now be formulated as
\begin{equation}
\label{eq:xpkz}
\bnpk^2\left[1+\xpk(r,z)\right]=
\int\!\!\frac{d^3\vk}{(2\pi)^3}\int\!\!d^3\vr'\,e^{i\vk\cdot(\vr-\vr')}
\int\!\!d^6\zeta_1d^6\zeta_2\,\npk(\vx_1')\npk(\vx_2')
P_2(\vy_1,\vy_2;\vr',z_i)
\exp\left(-\frac{1}{2}\vj^\dagger\Sigma\vj+i\vj^\dagger\Xi\right)\;,
\end{equation}
where $\vj=\sigma_v(\vk,-\vk)$.  The task of computing the redshift
evolution of the peak correlation function boils  down to the
evaluation of the 6-dimensional covariance  matrix $\Sigma$ and mean
vector $\Xi$. 

\subsection{A simple illustration: the mass correlation function}

To understand the physical meaning of nonlinear corrections as well as
emphasize the relation with, e.g., RPT, it is instructive to consider 
first the {\it unbiased} case, i.e. the evolution of the matter correlation 
function. Therefore, there is no peak constraint, so the two-particle 
probability density $P_2(\vups_1,\vups_2;\vr,z)d^3\vups_1d^3\vups_2$ is 
simply the probability to find a pair of dark matter particles separated 
by a distance $\vr$ and with velocities $\vups_1$ and $\vups_2$, 
respectively. As a consequence, $\Sigma$ is the covariance of matter 
velocity components, i.e. $\Sigma_{ij}=\la\ups_i\ups_j\ra$, and 
$\Xi\equiv 0$ because there is no net mean streaming between two randomly 
selected locations. After some simplifications, the mass correlation  
function evolved with the Zel'dovich ansatz takes the simple form
\begin{equation}
\label{eq:ximza}
1+\xim(r,z)=\int\!\!\frac{d^3\vk}{(2\pi)^3}\,e^{i\vk\cdot\vr}\,
e^{-\frac{1}{3}k^2\sigma_v^2} \int\!\!d^3\vr' e^{-i\vk\cdot\vr'}
\exp\left\{\frac{1}{3}\frac{\sigma_v^2}{\sigma_{-1}^2}
\left[\xi_0^{(-1)}+\xi_2^{(-1)}\right]k^2
-\frac{\sigma_v^2}{\sigma_{-1}^2}\xi_2^{(-1)}\left(\vk\cdot\rvh\right)^2
\right\}\;.
\end{equation}
This agrees with Eqs.(16) and (17) of \cite{1996ApJ...472....1B}
provided that his $\phi(r)$ corresponds to our  $\xi_0^{(-2)}(r)$, so
that $\Delta\phi(0)\equiv -\sigma_{-1}^2$. It should be noted that 
$\sigma_{-1}$ and $\xi_\ell^{(-1)}$ are evaluated at redshift $z_i\gg 1$. 
Hence, the ratio $\sigma_v/\sigma_{-1}$ is equal to $D(z)/D(z_i)$. After 
some manipulations, the last exponent can be re-expressed as 
\citep[in the notation of][]{2006PhRvD..73f3520C}
\begin{equation}
I(\vk,\vr)\equiv \left(\frac{D(z)}{D(z_i)}\right)^2
\biggl\{\frac{1}{3}\left[\xi_0^{(-1)}+\xi_2^{(-1)}\right]k^2-\xi_2^{(-1)}
\left(\vk\cdot\rvh\right)^2\biggr\}
=\left(\frac{D(z)}{D(z_i)}\right)^2\int\!\!\frac{d^3\vq}{(2\pi)^3}\,
\frac{\left(\vk\cdot\qvh\right)^2}{q^2} P_\delta(q)\,e^{i\vq\cdot\vr}\;,
\end{equation}
where we have used that
\begin{equation}
\frac{1}{4\pi}\int\!\!d\Omega_{\qvh}\,\qh_i\qh_j\,e^{i\vq\cdot\vr}
=\frac{1}{3}\Bigl[j_0(qr)+j_2(qr)\Bigr]\delta_{ij}-j_2(qr)\,\rh_i\rh_j\;.
\end{equation}
On taking the Fourier transform of Eq.(\ref{eq:ximza}), we recover the 
well-known expression for the nonlinear mass power spectrum in the 
Zel'dovich approximation,
\begin{align}
\label{eq:pkmza}
P_{\rm m}(k,z)
&= e^{-\frac{1}{3}k^2\sigma_{-1}^2(z)}\int\!\!d^3\vr'\,e^{i\vk\cdot\vr}\,
\sum_{n=1}^{\infty}\frac{\left[I(\vk,\vr)\right]^2}{n!} \\
&= e^{-\frac{1}{3}k^2\sigma_{-1}^2(z)}\sum_{n=1}^\infty (2\pi)^3\, n!
\left(\frac{D(z)}{D(z_i)}\right)^{2(n+1)}
\prod_{j=1}^n \biggl\{\int\!\!\frac{d^3\vq_j}{(2\pi)^3} 
P_\delta(q_j,z_i)\biggr\}\,\Bigl[F_n(\vq_1,\cdots,\vq_n)\Bigr]^2
\delta^{(3)}(\vk-\vq_{1\cdots n}) \nonumber\;,
\end{align}
where $\vq_{1\cdots n}=\vq_1+\cdots+\vq_n$ and the kernels $F_n$ 
\citep{1984ApJ...279..499F,1986ApJ...311....6G} are symmetric, homogeneous
functions of the wave vectors $\vq_1$,...,$\vq_n$ that describe the 
nonlinear evolution of the density field in the Zel'dovich approximation
\citep{1987ApJ...320..448G},
\begin{equation}
\label{eq:kernza}
F_n(\vq_1,\cdots\vq_n)=\frac{1}{n!}\frac{(\vk\cdot\qvh_1)}{q_1}
\cdots\frac{(\vk\cdot\qvh_n)}{q_n}\;.
\end{equation}
They are nearly independent of $\Omega_{\rm m}$ and $\Omega_\Lambda$.
Note that, in the exact dynamics, $F_2(\vq_1,\vq_2)=5/7
+1/2(q_1/q_2+q_2/q_1)\qvh_1\cdot\qvh_2+2/7(\qvh_1\cdot\qvh_2)^2$.
In the Zel'dovich approximation however, the coefficients are all 
equal to 1/2, reflecting the fact that momentum is only conserved at
first order.

In Eq.(\ref{eq:pkmza}), the sum represents all the mode-coupling
corrections, whereas the decaying exponential pre-factor corresponds to 
the propagator \citep{2006PhRvD..73f3520C}, which for Gaussian initial 
conditions describes the (imperfect) correlation between the nonlinear 
and linear density field. Hence, the Lagrangian formulation of 
\cite{1996ApJ...472....1B} furnishes an easy way to obtain the resummed 
RPT propagator \citep[see also][for a similar Lagrangian description]
{2006PhRvD..73f3520C,2008PhRvD..77f3530M}. At the second order, the 
mass correlation is
\begin{equation}
\label{eq:ximza2}
\xim(r,z)=\int\!\!\frac{d^3\vk}{(2\pi)^3}\,e^{i\vk\cdot\vr}\,
e^{-\frac{1}{3}k^2\sigma_{-1}^2(z)}\biggl\{P_\delta(k,z)+\frac{2}{(2\pi)^3}
\int\!\!d^3\vq_1 \int\!\!d^3\vq_2\,\Bigl[F_2(\vq_1,\vq_2)\Bigr]^2
P_\delta(q_1,z) P_\delta(q_2,z)\,\delta^{(3)}(\vk-\vq_1-\vq_2)\biggr\}\;.
\end{equation}
The second order kernel scales as $F_2(\vq_1,\vq_2)\propto k^2$ in the 
(squeezed) limit where the total momentum $\vk$ goes to zero as a 
consequence of mass-momentum conservation 
\citep{Zeldovich1965,1974A&A....32..391P}.

\subsection{Including the peak constraint}
\label{sub:xpkevol}

The computation of Eq.(\ref{eq:xpkz}) is rather involved when the peak 
constraint is taken into account. Here, we will explain the basic result
and show that it generalizes previous formulae based on the local bias
scheme. Details of the calculation can be found in Appendix 
\S\ref{app:xpkevol} where it is shown, among others, that it is quite 
convenient to work with the Fourier transforms of $\Sigma$ and $\Xi$.
We eventually arrive at
\begin{align}
\label{eq:xpkevol}
\xpk(r,z) 
&= \int\!\!\frac{d^3\vk}{(2\pi)^3}\,
\biggl\{e^{-\frac{1}{3}k^2\sigma_{\rm vpk}^2\!(z)} 
\biggl[\left(\frac{D(z)}{D(z_i)}\right) \bias{vpk}(k)
+\bias{I}(k,z_i)\biggr]^2 P_{\delta_S}(k,z_i)
+\pmc(k,z)\biggr\}\, e^{i\vk\cdot\vr} \\
&= \int\!\!\frac{d^3\vk}{(2\pi)^3}\,G^2(k,z)
\Bigl[\bias{I}^{\rm E}(k,z)\Bigr]^2 P_{\delta_S}(k,z_0)
\,e^{i\vk\cdot\vr}+\xmc(k,z)\nonumber \;.
\end{align}
The last equality follows upon making the replacement $z_i\to z_0$ 
(everywhere but in the exponential pre-factor). Here, $z_0$ is some 
fiducial redshift which we take to be the collapse redshift (because 
results are usually normalized to low redshift quantities). The 
function
\begin{equation}
\label{eq:damping}
G^2(k,z)=D_+^2(z)\, e^{-\frac{1}{3}k^2\sigma_{\rm vpk}^2\!(z)}\;,
\end{equation}
where $D_+(z)\equiv D(z)/D(z_0)$ is the linear growth rate normalized 
to its value at the epoch of collapse, is a damping term induced by 
velocity diffusion around the mean displacement 
\citep{1996ApJ...472....1B}. Namely,
$G(k,z)$ is the peaks propagator in the Zel'dovich approximation, and is
analogous to that introduced in \cite{2006PhRvD..73f3519C,
2006PhRvD..73f3520C} for the matter evolution (the latter depends on 
$\sigma_{-1}$). In the exponent, the factor of $1/3$ reflects the fact
that the dynamics governing $\xpk(r,z)$ is effectively one-dimensional
(because, on average, only the streaming along the separation vector of 
a peak pair matters). The first order Eulerian and Lagrangian peak biases
$\bias{I}^{\rm E}(k,z)$ and $\bias{I}(k,z_0)$, both
defined  with respect to $P_{\delta_S}(k,z_0)$, are then related
according to
\begin{equation}
\label{eq:bnIE}
\bias{I}^{\rm E}(k,z)\equiv \bias{vpk}(k)+D_+^{-1}\!(z)\,\bias{I}(k,z_0)\;.
\end{equation}
In the limit $z\to\infty$, we recover 
$\bias{I}^2(k,z_0)P_{\delta_S}(k,z_0)$ at the first order. The presence 
of $\bias{vpk}(k)$ reflects the fact that peaks stream towards (or move 
apart from) each other in high (low) density environments, but this effect 
is $k$-dependent owing to the statistical velocity bias. Still, the 
$k$-dependence of $\bias{vpk}(k)$ is such that the linear Eulerian peak 
bias $\bias{I}^{\rm E}$ is scale-independent in the limit $k\ll 1$, in 
agreement with the ``local bias theorem'' 
\citep{1993MNRAS.262.1065C,1998ApJ...504..607S,2008PhRvD..78h3519M}. On
writing the Eulerian linear peak bias as  
$\bias{I}^{\rm E}\equiv \bias{10}^{\rm E}+\bias{01}^{\rm E}k^2$, 
Eq.(\ref{eq:bnIE}) becomes
\begin{gather}
\label{eq:bnIE1}
\bias{10}^{\rm E}(z)\equiv 1+D_+^{-1}\!(z)\,\bias{10}(z_0),\qquad
\bias{01}^{\rm E}(z)\equiv D_+^{-1}\!(z)\,\bias{01}(z_0)
-\frac{\sigma_0^2}{\sigma_1^2}\;.
\end{gather}
The first relation is the usual formula for the Eulerian, linear
scale-independent bias \citep{1996MNRAS.282..347M}. It shows that, 
unsurprisingly, the bias of the peak distribution eventually relaxes 
to unity \citep{1996ApJ...461L..65F,1998ApJ...500L..79T,
2008PhRvD..77d3527H}.
The second relation implies that $\bias{01}^{\rm E}$ approaches the 
negative, $R_S$-dependent constant $-\sigma_0^2/\sigma_1^2$ as the
gravitational instability grows. 
A scale dependence in the linear peak bias $\bias{I}^{\rm E}$ 
thus persists in the long term if the linear velocities are 
statistically biased (otherwise $\bias{01}^{\rm E}\to 0$ as 
$z\to\infty$).

Thinking of the first order peak statistics as arising from the 
{\it continuous} bias relation Eq.(\ref{eq:pkbiasing}) furnishes a
straightforward derivation of Eq.(\ref{eq:bnIE1}). Namely, the 
linear continuity equation for the mass density field reads 
$\dot{\delta}(t)=-\nabla\cdot\vv(t)=\theta(t)$ whereas, for density 
peaks, $\dot{\delta}_{\rm pk}(t)=-\nabla\cdot\vv_{\rm pk}(t)
=\theta_{\rm pk}(t)=\bias{vpk} W\theta$. Recall that $\bias{vpk} W$ 
is an operator which, in Fourier space, is a multiplication by 
$(1-\sigma_{0S}^2/\sigma_{1S}^2k^2)W(k R_S)$. The solution is
\begin{equation}
\left(\begin{array}{c} \delta(t_0) \\ \dpk(t_0)
\end{array}\right)=\left(\begin{array}{cc} D_+^{-1}(t) & 0 \\
\bigl(D_+^{-1}(t)-1\bigr)\bias{vpk}\, W & 1 \end{array}\right)
\left(\begin{array}{c} \delta(t) \\ \dpk(t)
\end{array}\right)\;.
\end{equation}
As a result, the leading-order contribution to the cross correlation 
between the average overdensity of peaks defined on scale $R_S$ and 
the mass density field smoothed on an arbitrary scale $R_B$ satisfies
\begin{align}
\left\la\dpk(t_0)\delta_B(t_0)\right\ra &= 
\left\la D_+^{-1}(t)\delta_B(t)\Bigl[\bigl(D_+^{-1}(t)-1\bigr)
\bias{vpk} W \delta(t)+\dpk(t)\Bigr]\right\ra \\
&= D_+^{-1}(t)\Bigl[\bigl(D_+^{-1}(t)-1\bigr)\left\la\delta_B(t)
\bias{vpk}\delta_S(t)\right\ra+\left\la\delta_B(t)\dpk(t)\right\ra\Bigr] 
\nonumber \\
&= D_+^{-1}(t)\Biggl[\bigl(D_+^{-1}(t)-1\bigr)\left(\sigma_{0\times}^2(t)
-\frac{\sigma_{0S}^2}{\sigma_{1S}^2}\sigma_{1\times}^2(t)\right)
+\bias{10}(t)\sigma_{0\times}^2(t)+\bias{01}(t)\sigma_{1\times}^2(t)
\Biggr] \nonumber\;.
\end{align}
On writing the left-hand side as $\bias{10}(t_0)\sigma_{0\times}^2(t_0)
+\bias{01}(t_0)\sigma_{1\times}^2(t_0)$ and isolating the terms in 
$\sigma_{0\times}^2$ and $\sigma_{1\times}^2$, we arrive at
\begin{align}
\bias{10}(t_0)\sigma_{0\times}^2(t_0) &=
D_+^{-1}(t)\Bigl[\bigl(D_+^{-1}(t)-1\bigr)\sigma_{0\times}^2(t)
+\bias{10}(t)\sigma_{0\times}\sigma_{0\times}^2(t)\Bigr] \\
\bias{01}(t_0)\sigma_{1\times}^2(t_0) &=
D_+^{-1}(t)\biggl[-\bigl(D_+^{-1}(t)-1\bigr)\frac{\sigma_{0S}^2}
{\sigma_{1S}^2}\sigma_{1\times}^2(t)+\bias{01}(t)\sigma_{1\times}^2(t)
\biggr]\;.
\end{align}
Using the fact that the spectral moments of the linear mass density 
field scale as $\sigma_{n\times}(t_0)=D_+^{-1}(t)\sigma_{n\times}(t)$,
we can rewrite the above as
\begin{equation}
\bias{10}(t)=1+D_+^{-1}(t)\Bigl(\bias{10}(t_0)-1\Bigr),\qquad
\bias{01}(t)=-\frac{\sigma_{0S}^2}{\sigma_{1S}^2}+D_+^{-1}(t)
\biggl(\bias{01}(t_0)+\frac{\sigma_{0S}^2}{\sigma_{1S}^2}\biggr)\;,
\end{equation}
which is precisely Eq.(\ref{eq:bnIE1}). On linear scales, although the 
peak bias is deterministic in Fourier space, it is generally stochastic 
and scale dependent in configuration space \citep{2010PhRvD..81b3526D}. 
In principle, it would be possible to solve also for the cross-correlation 
coefficient between the peaks and the smoothed mass density field 
$\delta_B$ by considering the time evolution of the peak rms variance 
$\la\dpk^2(t)\ra$ for instance \citep[see, e.g.,][]{1998ApJ...500L..79T}.

With the peak constraint, the one-loop contribution to the mode-coupling 
power $\pmc(k,z)$ in the Zel'dovich approximation eventually becomes
\begin{align}
\label{eq:pmc}
\lefteqn{\pmc(k,z)=\frac{2}{(2\pi)^3} 
D_+^4(z)\, e^{-\frac{1}{3}k^2\sigma_{\rm vpk}^2}\int\!\!d^3\vq_1\!
\int\!\!d^3\vq_2\Bigl[\bias{II}^{\rm E}(\vq_1,\vq_2,z)\Bigr]^2
P_{\delta_S}(q_1) P_{\delta_S}(q_2)\,\delta^{(3)}(\vk-\vq_1-\vq_2)} \\
&\quad +\frac{6}{(2\pi)^3}\frac{1}{\sigma_{1S}^2} D_+^2(z)\, 
e^{-\frac{1}{3}k^2\sigma_{\rm vpk}^2}
\int\!\!d^3\vq_1\!\int\!\!d^3\vq_2\, q_1 q_2 (\kvh\cdot\qvh_1)
(\kvh\cdot\qvh_2)\bias{II}^{\rm E}(\vq_1,\vq_2,z)P_{\delta_S}(q_1) 
P_{\delta_S}(q_2)\,\delta^{(3)}(\vk-\vq_1-\vq_2) \nonumber \\
&\quad -\frac{5}{2(2\pi)^3}\frac{1}{\sigma_{2S}^2} D_+^2(z)\,
e^{-\frac{1}{3}k^2\sigma_{\rm vpk}^2}
\biggl(1+\frac{2}{5}\partial_\alpha\ln G_0^{(\alpha)}\!(\gamma_1,\gamma_1\nu)
\Bigl\rvert_{\alpha=1}\biggr)
\int\!\!d^3\vq_1\!\int\!\!d^3\vq_2\,q_1^2 q_2^2
\Bigl[3(\kvh\cdot\qvh_1)^2-1\Bigr] \nonumber \\ 
&\qquad \times \Bigl[3(\kvh\cdot\qvh_2)^2-1\Bigr]
\bias{II}^{\rm E}(\vq_1,\vq_2,z)P_{\delta_S}(q_1) P_{\delta_S}(q_2)\,
\delta^{(3)}(\vk-\vq_1-\vq_2) \nonumber \\ 
&\quad +\frac{25}{64(2\pi)^3}\frac{1}{\sigma_{2S}^4}\,
e^{-\frac{1}{3}k^2\sigma_{\rm vpk}^2}
\biggl(1+\frac{2}{5}\partial_\alpha\ln G_0^{(\alpha)}\!(\gamma_1,\gamma_1\nu)
\Bigl\rvert_{\alpha=1}\biggr)^2\int\!\!d^3\vq_1\!\int\!\!d^3\vq_2\, 
q_1^4 q_2^4\biggl\{11-30(\kvh\cdot\qvh_2)^2+27(\kvh\cdot\qvh_2)^4 \nonumber \\
&\qquad -6(\kvh\cdot\qvh_1)^2\times\Bigl[5-42(\kvh\cdot\qvh_2)^2
+45(\kvh\cdot\qvh_2)^4\Bigr]+9(\kvh\cdot\qvh_1)^4
\Bigl[3-30(\kvh\cdot\qvh_2)^2+35(\kvh\cdot\qvh_2)^4\Bigr]
\biggr\} \nonumber \\ 
&\qquad \times P_{\delta_S}(q_1) P_{\delta_S}(q_2)
\delta^{(3)}(\vk-\vq_1-\vq_2) \nonumber \\
&\quad +\frac{27}{8(2\pi)^3}\frac{1}{\sigma_{1S}^4}\,
e^{-\frac{1}{3}k^2\sigma_{\rm vpk}^2}\int\!\!d^3\vq_1\!\int\!\!d^3\vq_2\,
q_1^2 q_2^2\Bigl[3(\kvh\cdot\qvh_1)^2(\kvh\cdot\qvh_2)^2-2(\kvh\cdot\qvh_1)^2
+1\Bigr] P_{\delta_S}(q_1) P_{\delta_S}(q_2)\delta^{(3)}(\vk-\vq_1-\vq_2) 
\nonumber \\
&\quad -\frac{15}{4(2\pi)^3}\frac{1}{\sigma_{1S}^2\sigma_{2S}^2}\,
e^{-\frac{1}{3}k^2\sigma_{\rm vpk}^2}\int\!\!d^3\vq_1\!\int\!\!d^3\vq_2\,
q_1^3 q_2^3 (\kvh\cdot\qvh_1)(\kvh\cdot\qvh_2)\Bigl[15(\kvh\cdot\qvh_1)^2
(\kvh\cdot\qvh_2)^2-18(\kvh\cdot\qvh_1)^2+7\Bigr] \nonumber \\ 
&\qquad \times P_{\delta_S}(q_1)P_{\delta_S}(q_2)
\delta^{(3)}(\vk-\vq_1-\vq_2) \nonumber \;.
\end{align}
where all spectral moments $\sigma_n$ and power spectra $P_{\delta_S}(q)$ 
are evaluated at $z_0$. The second order Eulerian bias 
$\bias{II}^{\rm E}(\vq_1,\vq_2,z)$ is a symmetric function of $q_1$ and 
$q_2$,
\begin{equation}
\label{eq:bnIIE}
\bias{II}^{\rm E}(\vq_1,\vq_2,z) \equiv {\cal F}_2(\vq_1,\vq_2)
+\frac{1}{2}D_+^{-1}(z)\Bigl[{\cal F}_1(\vq_1)\bias{I}(q_2,z_0)
+{\cal F}_1(\vq_2)\bias{I}(q_1,z_0)\Bigr]
+\frac{1}{2}D_+^{-2}(z)\bias{II}(q_1,q_2,z_0) \;.
\end{equation}
Here, $\bias{II}(q_1,q_2,z_0)$ represents  the second order Lagrangian 
peak bias Eq. (\ref{eq:bnII}) and, in analogy with standard PT, we have 
introduced the kernel ${\cal F}_n$ which characterize the $n$th-order 
evolution of the peak correlation function in the Zel'dovich approximation,
\begin{equation}
\label{eq:peakkernels}
{\cal F}_n(\vq_1,\cdots\vq_n)\equiv \frac{1}{n!}\,
\frac{(\vk\cdot\qvh_1)}{q_1}\cdots
\frac{(\vk\cdot\qvh_n)}{q_n}\,
\bias{vpk}(q_1)\cdots \bias{vpk}(q_n)\;.
\end{equation}
${\cal F}_n$ is identical to the standard PT kernels except for the
velocity bias $\bias{vpk}(q)$. As can be seen,  the second order
mode-coupling power is simply obtained from Eq.(\ref{eq:xpkeasy}) upon
a replacement $\bias{II}\to \bias{II}^{\rm E}$ in Eq.(\ref{eq:xpkeasy}),  
a Fourier transformation and a multiplication by the diffusion damping 
pre-factor $\exp(-(1/3)k^2\sigma_{\rm vpk}^2)$. Therefore, the 
mode-coupling induced by gravity is Eq.(\ref{eq:pmc}) minus the second
order terms in Eq.(\ref{eq:xpkeasy}). We also note the plus sign in 
the second term of (\ref{eq:pmc}), which follows from the fact that the 
Fourier transform of $\xi_1^{(1/2)}(r)$ is $i q (\rvh\cdot\qvh)$.  
Finally, we simply Fourier transform $\pmc(k,z)$ to obtain the 
mode-coupling contribution in configuration space.

\begin{figure*}
\center
\resizebox{0.45\textwidth}{!}{\includegraphics{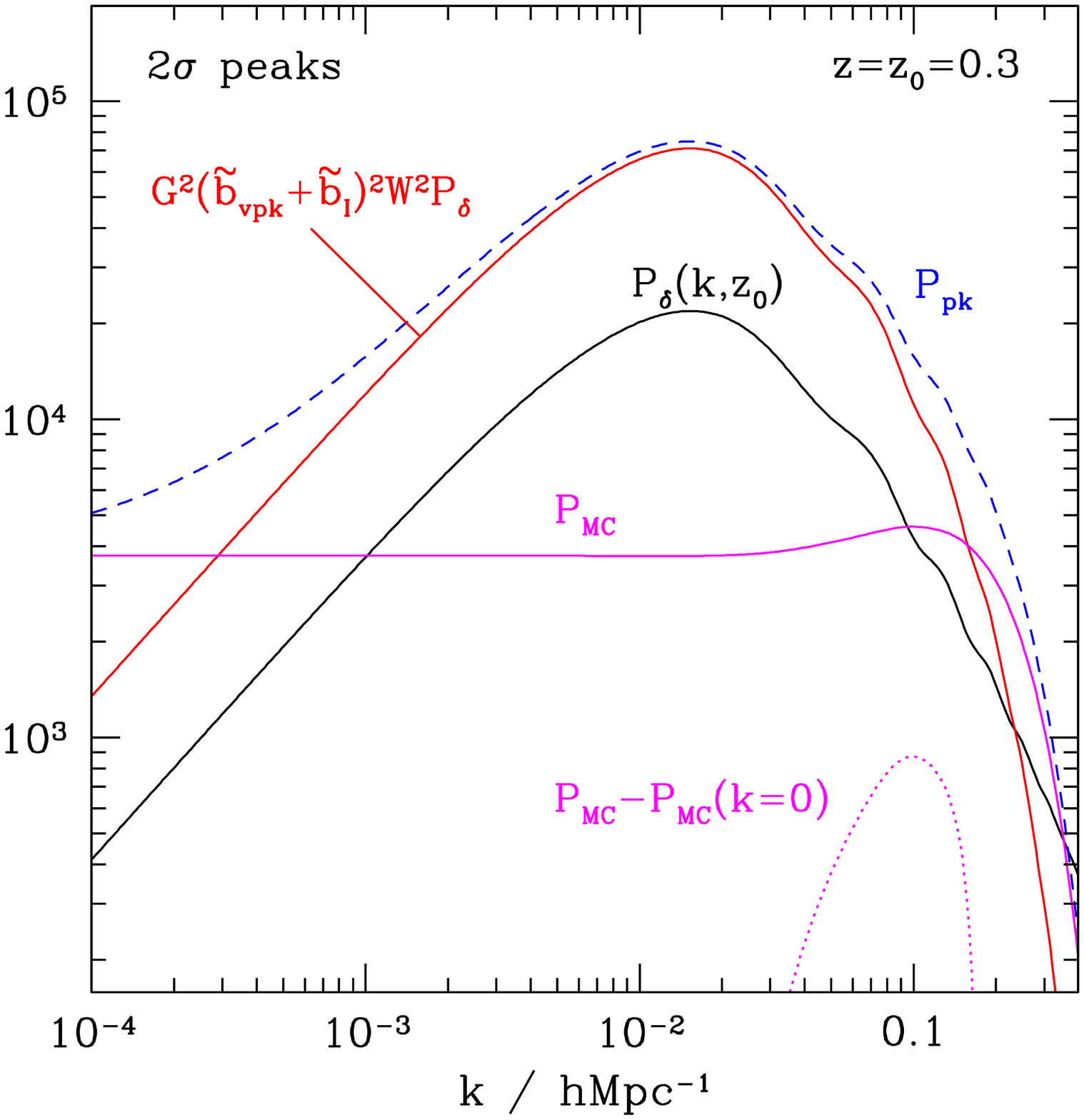}}
\resizebox{0.45\textwidth}{!}{\includegraphics{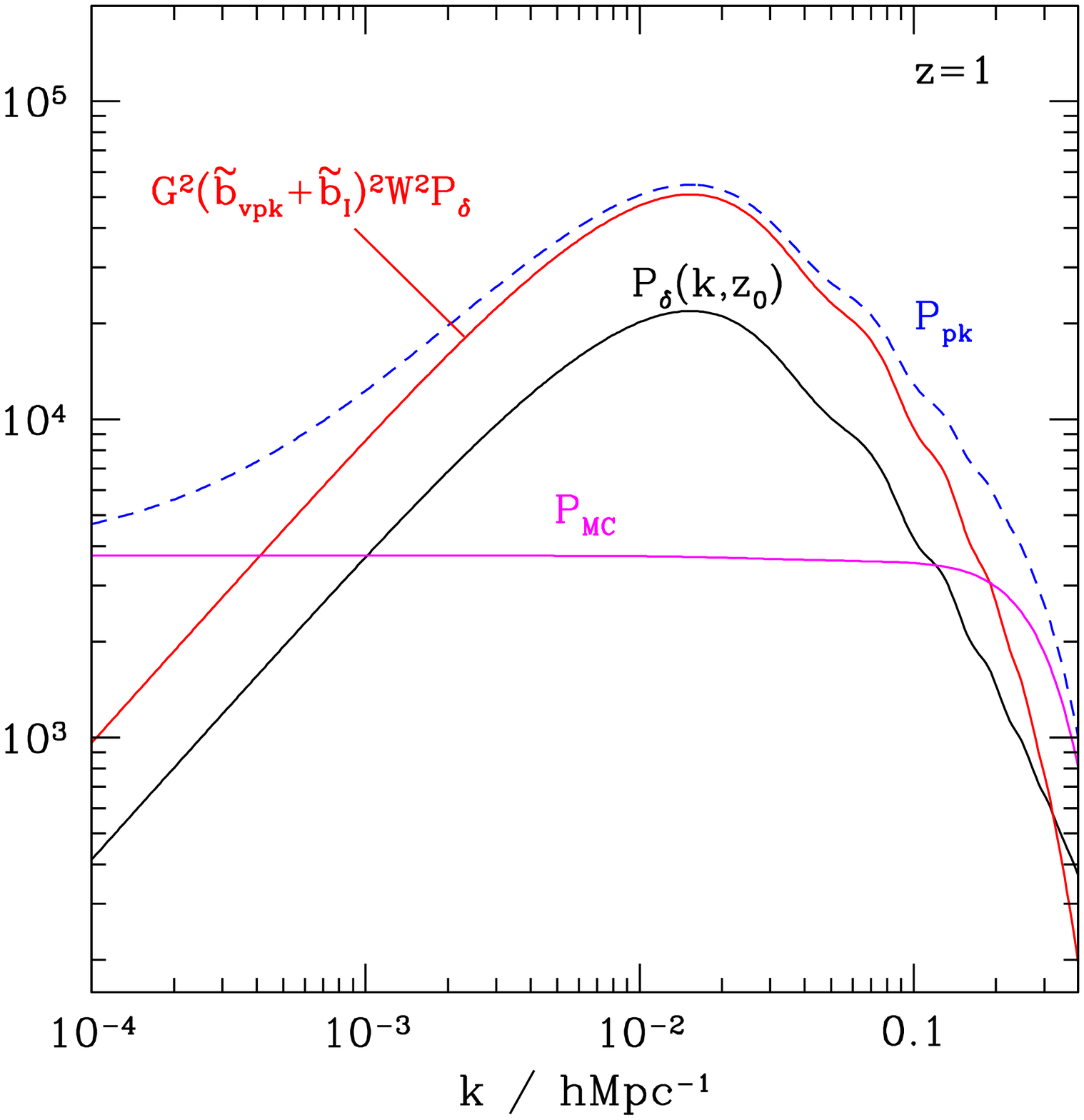}}
\caption{The evolved power spectrum $\ppk(k,z)$ (dashed curve) for the 
2$\sigma$ peaks as predicted by Eq.(\ref{eq:xpkevol}) at the redshift of 
collapse $z=z_0=0.3$ (left panel) and at $z=1$ (right panel). The first 
order term and the mode-coupling contribution are shown as solid curves 
together with the unsmoothed linear mass power spectrum $P_\delta(k,z_0)$. 
At low wavenumber, the scale-dependent contribution to the mode-coupling 
power, $\pmc(k)-\pmc(k=0)$, scales as $k^2$ since there is no mass-momentum
conservation. The Poisson expectation $1/\bnpk$ (not shown on this figure) 
is $\approx 10^4$.}
\label{fig:pmc}
\end{figure*}

In Eq.(\ref{eq:xpkz}), each power of $\vj$ brings a factor of ${\cal
F}_1(q)$.  At the second order, there is only one contribution
proportional to $\vj^4\sim {\cal F}_2^2$ but, contrary to the Eulerian
PT expression, it does not involve the first order bias. In fact,
taking the local bias  limit in which $\bias{I}=b_I$, $\bias{II}=b_{II}$ 
and ignoring the exponential damping pre-factor and a possible statistical 
velocity bias (i.e., $G\equiv 1$ and $\bias{vpk}\equiv 1$), we 
{\it do not recover} the familiar PT expression 
\citep[e.g.,][]{1998MNRAS.301..797H}
\begin{equation}
\label{eq:pmc_eul}
\pmc(k,z)=\frac{2}{(2\pi)^3}
\int\!\!d^3\vq_1\!\int\!\!d^3\vq_2
\left[b_I^{\rm E} F_2(\vq_1,\vq_2)+\frac{b_{II}^{\rm E}}{2}\right]^2 
P_{\delta_S}(q_1)P_{\delta_S}(q_2)\,\delta^{(3)}(\vk-\vq_1-\vq_2)\;,
\end{equation}
where $b_N^{\rm E}$ are Eulerian peak-background split bias factors, but 
rather (omitting sub-leading powers in the growth factors for simplicity)
\begin{equation}
\label{eq:pmc_pkloc}
\pmc(k,z)\sim 
\frac{2}{(2\pi)^3} \int\!\!d^3\vq_1\!\int\!\!d^3\vq_2
\biggl[F_2(\vq_1,\vq_2)+\frac{b_I}{2}\Bigl(F_1(\vq_1)+F_1(\vq_2)\Bigr)
+\frac{b_{II}}{2}\biggr]^2 P_{\delta_S}(q_1)P_{\delta_S}(q_2)\,
\delta^{(3)}(\vk-\vq_1-\vq_2)\;.
\end{equation}
The difference is not surprising. In the first expression, local bias is 
applied to the {\em evolved} density field and, thus, bias coefficients 
enter as multiplicative factors in each term of the perturbation series, 
including the mode-coupling contribution. In our approach, however, peaks 
are identified in the initial conditions and then evolved gravitationally
such that, in the long term, the Lagrangian bias factors drop to zero and 
the unbiased case is reproduced. In fact, our Eq.(\ref{eq:pmc_pkloc}) 
agrees with the local Lagrangian bias expression of 
\cite{2000MNRAS.318L..39C},
\begin{equation}
\pmc(k,z)\sim 
\frac{2}{(2\pi)^3} \int\!\!d^3\vq_1\!\int\!\!d^3\vq_2
\biggl[\left(1+b_0\right) F_2(\vq_1,\vq_2)
+\frac{b_I}{2}\Bigl(F_1(\vq_1)+F_1(\vq_2)\Bigr)+\frac{b_{II}}{2}\biggr]^2 
P_{\delta_S}(q_1)P_{\delta_S}(q_2)\,\delta^{(3)}(\vk-\vq_1-\vq_2)\;,
\end{equation}
except for a pre-factor $(1+b_0)$ multiplying the second order kernel 
$F_2(\vq_1,\vq_2)$ (in the peak model, this pre-factor is unity since 
the zeroth order bias is $b_0=0$). Therefore, in the local bias limit 
considered here, the one-loop contribution to the mode-coupling can be 
thought of as arising from the (nonlocal) relation 
\citep{1998MNRAS.297..692C}
\begin{equation}
1+\dpk(\vx,z)=\bigl[1+\dpk(\vq)\bigr]\bigl[1+\delta(\vx,z)\bigr]\;.
\end{equation}
Here, $\dpk(\vx,z)$ and $\delta(\vx,z)$ are the Eulerian peak and 
matter overdensity, $\dpk(\vq)=b_I\delta(\vq)+(1/2)b_{II}\delta(\vq)^2$
is the Lagrangian peak overabundance and $\vx=\vq+\vvs(\vq,z)$ is the 
mapping from Lagrangian to Eulerian coordinates. We expect also our 
Eq.(\ref{eq:pmc_pkloc}) to agree with the local Lagrangian bias 
expressions of \cite{2008PhRvD..78h3519M} if we include higher order 
perturbative corrections to the peak displacement. We defer the 
calculation of the mode-coupling at second order Lagrangian perturbation 
theory (2LPT) to a future work.

\begin{figure*}
\center
\resizebox{0.3\textwidth}{!}{\includegraphics{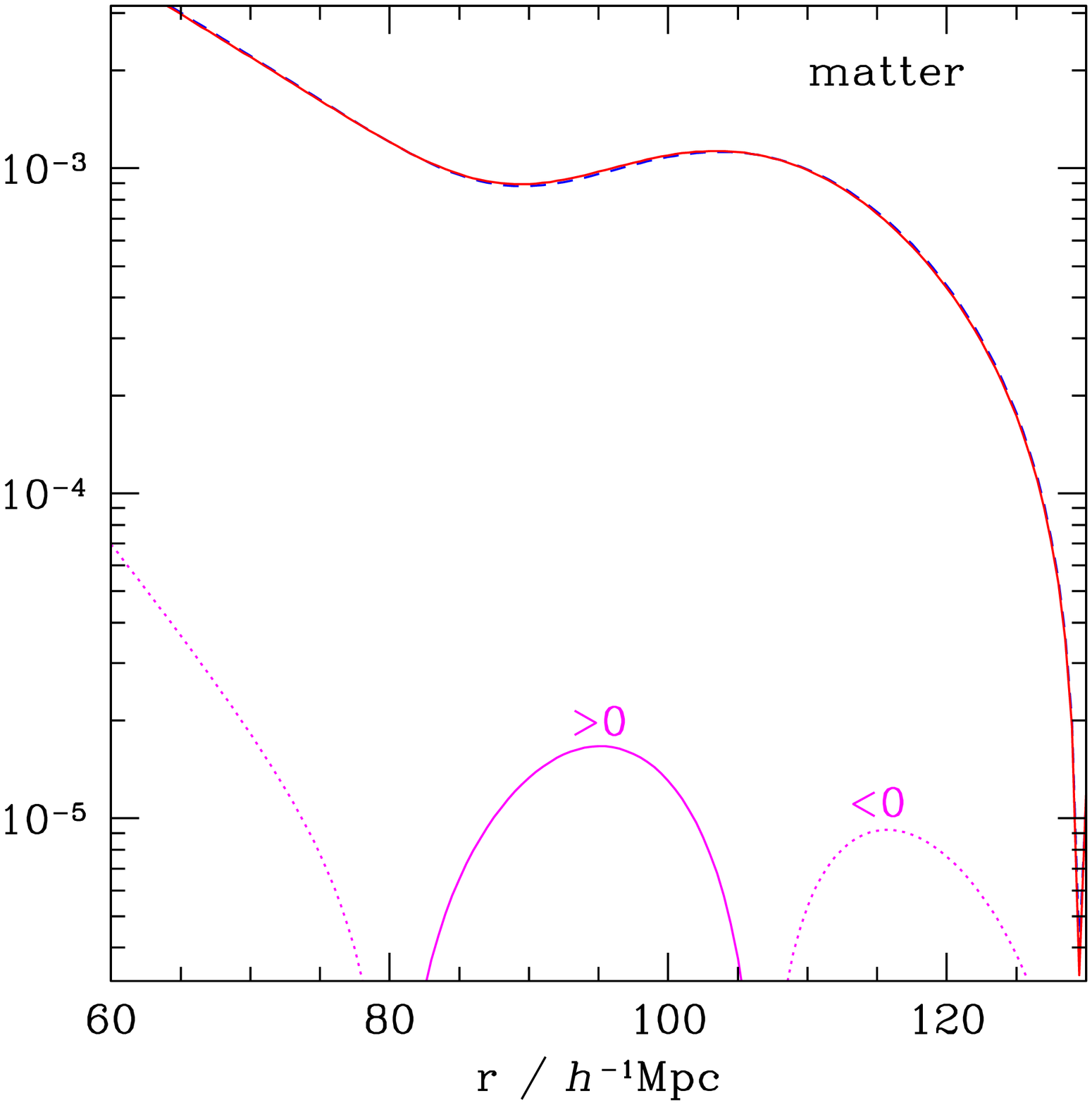}}
\resizebox{0.3\textwidth}{!}{\includegraphics{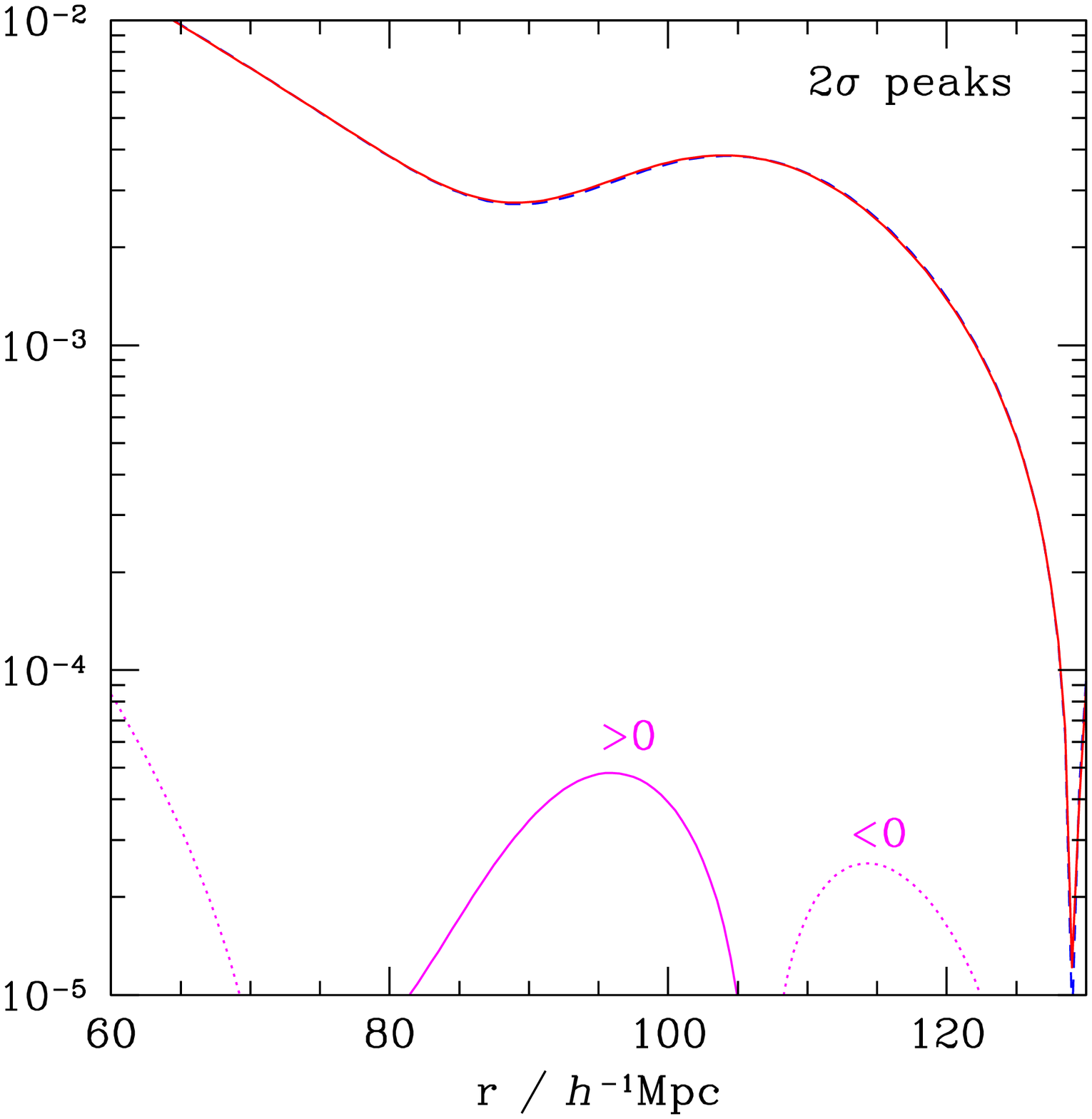}}
\resizebox{0.3\textwidth}{!}{\includegraphics{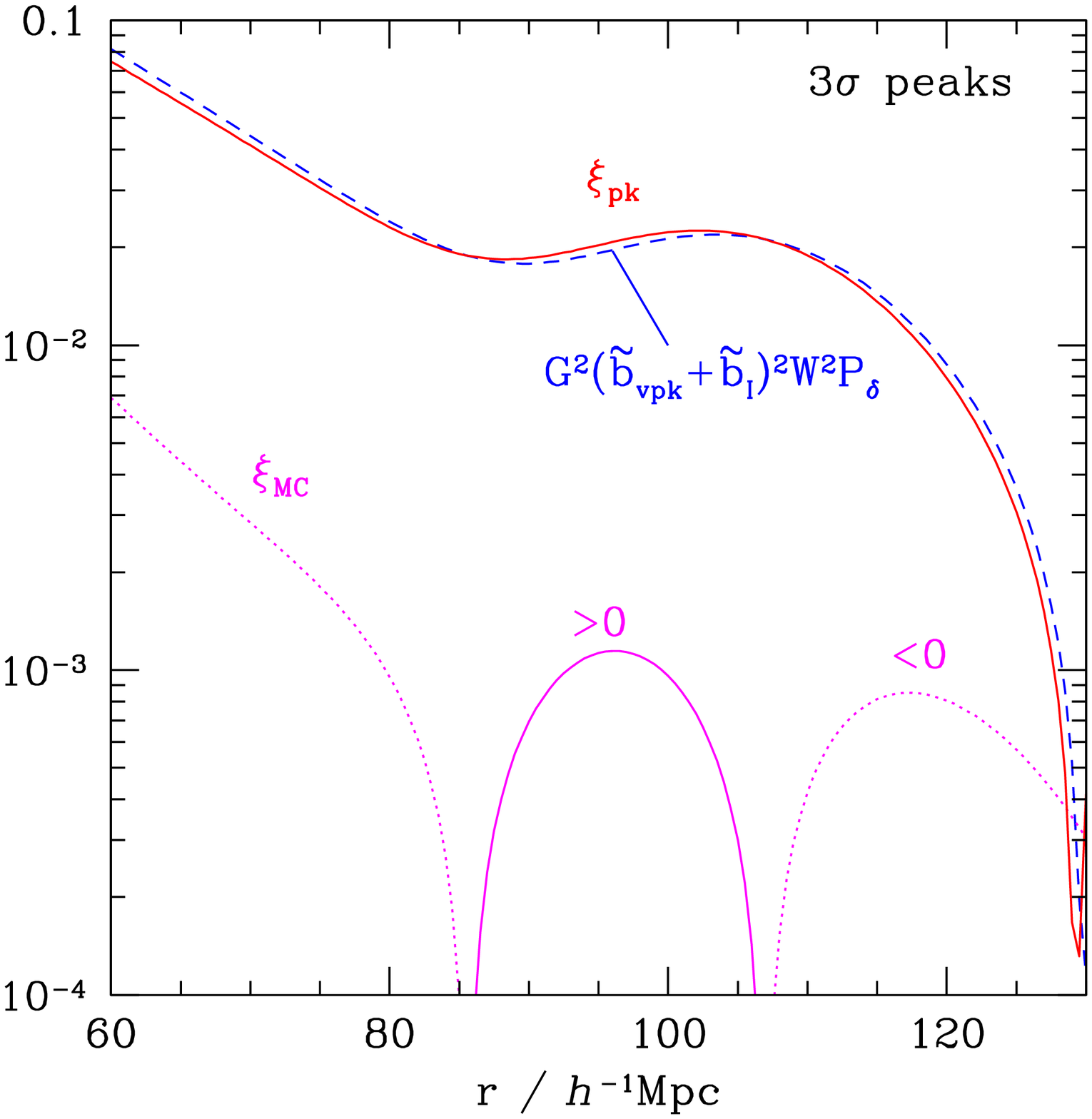}}
\caption{Matter and peak correlation functions at the redshift of 
collapse $z=z_0$ as predicted by Eqs.(\ref{eq:ximza2}) and 
(\ref{eq:xpkevol}), respectively.
In configuration space, the mode-coupling contribution $\xmc$ is 
always positive left to the BAO and negative right to the BAO, so
it introduces a shift in the acoustic scale. The magnitude of $\xmc$ 
depends strongly on the bias parameters.}
\label{fig:xmc}
\end{figure*}

\subsection{Mode-coupling power: shot-noise and shift in the baryonic acoustic scale}

To help visualize the results of the previous Section, Fig.~\ref{fig:pmc}
shows the evolved power spectrum $\ppk(k,z)$ of 2$\sigma$ density peaks 
as computed from  Eq.~(\ref{eq:xpkevol}) at redshift $z=z_0=0.3$ and 
$z=1$. The peak power spectrum is split into two distinct parts. The 
first piece is the linear mass power spectrum $P_\delta$ smoothed with 
the filter $W(R_S,k)^2$, amplified by a scale-dependent pre-factor
 $(\bias{vpk}(k)+D_+^{-1}(z)\bias{I}(k,z_0))^2$ and damped with a diffusion 
kernel. The second piece is a sum of the mode-coupling power present in 
the initial conditions (induced by the peak biasing) and of that generated 
during the nonlinear evolution (induced by gravity). 

At small wavenumber, the mode-coupling power arising from the peak biasing 
dominates and, in the limit $k\ll 1$, contribute a pure white noise term 
whose amplitude is independent of redshift. Namely, the low-$k$ white noise 
tail is generated by the peak biasing and not by gravity. While it can be 
shown that the redistribution of the matter caused by the nonlinear 
interactions cannot build a white noise tail 
$P_\delta(k)\propto k^0$ in the mass power spectrum \citep[see, e.g.,][]
{1980lssu.book.....P}, it is interesting that this holds also for tracers 
of the mass distribution (such as density peaks) which do not conserve 
momentum. In Eq.(\ref{eq:pmc}), the redshift independence of $\pmc(k=0,z)$ 
is ensured by the fact that all the $D(z)$-dependent terms come at least 
with a factor of ${\cal F}_{n=1,2}$ which vanishes in the limit $k\to 0$. 
The amplitude of $\pmc(k=0)$, however, strongly depends upon the peak 
height. A quick calculation shows that, as $k\to 0$,
\begin{align}
\label{eq:pmc_lowk}
\pmc(k,z) &\to \frac{1}{(2\pi)^2}\int_0^\infty\!\!dq\,q^2
\Bigl[\bias{II}(q,q,z_0)\Bigr]^2 P_{\delta_S}(q)^2
-\frac{1}{2\pi^2}\int_0^\infty\!\!dq\,q^4\Biggl[\frac{1}{\sigma_1^2}
+\frac{q^2}{\sigma_2^2}\biggl(1+\frac{2}{5}\partial_\alpha\ln 
G_0^{(\alpha)}\!(\gamma_1,\gamma_1\nu)\Bigl\rvert_{\alpha=1}\biggr)
\Biggr] \\
&\qquad \times \bias{II}(q,q,z_0) P_{\delta_S}(q)^2
+\frac{1}{(2\pi)^2}\int_0^\infty\!\!dq\,q^6 \Biggl[\frac{63}
{10\sigma_1^4}+\frac{46 q^2}{7\sigma_1^2\sigma_2^2}+\frac{3525 q^4}
{448\sigma_2^4}\biggl(1+\frac{2}{5}\partial_\alpha\ln 
G_0^{(\alpha)}\!(\gamma_1,\gamma_1\nu)\Bigl\rvert_{\alpha=1}\biggr)^2
\Biggr]P_{\delta_S}(q)^2 \nonumber \;.
\end{align}
All three integrals converge owing to the filtering of the mass density
field. However, the first and third terms are always positive whereas 
the second term can be negative. For the 2$\sigma$ peaks, these are 
3.2, -1.4 and $1.9\times 10^3$, respectively. As the threshold height is 
raised, the first term increasingly dominates (we find 10.6, -2.1 and 
$1.0\times 10^4$ for the 3$\sigma$ peaks). Therefore, at second order 
in the expansion, the shot-noise, which we define as the sum of 
Eq.(\ref{eq:pmc_lowk}) and the Poisson expectation $1/\bnpk$, is 
super-Poisson for $\nu>2$. This prediction seems to be at odds with 
recent lines of evidence suggesting that the shot-noise of massive halos
is sub-Poisson \citep{2002MNRAS.333..730C,2007PhRvD..75f3512S,
2010PhRvD..82d3515H}. However, we caution that this super-Poisson
behavior may be an artifact of truncating the computation of the peak
power spectrum at second order. We expect that, as higher powers of 
$\xi_\ell^{(n)}$ are included in the description of the peak correlation, 
small scale exclusion will increase and eventually make the large scale 
shot-noise correction sub-Poisson \citep[see][for a rough estimate of 
the effect]{2007PhRvD..75f3512S}. 

\begin{figure*}
\center
\resizebox{0.45\textwidth}{!}{\includegraphics{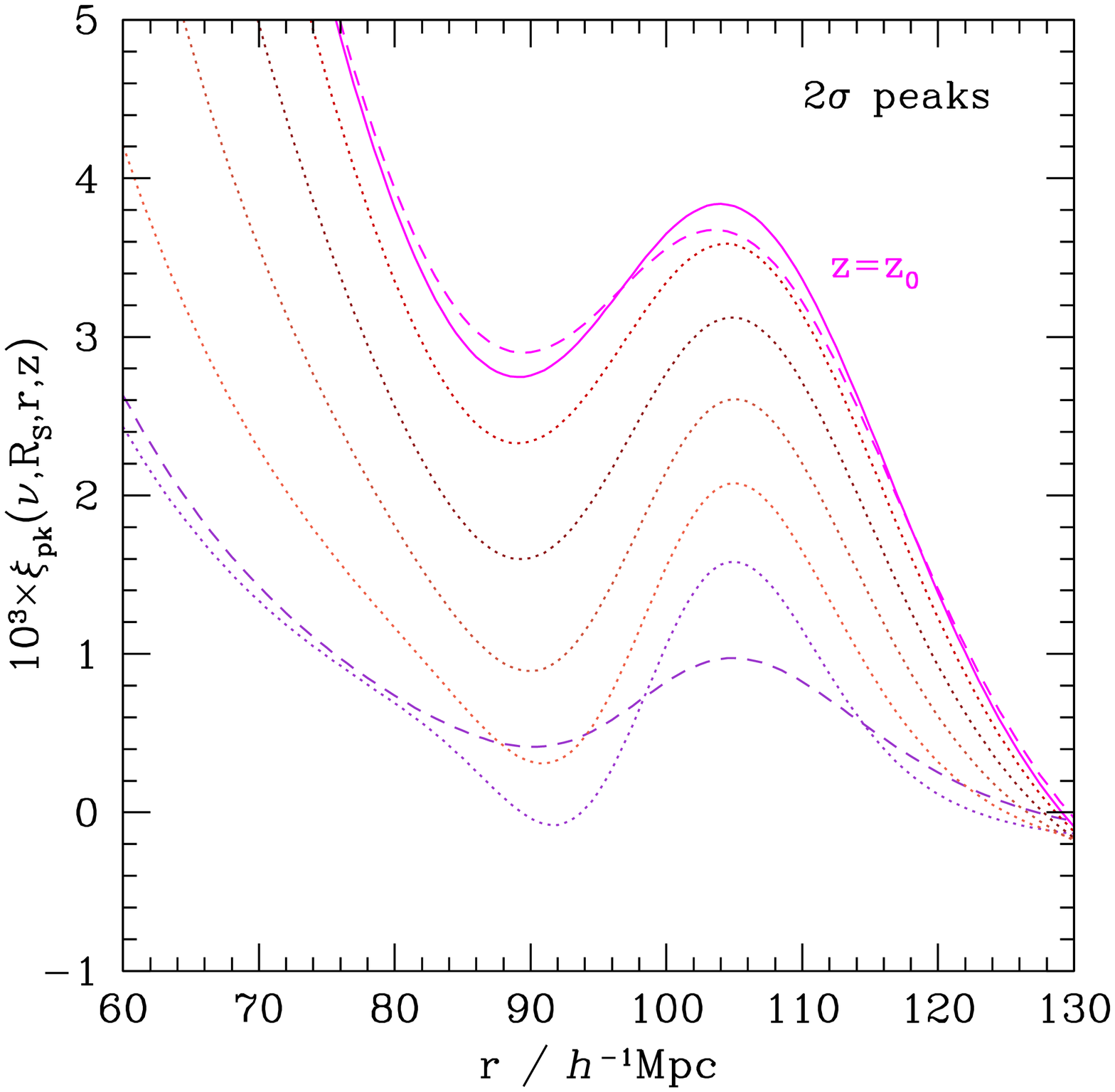}}
\resizebox{0.45\textwidth}{!}{\includegraphics{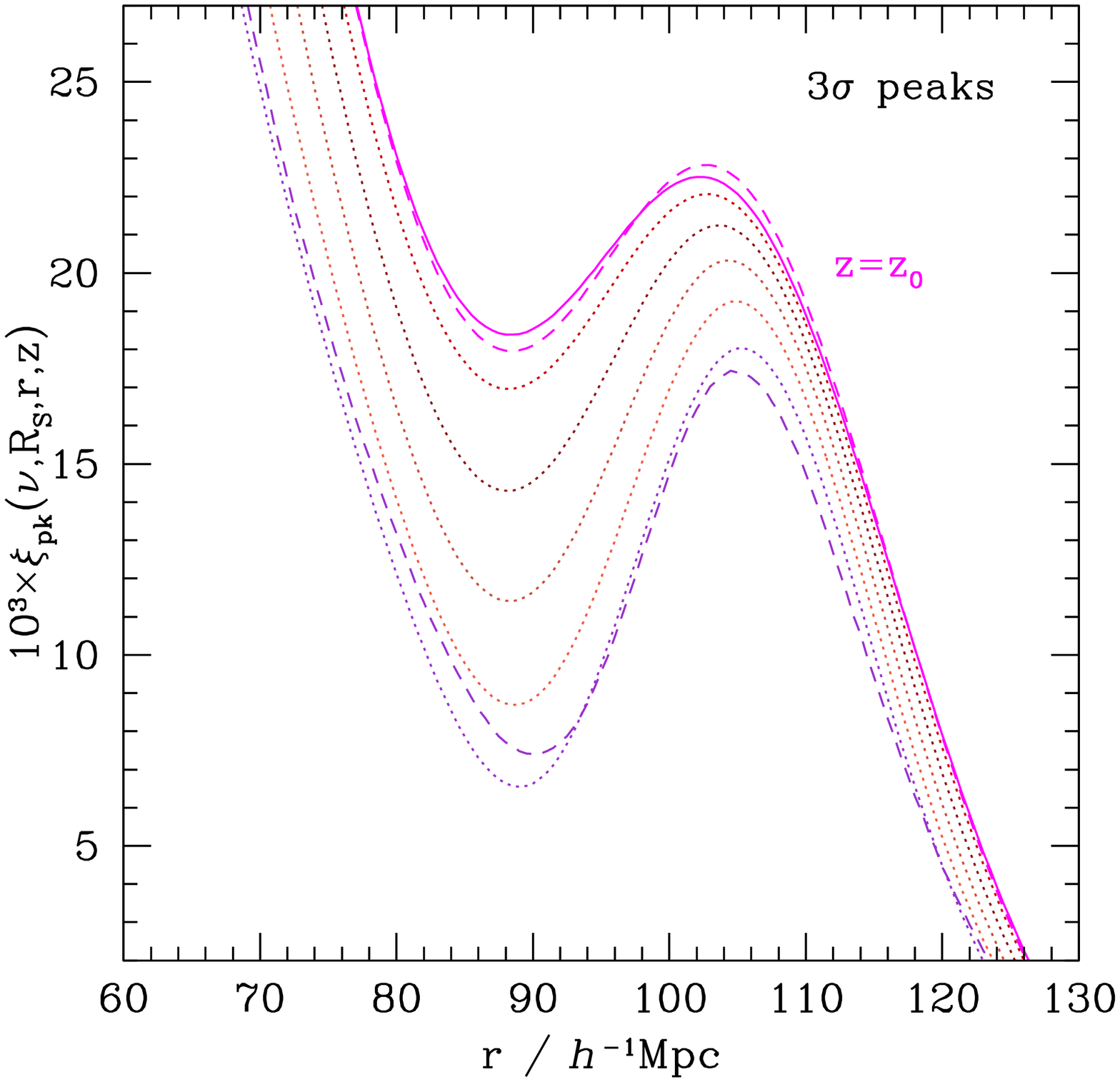}}
\caption{Redshift evolution of the correlation of 2$\sigma$ and 
3$\sigma$ peaks collapsing at $z_0=0.3$ as predicted by 
Eq.(\ref{eq:xpkevol}). The curves from bottom to top represent 
$\xpk(\nu,R_S,r,z)$ at redshift $z=\infty$, 5, 2, 1, 0.5 (dotted 
curves) and $z=z_0$ (solid curve). 
Only the correlation at the collapse epoch ($z=z_0$) can be measured 
in real data. For comparison, the dashed curves show the correlation 
at $z=\infty$ and $z=z_0$ in a local bias approximation (see text).}
\label{fig:baoz}
\end{figure*}

Before continuing, we note that any local nonlinear biasing introduces 
constant power at small wavenumber in addition to the conventional 
$1/\bar{n}$ shot-noise \citep{1998MNRAS.301..797H,1998ApJ...504..607S,
2000ApJ...537...37T,2006PhRvD..74j3512M,2007PhRvD..75f3512S,
2010arXiv1006.2343B}. At second order, this white noise contribution is 
always positive as it is given by the first integral of 
Eq.(\ref{eq:pmc_lowk}) with $\bias{II}(q,q,z_0)$ replaced by $b_{II}$ and 
$\delta$ smoothed on some arbitrary scale $R$. Therefore, in contrast to 
the prediction of the peak model, the magnitude of this term is not well 
defined in local biasing schemes owing to the freedom at filtering the 
mass density field.

Ignoring the $k^0$ tail, the next-to-leading term should scale as $k^2$
since there is no local conservation of momentum (which would otherwise 
enforce a $k^4$ behavior). One can easily check that this is indeed the 
case by noticing that, in the low-$k$ limit, 
\begin{equation}
F_1(\vq_1)=\frac{k}{q_1}\mu,\qquad F_1(\vq_2)=-\frac{k}{q_1}\mu
+\frac{k^2}{q_1^2}\left(1-2\mu^2\right)+{\cal O}(k^3),\qquad 
F_2(\vq_1,\vq_2)=\frac{k^2}{2q_1^2}\mu\Bigl[-\mu+\frac{k}{q_1}
\left(1-2\mu^2\right)\Bigr]+{\cal O}(k^4)\;.
\end{equation}
As $k\to 0$, this implies $F_1(\vq_1)+F_1(\vq_2)\approx (k/q_1)^2(1-2\mu^2)$,
so the first two terms of $\bias{II}^{\rm E}$ scale as $k^2$ at low
wavenumber.

\begin{figure*}
\center
\resizebox{0.45\textwidth}{!}{\includegraphics{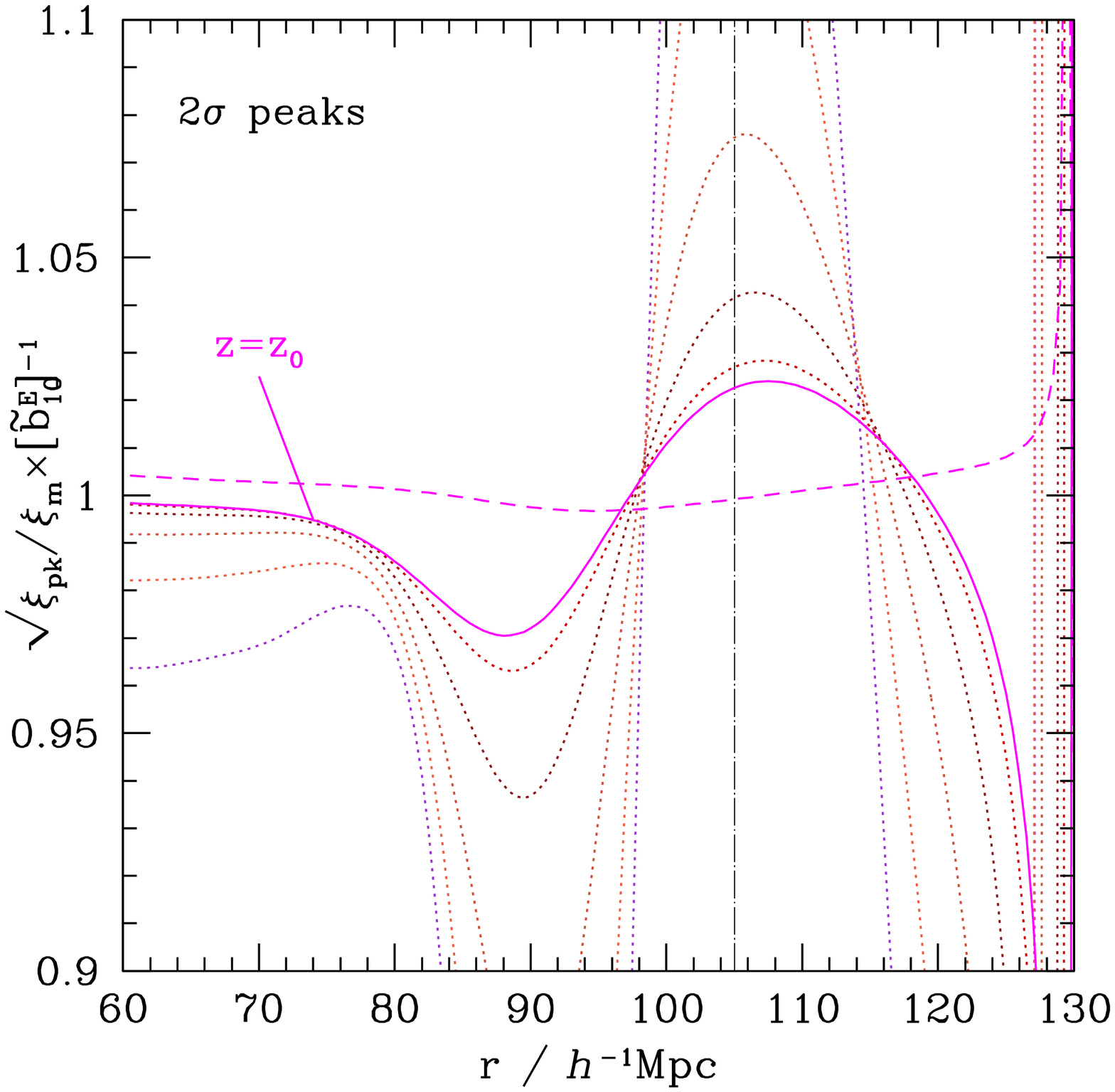}}
\resizebox{0.45\textwidth}{!}{\includegraphics{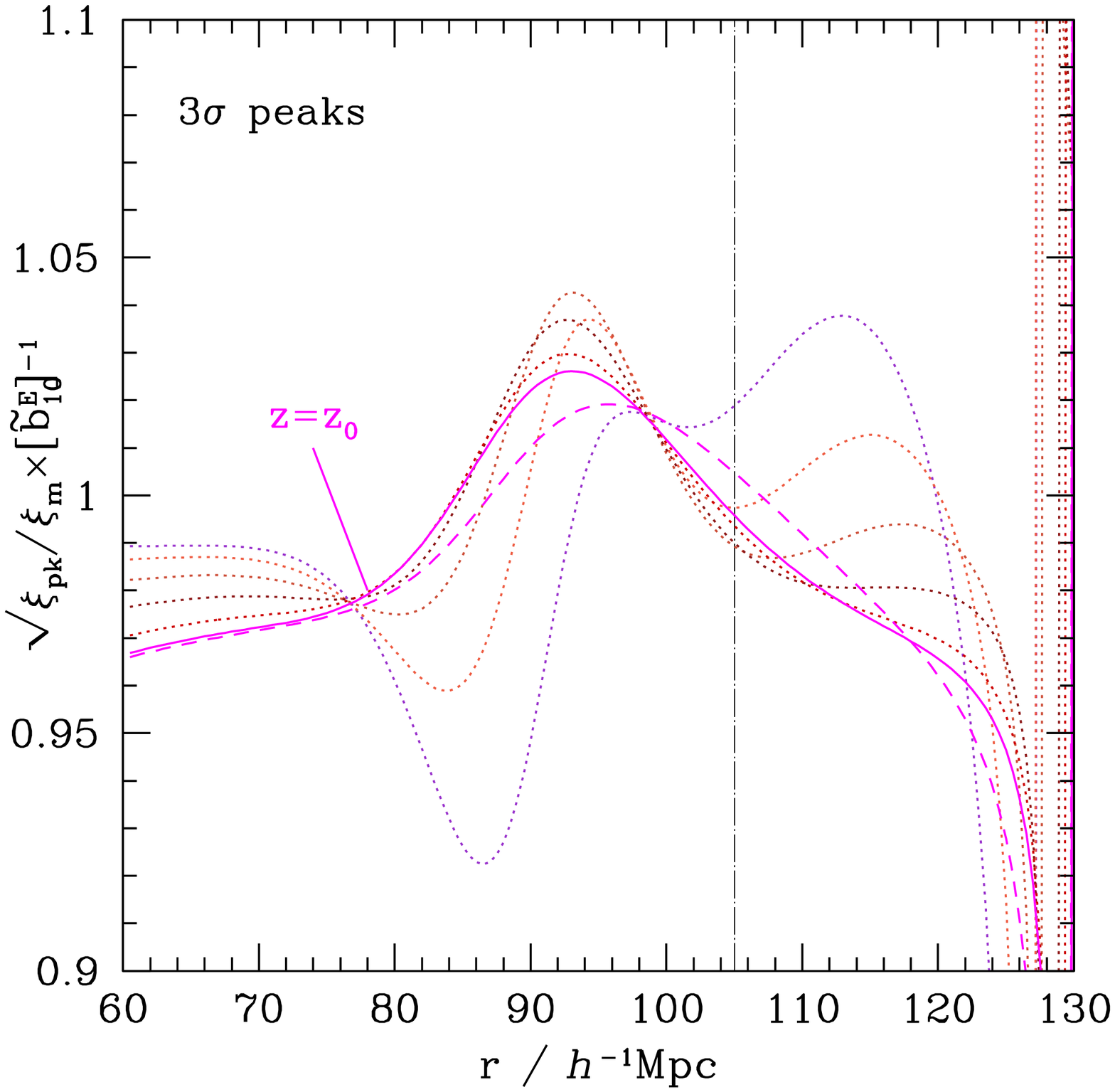}}
\caption{Ratio of the peak correlation $\xpk(\nu,R_S,r,z)$ to the 
mass correlation $\xim(r,z)$ at redshift $z=\infty$, 5, 2, 1, 0.5 
and $z_0$ (at separation $60\hmpc$, curves from bottom to top 
in the left panel). For convenience, the ratio is normalized to 
$\bias{10}^{\rm E}(z)$ so that the results approach unity as 
$r\to\infty$. The dashed curve shows the quantity at redshift of 
collapse $z_0=0.3$ in the local bias approximation. At $z=z_0$, 
the bias of density peaks exhibits a $\sim$5\% scale dependence 
across the baryonic acoustic feature. 
For the 2$\sigma$ peaks however, most of this scale dependence is
caused by the initial amplification of the BAO contrast relative
to that of the mass (see Fig.\ref{fig:bao}) whereas, for 3$\sigma$
peaks, this scale dependence is mainly generated by the gravity
mode-coupling. The vertical line denotes the position of the 
linear BAO feature.}
\label{fig:ratio}
\end{figure*}

In configuration space, the mode-coupling contribution $\xmc(r,z)$ 
changes sign across the acoustic scale $r_0\approx 105\hmpc$ because 
the dominant term involves the derivative $\xi_1^{(1/2)}(r)$ of the 
mass correlation which is positive (negative) to the left (right) of 
the acoustic peak \citep{2008PhRvD..77b3533C}. This leads to a shift 
$\Delta r_0$ of the inferred acoustic scale towards small scales
\citep{2008PhRvD..77d3525S,2008PhRvD..77b3533C,2009PhRvD..80f3508P}. 
Typically, a 1\% shift in the acoustic scale translates to a fairly
substantial $\sim$4\% bias in the estimated dark energy equation of
state \citep[e.g.,][]{2008MNRAS.383..755A}.

As shown in Fig.\ref{fig:xmc}, the magnitude of $\xmc(r,z)$ strongly 
depends on the first  and
second order bias parameters. For the 2$\sigma$ peaks, the strength 
of the mode-coupling contribution relative to the linear term 
$\xpk^{(1)}$ is slightly smaller than for the matter ($\sim$ 1 
percent), whereas for the 3$\sigma$ peaks, it is much larger ($\sim$
5 percent) due to the large, positive first and second order bias
factors. To estimate the 
``physical'' shift of the acoustic peak generated by mode-coupling, 
we must correct beforehand for the ``apparent'' shift induced by the 
convolution with the propagator $G^2$ \citep[we use the terminology of][]
{2008PhRvD..77d3525S,2008PhRvD..77b3533C}. The latter is usually taken 
into account in the data analysis and cosmological parameter forecast 
\citep{2007ApJ...664..660E,2008MNRAS.383..755A}. Here, we will simply
ignore the diffusion damping both in the linear and mode-coupling
piece and, therefore, calculate the shift of the acoustic peak in the
perfectly de-convolved correlation function
\begin{equation}
\xpk^{\rm un}(r,z_0)=\bigl[\,\bias{I}^{\rm E}\,\bigr]^2\otimes
\xi_0^{(0)}(r,z_0)+\xmc^{\rm un}(r,z_0)\;.
\end{equation}
It is important to note that, in the peak model, the ``physical'' shift 
is generated by second order mode-coupling {\it and} by scale dependence 
in the linear biases $\bias{I}$ and $\bias{vpk}$. The relative 
contribution of the second source of shift, which is not present in the 
local bias approximation, turns out to be equally important at high and
low peak height (This is somewhat counterintuitive since the 
scale dependence is more pronounced at low $\nu$). For the 2$\sigma$ and 
3$\sigma$ peaks, we find that the location of the acoustic peak in the 
linear  mass correlation is ``physically'' shifted by $\Delta r_0=-0.25$ 
and -1.65$\hmpc$, respectively. The relative contribution of the first 
order scale dependence is in both cases $\sim 20$\% only, mainly because 
$\bias{I}$ does not shift the Fourier phases. We believe these values 
should change somewhat at second order in the peak displacement (2LPT) 
since the cross term between the monopole and dipole 
(of $[\bias{II}^{\rm E}]^2$), which generates most of the shift, will not 
remain the same.

\subsection{scale dependence across the acoustic peak: theoretical predictions and comparison with simulations}

We will now present results for the evolved peak correlation function
and compare them to the auto-correlation of simulated dark matter halos.

To begin, Fig. \ref{fig:baoz} shows the redshift evolution of the 
correlation $\xpk(\nu,R_S,r,z)$ of 2$\sigma$ and 3$\sigma$ peaks from 
the initial conditions at $z=\infty$ (bottom dotted curve) until 
collapse at $z=z_0=0.3$ (top solid curve). The intermediate 
redshift values are $z=5$, 2, 1 and 0.5. It should be noted that only 
the correlation at the collapse epoch can be measured in real data
(assuming that the tracers are observed at the epoch their host dark
matter halos collapse). For comparison, the bottom and top dashed 
curves represent the initial and final correlation for the local bias 
approximation in which $(\bias{I},\bias{vpk})\equiv (b_I,1)$ and the 
mode-coupling power is given by Eq.(\ref{eq:pmc_pkloc}). As the redshift
decreases, gravitational instability generates coherent motions which 
amplify the large scale amplitude of the peak correlation, and random 
motions which increasingly smear out the initial BAO feature. Although
velocity diffusion due to large scale flows is less important for the 
peaks than for the locally biased tracers (owing to the fact that 
$\bias{vpk}<1$), the final correlations are noticeably more similar than 
they were initially. Still, mild differences subsist at $z=z_0$ between 
the peak and local bias predictions, especially around the baryonic 
acoustic feature.

In order to quantify these deviations, we take the square root of the 
ratio between the peak correlation $\xpk$ (Eq.\ref{eq:xpkevol}) and 
the mass correlation $\xim$ (Eq.\ref{eq:ximza2}). Both are consistently 
evolved at second order with the Zel'dovich approximation. 
$\sqrt{\xpk/\xim}$ measures the scale dependence of peak bias as a 
function of separation. For illustration purposes, we normalize this 
ratio to $\bias{10}^{\rm E}(z)$ so that, on scales much larger than 
the acoustic scale (not shown in this figure), the normalized ratio 
rapidly converges to unity. 
Results are shown in Fig.\ref{fig:ratio} for the 2$\sigma$ and 3$\sigma$
peaks. The dashed curve represent the local bias prediction at the 
collapse redshift $z=z_0$. Across the baryonic acoustic feature, the peak 
bias exhibits a residual scale dependence of $\sim$5\% amplitude. For the 
2$\sigma$ peaks, this scale dependence arises principally from the initial 
amplification of the BAO contrast (see Fig.\ref{fig:bao}) whereas, for 
the 3$\sigma$ peaks, it is mostly induced by the gravity mode-coupling. 
In stark contrast to the peak model, the local bias approximation 
predicts negligible scale dependence for 2$\sigma$ tracers (there is a
sharp upturn at $r\simeq 130\hmpc$ due to the fact that the zero 
crossings of $\xpk$ and $\xim$ are different). At 3$\sigma$ however, the 
discrepancy between both models is relatively smaller because the 
contribution of the mode-coupling generated during 
gravitational evolution, which is weakly sensitive to the bias factors 
$\bias{01}$, $\bias{11}$ and $\bias{02}$ in the limit $\nu\gg 1$, dominates 
the scale dependence of bias. The mode-coupling also contributes to 
suppress the peak bias by 2-3\% at separations $r\sim 60-80\hmpc$ 
\citep[see also][]{2009MNRAS.400.1643S}.

Clearly, a very large simulated volume is required to search for similar 
scale dependences in the bias of the most massive objects created by
gravitational collapse. Hence, we will present measurements of the 
baryonic acoustic feature in the auto-correlation of halos extracted 
from a single realization of 2048$^3$ particles in a cubical box of 
side $7.68\hgpc$. The simulated volume thus is more than  
sixteen times the Hubble volume. Halos were subsequently found using
a friends-of-friends algorithm with a linking length of 0.2 times the 
mean interparticle distance, leading to a final catalog of more than
15 million halos of mass larger than $\approx 7\times 10^{13}\hmsun$
(with 20 particles or more). This N-body run and the associated halo 
catalog is a part of the {\small MICE} simulations project 
\citep[see][for an exhaustive description of the runs and 
http://www.ice.cat/mice for publicly available data]
{2008MNRAS.391..435F,2009MNRAS.393.1183C,2010MNRAS.403.1353C}. A more 
detailed clustering analysis of this simulation will be presented 
elsewhere~\footnote{M. Crocce, P. Folsalba, F.J. Castander and E. 
Gazta\~{n}aga, in prep}.

\begin{figure*}
\center
\resizebox{0.45\textwidth}{!}{\includegraphics{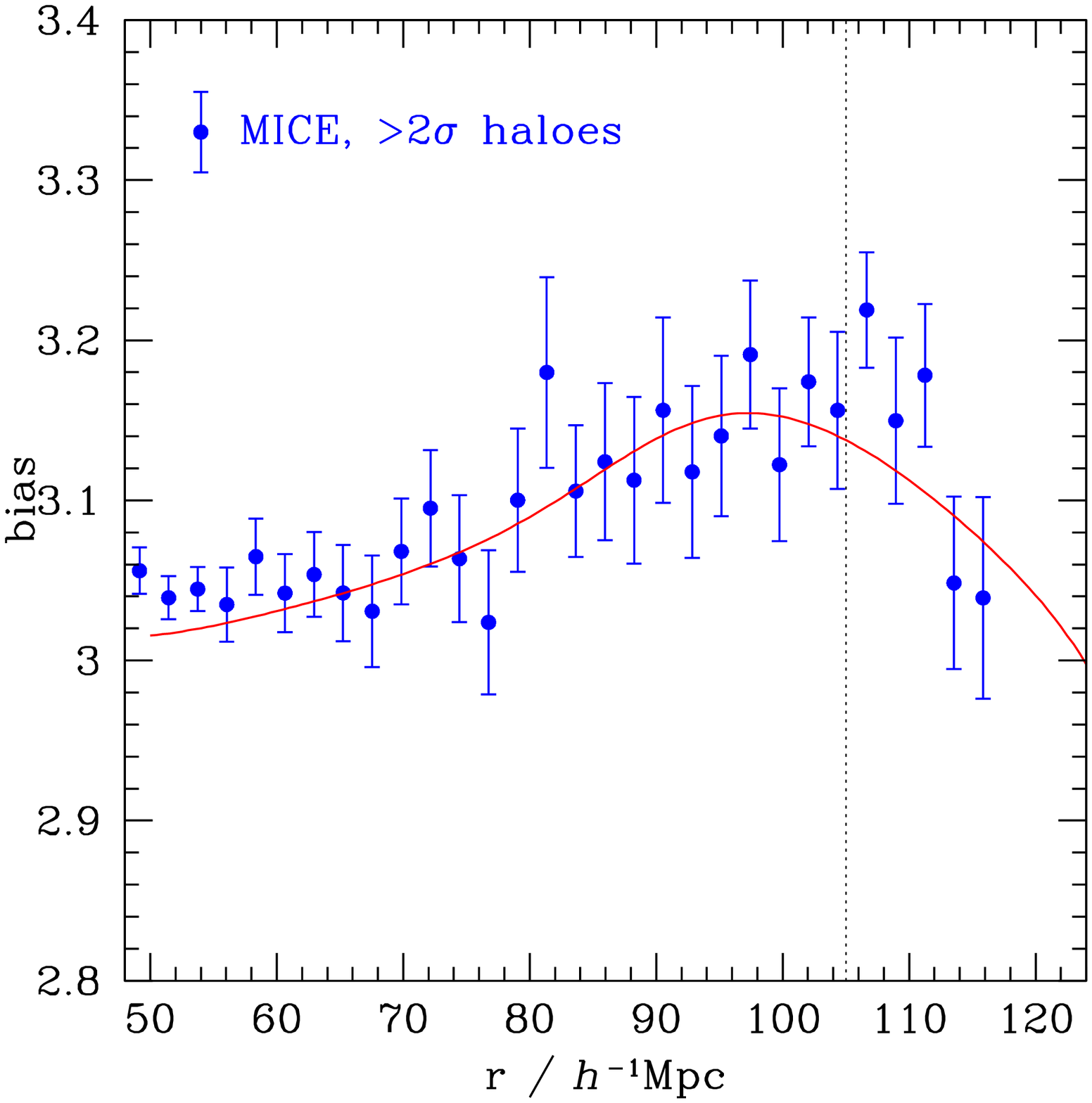}}
\resizebox{0.45\textwidth}{!}{\includegraphics{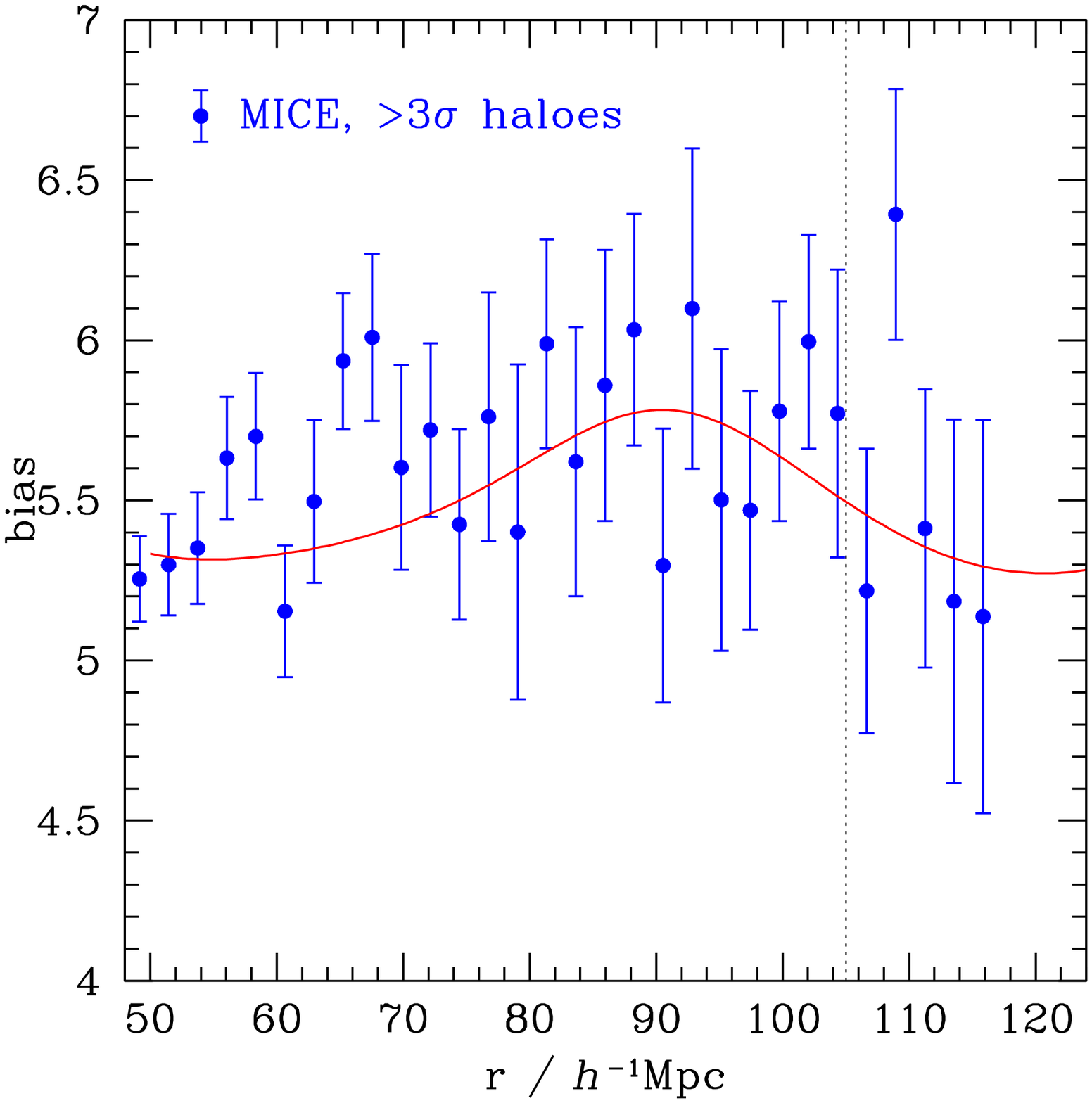}}
\caption{Scale-dependence of bias around the baryonic acoustic feature 
of the halo auto-correlation. Filled symbols with error bars represent the 
measurement for $>2\sigma$ and $>3\sigma$ halos extracted from the 
{\small MICE} N-body simulation (see text) at $z=0$. The corresponding 
mass cut are $M>1.2\times 10^{14}$ and $M>7.0\times 10^{14}\hmsun$, 
respectively.
The solid curve corresponds to a prediction based on the peak model 
with linear peak-background split bias $b_I$ similar to that of the 
simulated halo catalogs. For the low mass cut, the threshold height
is $\nu_t=2.15$ whereas, for the high mass cut, it is $\nu_t=3$ 
(see Eq.\ref{eq:approxxih}). A vertical line denotes the position of
the linear BAO feature.}
\label{fig:nbody}
\end{figure*}

The filled symbols in Fig.\ref{fig:nbody} show the measured bias 
(defined as the squared root ratio of their auto-correlation function 
to that of the dark matter field) for halos with significance larger 
than $\nu_t=2$ and 3. Here, we exceptionally adopted a tophat filter to 
define the peak significance $\dsc(z)/\sigma_0$. Therefore, the $\nu_t$ 
threshold corresponds to a mass cut $M_t=2\times 10^{14}$ and 
$7\times 10^{14}\hmsun$, respectively (these halos have at least 35 
and 200 particles respectively). The correlation was computed by direct 
pair counting using the estimator of \citep{1993ApJ...412...64L}.
Measurements were done first dividing the full boxsize of $452\hhhgpc$ 
into 27 non-overlapping regions of equal volume. 
The bias shown is the average over these regions and the error 
bars those associated with the corresponding variance of the auto 
correlation measurements (we actually depict the error on the mean, 
i.e., $\sigma_b/\sqrt{27}$). 

From Fig.\ref{fig:nbody}, it is clear that, while the $>3\sigma$ halo sample 
is too sparse to furnish useful information, the correlation of $>2\sigma$ 
halos shows unambiguous evidence for a scale dependence of bias of $\sim$5\% 
relative magnitude around the baryonic acoustic feature. To the best of our 
knowledge, this is the first time that such a scale dependence is reported
from N-body simulations \citep[see, however,][]{2009ApJ...701.1547K}.

To compare predictions from the peak model with the measurements from
the N-body simulation, we remark that the set of dark matter halos 
with significance larger than $\nu_t$ includes all halos of mass larger 
than some threshold value $M_t$. As an attempt to reproduce the number 
counts of halos above a mass cut $M_t$, one may want to smooth the 
mass density field with a range of filter radii $R_S$ and identify peaks 
of height $\dsc(z_0)/\sigma_0(R_S)$ with virialized halos of mass 
$M=M_S$. We adopt a sharp threshold even though the actual selection 
function $w(M|\nu,R_S)$, which gives the probability that a peak of height 
$\nu$ in the linear density field smoothed on scale $R_S$ forms a 
virialized halo of mass $M$, may be fairly smooth \citep[see, e.g.,]
[for more realistic selection functions]{1986ApJ...304...15B}. 
Consequently, we approximate the correlation function of dark matter 
halos as
\begin{equation}
\label{eq:approxxih}
\xi_{\rm h}(r,z)=\frac{1}{N_{\rm h}(\nu_t)}
\int_{\nu_t}^\infty\!\!d\nu\,\bnpk\bigl(\nu,R_S(\nu)\bigr)\,\xpk(r,z)
\qquad \mbox{where}\qquad
N_{\rm h}(\nu_t)=
\int_{\nu_t}^\infty\!\!d\nu\,\bnpk\bigl(\nu,R_S(\nu)\bigr)\;.
\end{equation}
Here, $R_S(\nu)$ is the filtering radius at which 
$\dsc(z)/\sigma_0(R_S,z)\equiv\nu$. The fraction of mass $f(\nu)d\nu$ 
(or mass function) in peaks of height $\nu$ thus is 
\begin{equation}
\label{eq:approxfnu}
f(\nu)d\nu\equiv 
\left(M_S/\bar{\rho}\right)\bnpk\bigl(\nu,R_S(\nu)\bigr)d\nu\;,
\end{equation}
and $M_S/\bar{\rho}=(2\pi)^{3/2}R_S^3$ is the mass enclosed in the 
Gaussian filter. In our simulation, the measured cumulative number 
density for the $>2\sigma$ and $>3\sigma$ halos is 
$N_{\rm h}\approx 1.45\times 10^{-5}$ and 
$\approx 5.10\times 10^{-7}\hhhmpc$, respectively. For these threshold
heights, Eq.(\ref{eq:approxxih}) predicts slightly larger densities, i.e., 
$N_{\rm h}\approx 1.77\times 10^{-5}$ and $\approx 8.40\times 10^{-7}$. 
To further improve the agreement with the simulated mass function, one 
shall ensure that a peak of height $\dsc$ on a smoothing scale $R_S$ is not
embedded in a region of height $\dsc$ on any larger smoothing scale. 
Carefully accounting for clouds in clouds in the peak formalism is a 
nontrivial problem even though, in a first approximation, we may simply 
enforce that the height of the peak be less than $\dsc$ on scale $R_S+dR_S$ 
\citep{1990MNRAS.245..522A,1995ApJ...453....6M}. 
Fig.1 of \cite{2001NYASA.927....1S} demonstrates that this substantially 
improves the agreement with the measured halo abundances at $\nu\gtrsim 3$
but underestimates the counts at $\nu\lesssim 2$. Furthermore, adding this 
extra constraint also modifies the peak bias factors and, possibly, the 
$\bias{II}$-independent terms in Eq.(\ref{eq:xpkeasy}).
For these reasons, we hereafter stick to the naive albeit reasonably good 
approximation Eq.(\ref{eq:approxxih}), and defer a more detailed modeling 
of the halo mass function to a future work. 

The predicted halo correlation Eq.(\ref{eq:approxxih}) is shown in Fig.
\ref{fig:nbody} as the solid curve. For the low mass cut (left panel), the 
theoretical prediction assumes in fact $\nu_t=2.15$, which furnishes a 
better match to the linear, scale-independent halo bias measured in the 
simulation. For the high mass cut, $\nu_t=3$ as in the halo sample. 
Overall, since we consider the clustering of high peaks at $z=0$, the
scale dependence induced by gravity mode-coupling is important. 
Consequently, all the theoretical curves exhibit a broad peak structure 
centered at $r\sim 90-100\hmpc$. By contrast, Lagrangian peak biasing 
solely would induces a ``sombrero''-like scale dependence with a maximum 
close to the acoustic scale (see Fig.\ref{fig:ratio}). 
Focusing on the low-$\nu$ halo sample, it is difficult to assess the 
quality of the fit without knowing the data covariances. Nevertheless, 
the measurements strongly suggest that our prediction based on the peak
model is a very good approximation. An improved description of peak
motions, collapse and mass function will somewhat modify the theoretical 
curve but, because the predicted abundances are close to the measured 
ones, we believe the change should not be dramatic. It will be interesting
to search for a similar effect in the clustering of low mass and/or high
redshift halos, for which the contribution of the peak biasing should 
be larger.

\section{Conclusion}
\label{sec:conclusion}

In standard approaches to clustering, biased tracers are treated as
a continuous field. Their overabundance is commonly assumed to be a 
local function of the smoothed mass density field. Gravitational 
motions can be straightforwardly included through a spherical 
collapse prescription or perturbation theory to predict galaxy and
dark matter halo correlation functions.

In the peak model where the tracers form a well-behaved point process,
the peak constraint renders the calculation almost prohibitive. For
this reason, all analytic studies of peak correlations thus far 
\citep{1986ApJ...304...15B,1995MNRAS.272..447R,1999ApJ...525..543M,
2008PhRvD..78j3503D,2010PhRvD..81b3526D} have obtained results which 
are strictly valid at leading order and in the (Gaussian) initial 
conditions solely .

To make predictions which can be directly compared with observational 
data or the outcome of numerical experiments, we have 
calculated the correlation of density peaks in the Gaussian linear 
mass field and computed its redshift evolution, consistently within
the Zel'dovich approximation. We have then used our model to explore 
the effect of peak biasing on the baryonic acoustic feature (BAO).

In the first part of this work (Sec.\S\ref{sec:xipk2nd}), we obtained 
a compact expression for the correlation of initial density maxima up 
to second order. We demonstrated that the $k$-independent pieces of the 
peak bias factors -- $\bias{10}$, $\bias{20}$, etc. -- are equal 
to the peak-background split biases derived from the peak number density 
$\bnpk(\nu,R_S)$, even though the later is not universal. Furthermore, 
we also showed that the peak-background split approach can be 
applied to derive the $k$-dependent bias factors -- $\bias{01}$,
$\bias{11}$, etc. This is especially interesting as it illustrates
how the peak-background split should be implemented in any other model
where the abundance of tracers depends on variables other than the 
local mass overdensity. So far however, we have not found a 
straightforward derivation of the terms linear and independent of 
$\bias{II}$ in the initial peak correlation.

In the second part (Sec.\S\ref{sec:evol}), we computed the gravitational 
evolution of the peak correlation under the assumption that the peaks 
locally flow with the matter. For simplicity, we considered the 
Zel'dovich approximation in which the initial density peaks execute a 
motion along straight line trajectories. Using the approach laid out 
in \citep{1996ApJ...472....1B}, we showed that the evolved peak 
correlation function can be expressed, like in 
renormalized perturbation theory \citep{2006PhRvD..73f3519C,
2008PhRvD..77f3530M} as the sum of a linear contribution dominant at
large scales, and a nonlinear contribution which describes the generation 
of power at a given scale due to the coupling between two modes (at 
one-loop level in our calculation) at different scales. The linear
term can be easily derived from a continuity equation argument for 
the average peak overabundance if one includes the peak velocity
bias $\bias{vpk}$. If one ignores the peak constraint, then our results 
reduce to those predicted by previous local Lagrangian bias calculations.
Thus, the peak model is a useful generalization of this model.
In particular, local Lagrangian biasing provides a good approximation
to our results only in the limit of large peak height and separation, 
and only if velocity bias is ignored.

We applied our model to predict the shape and amplitude of the 
baryon acoustic oscillation in the correlation of peaks at the epoch
of collapse. Most of the initially strong scale-dependent bias across
the BAO reported in \cite{2008PhRvD..78j3503D} is washed out by velocity 
diffusion -- which manifests itself as an exponential damping kernel. 
Still, for $2\sigma$ peaks collapsing at $z=0.3$, our model predicts a 
residual $\sim$5\% scale-dependent bias which should be quite asymmetric
about the acoustic scale. This prediction stands in stark contrast to 
the negligible scale dependence predicted by the local bias 
approximation for the same peak height. For 3$\sigma$ peaks, the 
mode-coupling power dominates so that the predictions of both models 
are quite similar: the bias exhibits a broad bump at distances 
$r\sim 90-100\hmpc$ smaller than the acoustic scale.

We then measured the clustering of massive halos extracted from a very
large N-body simulation. For halos with mass $M>2\times 10^{14}\hmsun$ -- 
which corresponds roughly to a peak significance $\nu>2$ -- the large 
scale bias shows strong evidence for a $\sim$5\% scale dependence around 
the baryonic acoustic feature. We have tried to reproduce this measurement 
using our model combined with a simple prescription to account for the 
observed halo counts. Our prediction, which is a weighted contribution
of the scale dependence induced by Lagrangian peak bias and gravity
mode-coupling, is in good agreement with the data. However, the model
slightly overestimates the cumulative halo abundance.

Clearly, there are a number of important missing ingredients -- such 
as the second order contribution to the peak displacement and a better 
treatment of nonlinear collapse -- which must be accounted to achieve
more accurate predictions. In this regards, all the machinery developed 
for the local bias scheme can be applied to the peak model, with the 
caveat that the bias factors, including the linear one, are $k$-dependent.
Numerical simulations will be essential to calibrate some of the 
parameters -- such as the diffusion scale appearing the propagator 
\citep[see, e.g.,][]{2010ApJ...720.1650S} -- or determine what is the 
actual shape of the filtering kernel. Future work should also include 
redshift-space distortions, non-Gaussian initial conditions and massive 
neutrino species, and investigate the effect of small scale exclusion and 
stochasticity on the peak power spectrum and correlation function.

\section*{Acknowledgments}

V.D. wish to thank the Center for Cosmology and Particle Physics at
New-York University, the Center for Particle Cosmology at University 
of Pennsylvania and the Institut de Ci\`{e}ncies de l'Espai at 
Universitat Aut\`{o}noma de Barcelona for their hospitality 
during the redaction of this paper. VD, RS and RKS would like
to thank the Centro de Ciencias de Benasque ``Pedro Pascual'' and 
the organizers of the Benasque 2010 workshop on modern cosmology for 
a very enjoyable and productive meeting. 
We are grateful to Takahiko Matsubara for correspondence, Cristiano 
Porciani, Eniko Reg\"os and Robert Smith for interesting discussions.
We acknowledge the use of numerical simulations from the {\small MICE}
collaboration (http://www.ice.cat/mice). We acknowledges partial 
support from FK UZH 57184001 and the Swiss National Foundation under 
contract 200021-116696/1 (VD); from Spanish Ministerio de Ciencia e 
Innovacion (MICINN), projects AYA2009-13936 Consolider-Ingenio 
CSD2007-00060, Juan de la Cierva program, and project 2009SGR1398 
from Generalitat de Catalunya (MC); from NASA NNX10AI71GS02 and 
NSF AST-0607747 (RS) and from NSF AST-0908241 (RKS).

\appendix

\section{Correlation of initial density peaks at the second order}
\label{app:xpk2L}

In this Appendix, we present the derivation of the correlation of
initial density maxima Eq.~(\ref{eq:2ptkacrice}) at the second order
in the mass correlation $\xi_0^{(0)}$ and its derivatives.

\subsection{SVT (scalar-vector-tensor) decomposition of the covariance matrix}

Computing $\xpk(r)$ requires knowledge of the joint probability
distribution  $P_2(\vy(\vx_1),\vy(\vx_2);\vr)$ where
$\vy^\top=(\eta_i,\nu,\zeta_A)$ is a ten-dimensional vector whose
components $\zeta_A, A=1,\cdots,6$ symbolize the independent entries
$ij=11,22,33,12,13,23$ of $\zeta_{ij}$.

We shall proceed in a way analogous to cosmological perturbation
theory \citep[e.g.][and references
therein]{Lifshiftz1946,1970ApJ...162..815P,
1980PhRvD..22.1882B,1984PThPS..78....1K,1992PhR...215..203M,
1994FCPh...15..209D,1995ApJ...455....7M,2004astro.ph..2060H}, and
decompose the variables into irreducible components according to their
transformation properties under spatial rotations. Although this
decomposition is not adequate for imposing the peak constraint, it
greatly simplifies the structure of the covariance matrix $C(r)$ of
the joint probability density $P_2$ \citep[see,
e.g.,][]{1995MNRAS.272..447R}. In this work, we choose the {\it
covariant} helicity basis $(\ve_+,\rvh,\ve_-)$ as reference frame,
where
\begin{equation}
\ve_+\equiv\frac{i\evh_\phi-\evh_\theta}{\sqrt{2}},~~~ \rvh\equiv
\vr/r,~~~ \ve_-\equiv\frac{i\evh_\phi+\evh_\theta}{\sqrt{2}}
\label{eq:hbasis}
\end{equation}
and $\evh_\theta$, $\evh_\phi$ are orthonormal vectors in spherical
coordinates $(\theta,\phi)$. The orthogonality relations between these
vectors are $\ve_{\pm}\cdot\ve_{\pm}=\rvh\cdot\rvh=1$ and
$\ve_+\cdot\ve_-=\ve_{\pm}\cdot\rvh=0$, where the inner product
between two vectors $\vu$ and $\vv$ is defined as $\vu\cdot\vv\equiv
u_i \ov{v}_i\equiv u_i v^i$. An overline will denote complex
conjugation throughout this Section. The property
$\ov{\vu\cdot\vv}=\vv\cdot\vu$ follows from our definition of the
inner product.  For the first derivative of the density field, the SVT
decomposition is
\begin{equation}
\veta\equiv \veta^{(S)} + \veta^{(V)}\equiv \eta^{(0)}\rvh  +
\eta^{(+1)}\evh_+ + \eta^{(-1)}\ve_-\;.
\end{equation}
Here, $\eta^{(0)}\equiv\veta\cdot\rvh$ is the {\it contravariant}
component of an irrotational vector  (spin-0),
$\rvh\wedge\veta^{(S)}=0$, whereas $\eta^{(\pm
1)}\equiv\veta\cdot\ve_{\pm}$ are the two independent {\it
contravariant} components of the transverse vector $\veta^{(V)}$
(spin-1), $\veta^{(V)}\cdot\rvh=0$. The correlation properties of
$\eta^{(0)}$ and $\eta^{(\pm 1)}$ can be obtained by projecting out
the scalar and vector parts of the correlation of the Cartesian
components $\eta_i$ which, because of statistical isotropy and
symmetry in $i,j$, is of the form
\begin{equation}
\la \eta_i(\vx_1)\eta_j(\vx_2)\ra = H_{||}(r) \rh_i\rh_j 
+ H_\perp(r) P_{ij}\;,
\label{eq:vdecomp}
\end{equation}
where
\begin{equation}
P_{ij}\equiv\left(\vii_3-\rvh\otimes\rvh\right)_{ij} = 
e_{+i}\ov{e}_{+j} + e_{-i}\ov{e}_{-j}
\label{eq:projector}
\end{equation}
is the projection operator onto the plane perpendicular to $\rvh$, and
$\vii_n$ is the $n\times n$ identity matrix. The vectors
$\ov{\ve}_{\pm}$ will be called {\it contravariant} basis
vectors. They satisfy $\ve_{\pm}\cdot\ov{\ve}_{\pm}= e_{\pm i} e_{\pm
i}\equiv 0$, which follows from the relations   $\ov{e}_{\pm}^i\equiv
e_{\pm i}$ and $\ov{e}_{\pm i}\equiv e_{\pm}^i$ between contravariant
and covariant basis. The {\it covariant} components of $\veta^{(V)}$
thus are  $\ov{\eta}^{\pm 1}\equiv \veta\cdot\ov{\ve}_{\pm}$, whereas
$\ov{\eta}^{(0)}=\eta^{(0)}$.  The
functions $H_{||}=(\xi_0^{(1)}-2\xi_2^{(1)})/3\sigma_1^2$ and
$H_\perp=(\xi_0^{(1)}+\xi_2^{(1)})/3\sigma_1^2$ are radial and
transverse correlations \citep[see, e.g.,][]{1988ApJ...332L...7G}.
Using $\eta^{(\pm 1)}=\eta_i e_{\pm}^i$  and  $\ov{\eta}^{(\pm
1)}=\eta_i \ov{e}_{\pm}^i$, we find that the correlations of the
spin-0 and  spin-1 components read
\begin{equation}
\la \eta^{(0)}_1\eta^{(0)}_2\ra = H_{||}(r),~~~ \la \eta^{(\pm
1)}_1\ov{\eta}^{(\pm 1)}_2\ra = H_\perp(r),~~~ \la \eta^{(\pm
1)}_1\ov{\eta}^{(\mp 1)}_2\ra = 0\;.
\label{eq:vcorr}
\end{equation}
Here and henceforth, the subscripts ``1'' and ``2'' will denote
variables  evaluated at position $\vx_1$ and $\vx_2$ for shorthand
convenience. Let us now consider the symmetric tensor $\zeta_{ij}$.
Writing $\zeta_{ij}$ as the sum of a traceless symmetric matrix
$\tilde{\zeta}_{ij}$  and its orthogonal complement, $\zeta_{ij}\equiv
\tilde{\zeta}_{ij}-(u/3)\vii_3$ where  $u\equiv -\tr\zeta=-\zeta_i^i$,
a suitable parameterization of the SVT decomposition for $\zeta_{ij}$
is \cite[e.g.,][]{1998PhRvD..57.3199D}
\begin{equation}
\zeta_{ij} \equiv 
-\frac{1}{3}u\,\delta_{ij} + S_{ij}\zeta^{(S)}  + \sqrt{\frac{1}{3}}
\left(\zeta^{(V)}_i\rh_j+\zeta^{(V)}_j\rh_i\right)  +
\sqrt{\frac{2}{3}} \zeta_{ij}^{(T)} \;.
\label{eq:tdecomp}
\end{equation}
The functions $u\equiv -\tr\zeta=-\zeta_i^i$ and
$\zeta^{(S)}\equiv\zeta^{(0)}$ are the longitudinal and transverse
spin-0 modes, $\zeta^{(V)}_i$ are the components of a spin-1 vector,
$\vzeta^{(V)}\cdot\rvh=0$, and $\zeta_{ij}^{(T)}$ is a symmetric,
traceless, transverse (spin-2) tensor,
$\delta^{ij}\zeta_{ij}^{(T)}=\zeta_{ij}^{(T)}\rh^j=0$.  Explicit
expressions for these functions are
\begin{align}
\zeta^{(S)} &\equiv \frac{3}{2} S^{lm}\zeta_{lm} = \frac{1}{2}
\left(3\rh^l\rh^m-\delta^{lm}\right)\zeta_{lm} \\ \zeta^{(V)}_i
&\equiv \sqrt{3}V_i^{lm}\zeta_{lm} =
\sqrt{3}\left(\delta_i^l-\rh_i\rh^l\right)\rh^m\zeta_{lm} \\
\zeta_{ij}^{(T)} &\equiv \sqrt{\frac{3}{2}}T_{ij}^{lm}\zeta_{lm} =
\sqrt{\frac{3}{2}}\left(P_i^l P_j^m - \frac{1}{2} P_{ij} P^{lm}\right)
\zeta_{lm}\;.
\end{align}
We have introduced factors of $\sqrt{1/3}$ and $\sqrt{2/3}$ in the
decomposition (\ref{eq:tdecomp}) such that the zero-point moments of
the spin-0, spin-1 and spin-2 tensor variables all equal $1/5$ (see
equations (\ref{eq:tspin0}), (\ref{eq:tspin1}) and (\ref{eq:tspin2})
below). Note that $P_i^j\equiv e_{+i} e_+^j+e_{-i} e_-^j$ and
$P^{ij}\equiv e_+^i\ov{e}_+^j+e_-^i\ov{e}_-^j$ to ensure nilpotency
of the operator $P$.  The 2-point correlation function of any
isotropic, symmetric tensor field $\zeta_{ij}$ is of the form
\citep[e.g.,][]{1975ctf..book.....L}
\begin{equation}
\label{eq:tcorr}
\begin{split}
\la \zeta_{ij}(\vx_1)\zeta_{lm}(\vx_2)\ra &= Z_1(r)\,\rh_i \rh_j \rh_l
\rh_m  +Z_2(r)\,\left(\rh_i\rh_l\delta_{jm}+\rh_i \rh_m\delta_{jl}
+\rh_j \rh_l\delta_{im}+\rh_j\rh_m\delta_{il}\right) \\ &\quad
+Z_3(r)\,\left(\rh_i\rh_j\delta_{lm}+\rh_l \rh_m\delta_{ij}\right)
+Z_4(r)\,\delta_{ij}\delta_{lm}
+Z_5(r)\left(\delta_{il}\delta_{jm}+\delta_{im}\delta_{jl}\right)\;.
\end{split}
\end{equation}
In the case of the tensor $\zeta_{ij}=\partial_i\partial_j\delta$
considered here, we have $Z_2=Z_3$ and $Z_4=Z_5$. To project out the
transverse  spin-0, spin-1 and spin-2 parts of the 4-rank correlation
tensor eq.(\ref{eq:tcorr}),  we act with the scalar, vector and tensor
projection operators $S_{ab}$, $V_a^{bc}$, $T_{ab}^{cd}$ on 
$\la\zeta_{ij}(\vx_1)\zeta_{lm}(\vx_2)\ra$ and obtain
\begin{align}
\la\zeta^{(S)}(\vx_1)\,\zeta^{(S)}(\vx_2)\ra &= Z_1(r)+4 Z_3(r)
+ 3Z_5(r) \\ 
\la\zeta_i^{(V)}(\vx_1)\,\zeta_j^{(V)}(\vx_2)\ra &= 
3 P_{ij} \left[Z_3(r)+Z_5(r)\right]\\
\la\zeta_{ij}^{(T)}(\vx_1)\,\zeta_{lm}^{(T)}(\vx_2)\ra &= 
3 T_{ijlm} Z_5(r)\;,
\end{align}
where 
\begin{gather}
Z_1+4 Z_3+3 Z_5 = \frac{1}{\sigma_2^2} \left(\frac{1}{5}\xi_0^{(2)}
-\frac{2}{7}\xi_2^{(2)}+\frac{18}{35}\xi_4^{(2)}\right),~~~
Z_3+Z_5 = \frac{1}{\sigma_2^2}\left(\frac{1}{15}\xi_0^{(2)}
-\frac{1}{21}\xi_2^{(2)}-\frac{4}{35}\xi_4^{(2)}\right), \\
Z_5 = \frac{1}{\sigma_2^2}\left(\frac{1}{15}\xi_0^{(2)}
+\frac{2}{21}\xi_2^{(2)}+\frac{1}{35}\xi_4^{(2)}\right)\nonumber \;.
\end{gather}
The correlation for the spin-0 variable $\zeta^{(0)}$ thus is
\begin{equation}
\la \zeta_1^{(0)} \zeta_2^{(0)}\ra = Z_1 + 4 Z_3 + 3Z_5\;.
\label{eq:tspin0}
\end{equation}
Taking the 2 independent components of $\vzeta^{(V)}$ and their
complex conjugate as being $\zeta^{(\pm 1)}\equiv
\vzeta^{(V)}\cdot\ve_{\pm}=\sqrt{3}\,e_{\pm}^i\rh^j\zeta_{ij}$ and
$\czeta^{(\pm 1)}\equiv\vzeta^{(V)}\cdot\ov{\ve}_{\pm}
=\sqrt{3}\,\ov{e}_{\pm}^i\rh^j\zeta_{ij}$ yields
\begin{equation}
\la \zeta_1^{(\pm 1)}\czeta_2^{(\pm 1)}\ra = 3(Z_3+Z_5),~~~
\la \zeta_1^{(\pm 1)}\czeta_2^{(\mp 1)}\ra = 0\;.
\label{eq:tspin1}
\end{equation}
Similarly, we choose $\zeta^{(\pm 2)}\equiv \zeta_{ij}^{(T)} e_{\pm}^i
e_{\pm}^j=\sqrt{3/2}\,e_{\pm}^i e_{\pm}^j\zeta_{ij}$ and  $\czeta^{(\pm
2)}\equiv \zeta_{ij}^{(T)} \ov{e}_{\pm}^i
\ov{e}_{\pm}^j=\sqrt{3/2}\,\ov{e}_{\pm}^i\ov{e}_{\pm}^j\zeta_{ij}$ 
for the two independent polarizations and their complex conjugate, 
respectively. The correlation properties of these variables are
\begin{equation}
\la \zeta_1^{(\pm 2)}\czeta_2^{(\pm 2)}\ra = 3 Z_5,~~~
\la \zeta_1^{(\pm 2)}\czeta_2^{(\mp 2)}\ra = 0\;.
\label{eq:tspin2}
\end{equation}
As expected, these are the only spin-2 degrees of freedom since
$\zeta_{ij}^{(T)}e_+^i \ov{e}_-^j\equiv 0$. Therefore, the second rank
tensor $\zeta_{ij}$ is fully characterized by the set of variables
$\{u,\zeta^{(0)},\zeta^{(\pm 1)},\zeta^{(\pm 2)}\}$. Furthermore,  all
the correlations are real despite the fact that some of these
variables are complex.  We note that, in the particular case
$\rvh\equiv\zvh$, our variables are directly related to the variables
$y_{lm}^n$ defined in \cite{1995MNRAS.272..447R}, who work in the
spherical basis
\begin{equation}
\left(\frac{i\yvh-\xvh}{\sqrt{2}},\zvh,\frac{i\yvh+\xvh}{\sqrt{2}}
\right)\;.
\label{eq:sbasis}
\end{equation}
The transformation from the helicity to the spherical basis vectors is
performed by the rotation operator $D^1\!(0,\theta,\phi)$
(whose  matrix elements are the Wigner $D$-functions $D^{\ell=1}_{m
m'}\!(0,\theta,\phi)$).  For the traceless tensor
$\tilde{\zeta}_{ij}$, the spin-0, spin-1 and  spin-2 components in the
spherical basis simplify to
\begin{gather}
\zeta^{(0)}=\frac{1}{2}\left(2\zeta_{33}-\zeta_{11}-\zeta_{22}\right),~~~
\zeta^{(\pm 1)}=\mp\sqrt{\frac{3}{2}}\left(\zeta_{13}\pm
i\zeta_{23}\right),~~~ \zeta^{(\pm
2)}=\sqrt{\frac{3}{8}}\left(\zeta_{11}-\zeta_{22}\pm
2i\zeta_{12}\right)\;.
\label{eq:zeta_sbasis}
\end{gather}
The relationship between the two sets of variables thus is
$\zeta^{(m)}\!(\vx)=y_{2m}^0(\vx)/\sqrt{5}$ (see eq. (22) of
\cite{1995MNRAS.272..447R}). For the gradient $\veta(\vx)$ of the
density field, the correspondence is
$\eta^{(m)}\!(\vx)=y_{1m}^0(\vx)/\sqrt{3}$. 

The 20-dimensional covariance matrix $\vcc(r)\equiv\la\vy\vy^\dagger\ra$,
where $\vy=(\vy_1,\vy_2)$ and $\vy^\dagger$  is its conjugate
transpose, describes the correlations of the fields at position
$\vx_1$ and $\vx_2$. For simplicity, we will assume in  what follows
that these are smoothed on the same mass scale.  $\vcc(r)$ may be
partitioned into four $10\times 10$ block matrices, the zero-point
contribution $\vmm$ in the top left and bottom right corners, and the
cross-correlation matrix $\vbb(r)$ and its transpose in the bottom
left and top right corners, respectively.  In terms of the variables
$\vy_i=(\eta_i^{(0)},\nu_i,u_i,\zeta_i^{(0)},
\eta_i^{(+1)},\zeta_i^{(+1)}, \eta_i^{(-1)},\zeta_i^{(-1)},$
$\zeta_i^{(+2)},\zeta_i^{(-2)})$, $\vmm$ and $\vbb$ have the block
diagonal decomposition $\vmm={\rm
diag}(\vmm^{(0)},\vmm^{(1)},\vmm^{(1)}, \vmm^{(2)},\vmm^{(2)})$ and
$\vbb={\rm diag}(\vbb^{(0)},\vbb^{(1)},$
$\vbb^{(1)},\vbb^{(2)},\vbb^{(2)})$. Explicitly,
\begin{equation}
\vmm^{(0)}=\left(\begin{array}{cccc} 1/3 & 0 & 0 & 0 \\ 0 & 1 &
\gamma_1 & 0 \\ 0 & \gamma_1 & 1 & 0 \\ 0 & 0 & 0 & 1/5
\end{array}\right),~~~ \vmm^{(1)}=\left(\begin{array}{cc} 1/3 & 0 \\ 0
& 1/5 \end{array}\right),~~~ \vmm^{(2)}=\frac{1}{5}\;,
\label{eq:covm}
\end{equation}
and
\begin{gather}
\label{eq:covb}
\vbb^{(0)}=\left(\begin{array}{cccc} 
\frac{1}{3\sigma_1^2}\bigl(\xi_0^{(1)}-2\xi_2^{(1)}\bigr) & 
-\frac{1}{\sigma_0\sigma_1}\xi_1^{(1/2)} & 
-\frac{1}{\sigma_1\sigma_2}\xi_1^{(3/2)} & 
-\frac{1}{5\sigma_1\sigma_2}\bigl(3\xi_3^{(3/2)}-2\xi_1^{(3/2)}\bigr) \\
\frac{1}{\sigma_0\sigma_1}\xi_1^{(1/2)} & 
\frac{1}{\sigma_0^2}\xi_0^{(0)} & 
\frac{1}{\sigma_0\sigma_2}\xi_0^{(1)} & 
\frac{1}{\sigma_0\sigma_2}\xi_2^{(1)} \\
\frac{1}{\sigma_1\sigma_2}\xi_1^{(3/2)} & 
\frac{1}{\sigma_0\sigma_2}\xi_0^{(1)} & 
\frac{1}{\sigma_2^2}\xi_0^{(2)} & 
\frac{1}{\sigma_2^2}\xi_2^{(2)} \\
\frac{1}{5\sigma_1\sigma_2}\bigl(3\xi_3^{(3/2)}-2\xi_1^{(3/2)}\bigr) & 
\frac{1}{\sigma_0\sigma_2}\xi_2^{(1)} & 
\frac{1}{\sigma_2^2}\xi_2^{(2)} & 
\frac{1}{5\sigma_2^2}\bigl(\xi_0^{(2)}-\frac{10}{7}\xi_2^{(2)}
+\frac{18}{7}\xi_4^{(2)}\bigr) 
\end{array}\right)\;, \\ 
\vbb^{(1)}=\left(\begin{array}{cc} 
\frac{1}{3\sigma_1^2}\bigl(\xi_0^{(1)}+\xi_2^{(1)}\bigr) & 
-\frac{\sqrt{3}}{5\sigma_1\sigma_2}\bigl(\xi_1^{(3/2)}+\xi_3^{(3/2)}\bigr) \\ 
\frac{\sqrt{3}}{5\sigma_1\sigma_2}\bigl(\xi_1^{(3/2)}+\xi_3^{(3/2)}\bigr) & 
\frac{1}{5\sigma_2^2}\bigl(\xi_0^{(2)}-\frac{5}{7}\xi_2^{(2)}
-\frac{12}{7}\xi_4^{(2)}\bigr)
\end{array}\right) \\
\vbb^{(2)}=\frac{1}{\sigma_2^2}\left(\frac{1}{5}\xi_0^{(2)}
+\frac{2}{7}\xi_2^{(2)}+\frac{3}{35}\xi_4^{(2)}\right) \;.
\end{gather}
The second and third entries of the first row of $\vbb^{(0)}$ are the
correlations of  $\eta_2^{(0)}$ with $\nu_1$ and $u_1$,
respectively. As expected, these correlations are negative if
$\nu_1>0$ or $u_1>0$ since, in this case, the line-of-sight derivative
is  preferentially directed towards $\vx_1$.  The determinant of
$C(r)$ reads as
\begin{equation}
\det\vcc(r) = \det\vmm^{(0)}\det\!\left[\vmm^{(0)}
-\vbb^{(0)\top}\left(\vmm^{(0)}\right)^{-1}
\vbb^{(0)}\right]\prod_{s=1,2}\left(\det\vmm^{(s)}\right)^2
\det\!\left[\vmm^{(s)}-\vbb^{(s)\top}\left(\vmm^{(s)}\right)^{-1}
\vbb^{(s)}\right]^2\;.
\end{equation}
It is worth noticing that, although $\vcc(r)$ does not depend on the
direction $\rvh$ of the separation vector $\vr$,  it is  {\it not}
equal to the angular average covariance matrix $\wi{\vcc}(r)\equiv
(1/4\pi)\int\!d\Omega_{\rvh}\,\vcc(\vr)$. The latter follows upon
setting  $\xi_j^{(n)}\equiv 0$ whenever $j\neq 0$ in  equations
(\ref{eq:covm}) and (\ref{eq:covb}). Furthermore, whereas the 2-point
probability distribution $P_2(\vy_1,\vy_2;\vr)$ associated to the
correlation matrix $\vcc(r)$ cannot be easily expressed in closed
form, the joint probability density $\wi{P}_2(\vy_1,\vy_2;r)$ of
covariance $\wi{\vcc}(r)$ may be  exactly written as the product
\begin{equation}
\wi{P}_2(\vy_1,\vy_2;r)=\wi{P}_2(\nu_1,u_1,\nu_2,u_2;r)\,
\wi{P}_2(\veta_1,\veta_2;r)\,\wi{P}_2(\tilde{\zeta}_1,
\tilde{\zeta}_2;r)\;.
\end{equation}
This factorization property reflects the fact that, upon angle 
averaging, the $\ell=0$ ($\nu_i,u_i$), $\ell=1$ ($\eta_i$) and $\ell=2$ 
($\tilde{\zeta}_{ij}$) representations of SO(3) decouple from each other. 
Consequently, it should be possible to cast the 2-point probability 
densities  in terms of rotational invariants such as the scalar product 
and the matrix trace.  After some manipulations, we find the joint
probability density for the $\ell=1$ and $\ell=2$ variables is
\begin{align}
\label{eq:p2eta}
\wi{P}_2(\veta_1,\veta_2;r) &= \left(\frac{3}{2\pi}\right)^3
\left(1-\Upsilon_1^2\right)^{-3/2}
\exp\left[-\frac{3\veta_1^2+3\veta_2^2-6\Upsilon_1\veta_1\cdot\veta_2}
{2\left(1-\Upsilon_1^2\right)}\right] \\
\label{eq:p2zeta}
\wi{P}_2(\tilde{\zeta}_1,\tilde{\zeta}_2;r) &=
\frac{1}{20}\frac{15^6}{(2\pi)^5} \left(1-\Upsilon_2^2\right)^{-5/2}
\exp\left[-\left(\frac{15}{4}\right)\frac{\tr\left(\tilde{\zeta}_1^2\right)
+\tr\left(\tilde{\zeta}_2^2\right)-2\Upsilon_2\tr\left(\tilde{\zeta}_1
\tilde{\zeta}_2\right)}{\left(1-\Upsilon_2^2\right)}\right]\;,
\end{align}
where we have defined $\Upsilon_n(r)=\xi_0^{(n)}(r)/\sigma_n^2$ for sake
of conciseness. The following relations
\begin{gather}
\label{eq:spin1dot}
\eta_1^{(0)}\eta_2^{(0)}+\eta_1^{(+1)}\ov{\eta}_2^{(+1)}+\eta_1^{(-1)}
\ov{\eta}_2^{(-1)} = \veta_1\cdot\veta_2 \\
\zeta_1^{(0)}\zeta_2^{(0)}+\sum_{s=1,2}\left(\zeta_1^{(+s)}\czeta_2^{(+s)}+
\zeta_1^{(-s)}\czeta_2^{(-s)}\right) = 
\frac{3}{2}\tr\left(\tilde{\zeta}_1\tilde{\zeta}_2\right)
\end{gather}
can be useful to derive equations (\ref{eq:p2eta}) and
(\ref{eq:p2zeta}). When the peak constraint is enforced, 
$\wi{P}_2(\veta_1,\veta_2;r)$ reduces to a simple multiplicative factor.
Finally, the joint probability density $\wi{P}_2(\nu_1,u_1,\nu_2,u_2;r)$ 
for the $\ell =0$ degrees of freedom evaluates to
\begin{equation}
\wi{P}_2(\nu_1,u_1,\nu_2,u_2;r)=
\frac{e^{-\Phi(\nu_1,u_1,\nu_2,u_2;r)}}
{(2\pi)^2\sqrt{\left(1-\gamma_1^2\right)\Delta_{\rm P}}}\;,
\end{equation}
Here, $\Phi(\nu_1,u_1,\nu_2,u_2;r)$ is the quadratic form associated
to the  inverse covariance matrix $\vcc_{\nu u}^{-1}(r)\equiv
(\vpp,\vrr^\top;\vrr,\vpp)$.  The $2\times 2$ matrix $\vpp$ in the top
left and bottom right corners reads as
\begin{equation}
\vpp= \frac{1}{\Delta_{\rm P}} \left(\begin{array}{cc}\vpp_{11} & \vpp_{12}
\\ \vpp_{12} & \vpp_{22}
       \end{array}\right)
=\frac{1}{\Delta_{\rm P}}
\left(\begin{array}{cc} 
1-\frac{\Upsilon_2^2-2\gamma_1^2\Upsilon_1\Sigma_2+\gamma_1^2\Upsilon_1^2}
{1-\gamma_1^2} &
-\gamma_1+\frac{\gamma_1\Upsilon_1\left(\Upsilon_0-\gamma_1^2\Upsilon_1\right)
-\gamma_1\Upsilon_2\left(\Upsilon_0-\Upsilon_1\right)}{1-\gamma_1^2} \\
-\gamma_1+\frac{\gamma_1\Upsilon_1\left(\Upsilon_0-\gamma_1^2\Upsilon_1\right)
-\gamma_1\Upsilon_2\left(\Upsilon_0-\Upsilon_1\right)}{1-\gamma_1^2} &
1-\frac{\Upsilon_0^2-2\gamma_1^2\Upsilon_0\Upsilon_1+\gamma_1^2\Upsilon_1^2}
{1-\gamma_1^2}
\end{array}\right)\;,
\end{equation}
whereas the matrix $\vrr$ in the top right and bottom left corners is
\begin{equation}
\vrr=-\frac{1}{\left(1-\gamma_1^2\right)\Delta_{\rm P}}
\left(\begin{array}{cc}
\left(\Upsilon_0-\gamma_1^2\Upsilon_1\right)\vpp_{11}
-\gamma_1\left(\Upsilon_2-\Upsilon_1\right)\vpp_{12} &
\left(\Upsilon_0-\gamma_1^2\Upsilon_1\right)\vpp_{12}
-\gamma_1\left(\Upsilon_2-\Upsilon_1\right)\vpp_{22} \\
\left(\Upsilon_0-\gamma_1^2\Upsilon_1\right)\vpp_{12}
-\gamma_1\left(\Upsilon_2-\Upsilon_1\right)\vpp_{22} &
\left(\Upsilon_2-\gamma_1^2\Upsilon_1\right)\vpp_{22}
-\gamma_1\left(\Upsilon_0-\Upsilon_1\right)\vpp_{12}
\end{array}\right)\;.
\end{equation}
Notice that the determinant $\Delta_{\rm P}=\vpp_{11}\vpp_{22}-\vpp_{12}^2$ 
asymptotes to $1-\gamma_1^2$ in the limit  $r\rightarrow\infty$. 

\subsection{Series expansion of the joint probability density}

To calculate the correlation function of initial density peaks at the
second order, we first separate the covariance matrix into 
$\vcc(r)\equiv\wi{\vcc}(r)+\delta\vcc(r)$. The angular average 
$\wi{\vcc}(r)\equiv(\vmm,\wi{\vbb}^\top;\wi{\vbb},\vmm)$ contains the
zero-point moments and the cross-correlation entries with
$\xi_0^{(n)}$ solely, whereas $\delta\vcc(r)\equiv
(0_{10},\delta\vbb^\top;\delta\vbb,0_{10})$, where $0_{10}$ is the
$10\times 10$ zero matrix, encodes the cross-correlations $\xi_{j\neq
0}^{(n)}$.  We have for instance
$\wi{\vbb}^{(2)}=\xi_0^{(2)}/(5\sigma_2^2)$ and
$\delta\vbb^{(2)}=(10\xi_2^{(2)}+3\xi_4^{(2)})/(35\sigma_2^2)$. Using
the identity
$\det(\vii+\vxx)=1+\tr\vxx+(1/2)[(\tr\vxx)^2-\tr(\vxx^2)]+\cdots$, we
expand the joint density $P_2(\vy_1,\vy_2;r)$ in the small
perturbation $\delta\vcc(r)$ and arrive at
\begin{align}
P_2(\vy_1,\vy_2;\vr) &\approx
\wi{P}_2(\vy_1,\vy_2;r)
\biggl[1-\frac{1}{2}\tr\!\Bigl(\wi{\vcc}^{-1}\delta\vcc\Bigr)
+\frac{1}{4}\tr\!\Bigl(\wi{\vcc}^{-1}\delta\vcc\wi{\vcc}^{-1}\delta\vcc
\Bigr)\biggr] \\
&\qquad \times \biggl[1+\frac{1}{2}\vy^\dagger\wi{\vcc}^{-1}\delta\vcc
\wi{\vcc}^{-1}\vy+\frac{1}{8}\Bigl(\vy^\dagger\wi{\vcc}^{-1}\delta\vcc
\wi{\vcc}^{-1}\vy\Bigr)^2-\frac{1}{2}\vy^\dagger\wi{\vcc}^{-1}\delta\vcc
\wi{\vcc}^{-1}\delta\vcc\wi{\vcc}^{-1}\vy\biggr] \nonumber\;.
\end{align} 
at second order in $\delta\vcc$. Here, the product
$\wi{\vcc}^{-1}\delta\vcc\wi{\vcc}^{-1}$ is of order  ${\cal
O}(\xi)$ in the correlation functions $\xi_\ell^{(n)}$, whereas
$\tr\!(\wi{\vcc}^{-1}\delta\vcc)$,  $\tr\!(\wi{\vcc}^{-1}\delta\vcc
\wi{\vcc}^{-1}\delta\vcc)$ and  $\wi{\vcc}^{-1}\delta\vcc\wi{\vcc}^{-1}
\delta\vcc\wi{\vcc}^{-1}$ are of order ${\cal O}(\xi^2)$. Expressing 
the matrices $\wi{\vcc}$ and $\delta\vcc$ in terms of the auto- and 
cross-covariances yields
\begin{align}
\label{eq:p2expand}
P_2(\vy_1,\vy_2;\vr) &\approx \wi{P}_2(\vy_1,\vy_2;r)
\Biggl\{1+\vy_2^\dagger\vmm^{-1}\delta\vbb\,\vmm^{-1}\vy_1+\frac{1}{2}
\Bigl(\vy_2^\dagger\vmm^{-1}\delta\vbb\,\vmm^{-1}\vy_1\Bigr)^2 \\ 
&\qquad 
-\frac{1}{2}\left(\vy_1^\dagger\vqq\vy_1+\vy_2^\dagger\vqq\vy_2\right)
+\frac{1}{2}\tr\!\Bigl[\vmm^{-1}\wi{\vbb}\,\vmm^{-1}\bigl(\delta\vbb
+\delta\vbb^\top\bigr)\Bigr]
+\frac{1}{2}\tr\!\Bigl(\vmm^{-1}\delta\vbb\vmm^{-1}\delta\vbb^\top\Bigr)
\Biggr\}\nonumber
\end{align}
at order ${\cal O}(\xi^2)$, where
\begin{equation}
\vqq\equiv 2\left(\vmm^{-1}\wi{\vbb}\,\vmm^{-1}\delta\vbb\,
\vmm^{-1}\right)+\vmm^{-1}\delta\vbb^\top\vmm^{-1}\delta\vbb\,\vmm^{-1}\;.
\end{equation}
Note that $\delta\vbb$ is {\it not} symmetric, so one must distinguish
between  $\delta\vbb$ and its transpose.

Let us consider the term
$\vy_2^\dagger\vmm^{-1}\delta\vbb\vmm^{-1}\vy_1$ linear in the
correlation functions.  Owing to the block-diagonal nature of $\vcc$,
it is a sum of contributions from the spin-0, spin-1 and spin-2
degrees of freedom. While the matrix
$(\vmm^{(s)})^{-1}\delta\vbb^{(s)}(\vmm^{(s)})^{-1}$ for $s=1,2$
generally has non-vanishing elements, it is easy to check that
$(\vmm^{(0)})^{-1}\delta\vbb^{(0)}(\vmm^{(0)})^{-1}$ has zero entries
for the elements $ij=22$, 23, 32 and 33. Furthermore, imposing the
constraint $\veta_1=\veta_2\equiv 0$ implies that
$\vy_2^{(s)\dagger}(\vmm^{(s)})^{-1}\delta\vbb^{(s)}(\vmm^{(s)})^{-1}
\vy_1^{(s)}$, $s=0,1,2$ contains only terms linear in $\zeta_1^{(s)}$,
$\czeta_2^{(s)}$ and products of the form
$\zeta_1^{(s)}\czeta_2^{(s)}$. At this point, we shall remember that
the principal axes of the tensors $\zeta_1\equiv\zeta(\vx_1)$ and
$\zeta_2\equiv\zeta(\vx_2)$ are not necessarily aligned with those  of
the coordinate frame. Without loss of generality, we can write
$\zeta_1=-\vrr\Lambda\vrr^\top$, where $\vrr$ is an orthogonal matrix
that contains the angular variables (e.g Euler angles) and $\Lambda$
is the diagonal matrix consisting of the three ordered eigenvalues
$\lambda_1\geq\lambda_2\geq\lambda_3$ of $-\zeta_1$. The value of
$u_i=-\tr\zeta_1$ is invariant under rotations of the principal axes,
while $\tilde{\zeta}_1$ transforms in the same manner as the  $\ell=2$
eigenfunctions of the (orbital) angular momentum operator,  i.e. the
spherical harmonics $Y_{\ell=2}^m(\rvh)$. Namely, on inspecting a
table of spherical harmonics in Cartesian coordinates, we can write
\begin{gather}
r^2 Y_2^0(\rvh)= \frac{1}{4}\sqrt{\frac{5}{\pi}}\left(3
z^2-r^2\right),~~~ r^2 Y_2^{\pm 1}(\rvh)=
\mp\frac{1}{2}\sqrt{\frac{15}{2\pi}}z\left(x\pm i y\right),~~~ r^2
Y_2^{\pm 2}(\rvh)= \frac{1}{4}\sqrt{\frac{15}{2\pi}}\left(x\pm i
y\right)^2\;.
\end{gather}
A comparison with eq.(\ref{eq:zeta_sbasis}) shows that
$\zeta^{(m)}\sim \sqrt{5/4\pi}\, Y_2^m(\rvh)$. Therefore, the
variables $\zeta^{(m)}$  must transform in accordance with
\begin{equation}
\zeta^{(m')}\!(\vx) = 
\sum_m D^2_{m m'}\!\left(\varphi,\vartheta,\psi\right)\zeta^{(m)}\!(\vx)
\end{equation}
under rotations of the principal axis frame. Here,  $D^2_{m
m'}\!(\varphi,\vartheta,\psi)$ are quadrupole Wigner $D$-functions
with the Euler angles $(\varphi,\vartheta,\psi)$ as argument, whereas
$\zeta^{(m)}$ and $\zeta^{(m')}$ are the components of $\tilde{\zeta}$
in the original and final eigenvector frames, respectively. Therefore,
averaging over distinct orientations of the principal axes gives
$\la\zeta^{(m)}\ra=0$.  Noticing that, at the zeroth order, the joint
density $\wi{P}_2(\vy_1,\vy_2;r)$ factorizes into the product
$\wi{P}_1(\vy_1)\wi{P}_1(\vy_2)$ of one-point probability
densities $\wi{P}_1(\vy_i)$ which do not depend upon $\vrr$, we
find the correlations $\xi_j^{(n)}$, $j\neq 0$ {\it do not contribute}
to the peak correlation at the first order. This is in agreement with
the findings of \cite{1995MNRAS.272..447R,2008PhRvD..78j3503D}.

The second order terms in the right-hand side of
Eq.(\ref{eq:p2expand}) will also yield  products in the variables
$\zeta_1^{(m)}$ and $\czeta_2^{(m')}$ whose angle average can be
reduced using the orthogonality conditions
\begin{gather}
\int_{\rm SO(3)}\!\!\!\!\!\!d\vrr\,
D_{m_1 m_1'}^{\ell_1}\!(\varphi,\vartheta,\psi)\,
D_{m_2 m_2'}^{\ell_2 \star}\!(\varphi,\vartheta,\psi)=
\frac{1}{2\ell_1+1}
\delta_{\ell_1 \ell_2}\delta_{m_1 m_2}\delta_{m_1' m_2'} \\
\int_{\rm SO(3)}\!\!\!\!\!\!d\vrr\,
D_{m_! m_1'}^{\ell_1}\!(\varphi,\vartheta,\psi)\,
D_{m_2 m_2'}^{\ell_2}\!(\varphi,\vartheta,\psi)=(-1)^{m_2-m_2'}
\frac{1}{2\ell_2+1}\delta_{\ell_1 \ell_2}\delta_{-m_1 m_2}
\delta_{-m_1' m_2'}\;.
\end{gather} 
Clearly, cross-products of the form $\zeta_1^{(m)}\czeta_2^{(m')}$
vanish upon averaging over the principal axis frames because they
involve distinct rotation operators. However, when the angle average
is taken at a single position $\vx_1=\vx_2$, we obtain
\begin{align}
\label{eq:ortho1}
\left\la\zeta^{(m_1')}\czeta^{(m_2')}\right\ra &= \sum_{m_1 m_2}
\zeta^{(m_1)}\czeta^{(m_2)}\left\la D^2_{m_1 m_1'} D^{2\star}_{m_2 m_2'}
\right\ra \\
&= \frac{1}{5}\sum_{m_1 m_2}\zeta^{(m_1)}\czeta^{(m_2)}\delta_{m_1 m_2}
\delta_{m_1' m_2'} \nonumber \\
&= \frac{3}{10}\,\delta_{m_1' m_2'}\,\tr\!\left(\tilde{\zeta}^2\right)
\nonumber \;,
\end{align}
where $\la\cdots\ra$ denotes the average over orientations and we have
also omitted the arguments of the Wigner $D$-functions for
brevity. Similarly, it is easy to show that
\begin{equation}
\label{eq:ortho2}
\left\la\zeta^{(m_1')}\zeta^{(m_2')}\right\ra =
\left\la\czeta^{(m_1')}\czeta^{(m_2')}\right\ra=
\frac{3}{10}\,(-1)^{m_2'}\delta_{-m_1' m_2'}\,
\tr\!\left(\tilde{\zeta}^2\right)\;.
\end{equation}
These relations can be used to integrate out the orientation of the
two  eigenframes in the series expansion of the joint probability
density $P_2(\vy_1,\vy_2;\vr)$. For example, let us consider the
contribution $(\vy_2^\dagger\vmm^{-1}\wi{\vbb}\,\vmm^{-1}\vy_1)
(\vy_2^\dagger\vmm^{-1}\delta\vbb\vmm^{-1}\vy_1)$. After some algebra
and with the aid of Eq.(22) of \cite{2008PhRvD..78j3503D}, we can write
\begin{multline}
\label{eq:1nd1nd}
(\vy_2^\dagger\vmm^{-1}\wi{\vbb}\,\vmm^{-1}\vy_1)(\vy_2^\dagger
\vmm^{-1}\delta\vbb\vmm^{-1}\vy_1) =
\left[5\zeta_1^{(0)}\zeta_2^{(0)}+5\sum_{s=1,2}\left(\zeta_1^{(+s)}
\czeta_2^{(+s)}+\zeta_1^{(-s)}\czeta_2^{(-s)}\right)+
\mbox{terms in $\nu_i$, $u_i$}\right] \\
\times \Biggl\{\left(\frac{5}{1-\gamma_1^2}\right) 
\left[\left(\frac{1}{\sigma_0\sigma_2}\xi_2^{(1)}
-\frac{\gamma_1}{\sigma_2^2}\xi_2^{(2)}\right)
\left(\nu_2\zeta_1^{(0)}+\nu_1\zeta_2^{(0)}\right)
+\left(\frac{1}{\sigma_2^2}\xi_2^{(2)}-\frac{\gamma_1}{\sigma_0\sigma_2}
\xi_2^{(1)}\right)\left(u_2\zeta_1^{(0)}+u_1\zeta_2^{(0)}\right)\right] \\ 
+\frac{5}{7\sigma_2^2}\left[\left(-10\xi_2^{(2)}+18\xi_4^{(2)}\right)
\zeta_1^{(0)}\zeta_2^{(0)}-\left(5\xi_2^{(2)}+12\xi_4^{(2)}\right)
\sum_{s=\pm 1}\zeta_1^{(s)}\czeta_2^{(s)}
+\left(10\xi_2^{(2)}+3\xi_4^{(2)}\right)\sum_{s=\pm 2}\zeta_1^{(s)}
\czeta_2^{(s)}\right]\Biggr\}\;,
\end{multline}
on enforcing the constraint $\veta_1=\veta_2\equiv 0$. To average over 
the orientation of the tensors $\zeta_1$ and $\zeta_2$, we note that
products of the form
$\zeta_1^{(m_1)}\czeta_2^{(m_1)}\zeta_1^{(m_2)}\czeta_2^{(m_2)}$
simplify to
\begin{equation}
\left\la\zeta_1^{(m_1)}\czeta_2^{(m_1)}\zeta_1^{(m_2)}\czeta_2^{(m_2)}
\right\ra=
\left(\frac{3}{10}\right)^2\delta_{-m_1 m_2}
\tr\!\left(\tilde{\zeta}_1^2\right)
\tr\!\left(\tilde{\zeta}_2^2\right)\;.
\end{equation}
After some further manipulation, this leads to the cancellation of the
term (\ref{eq:1nd1nd}). 

However, the matrix traces in Eq.(\ref{eq:p2expand}) do not vanish on
integrating over the angular variables, 
\begin{multline}
\label{eq:2ndtrace}
\tr\!\Bigl[\vmm^{-1}\wi{\vbb}\,\vmm^{-1}
\bigl(\delta\vbb+\delta\vbb^\top\bigr)\Bigr]
+\tr\!\Bigl(\vmm^{-1}\delta\vbb\vmm^{-1}\delta\vbb^\top\Bigr) =
\frac{6}{\sigma_1^2\sigma_2^2}
\Bigl[3\bigl(\xi_3^{(3/2)}\bigr)^2+2\bigl(\xi_1^{(3/2)}\bigr)^2\Bigr]
+\frac{6}{\sigma_1^4}\bigl(\xi_2^{(1)}\bigr)^2 \\
+\frac{10}{7\sigma_2^4}
\Bigl[5\bigl(\xi_2^{(2)}\bigr)^2+9\bigl(\xi_4^{(2)}\bigr)^2\Bigr]
+\left(\frac{6}{1-\gamma_1^2}\right)\biggl[\frac{1}{\sigma_0^2\sigma_1^2}
\bigl(\xi_1^{(1/2)}\bigr)^2-2\frac{\gamma_1^2}{\sigma_1^4}\xi_1^{(1/2)}
\xi_1^{(3/2)}+\frac{1}{\sigma_1^2\sigma_2^2}\bigl(\xi_1^{(3/2)}\bigr)^2
\biggr] \\
+\left(\frac{10}{1-\gamma_1^2}\right)\biggl[\frac{1}{\sigma_0^2\sigma_2^2}
\bigl(\xi_2^{(1)}\bigr)^2-2\frac{\gamma_1^2}{\sigma_1^2\sigma_2^2}
\xi_2^{(1)}\xi_2^{(2)}+\frac{1}{\sigma_2^4}\bigl(\xi_2^{(2)}\bigr)^2\biggr]\;,
\end{multline}
nor does the second order contribution
\begin{equation}
\label{eq:0nd2nd}
\frac{1}{2}\wi{P}_2(\vy_1,\vy_2;r)
\biggl[\left(\vy_2^\dagger\vmm^{-1}\delta\vbb\,\vmm^{-1}\vy_1\right)^2
-\left(\vy_1^\dagger\vqq\vy_1+\vy_2^\dagger\vqq\vy_2\right)\biggr]
\approx \frac{1}{2}\wi{P}_1(\vy_1)\wi{P}_1(\vy_2)
\biggl[\left(\vy_2^\dagger\vmm^{-1}\delta\vbb\,\vmm^{-1}\vy_1\right)^2
-\left(\vy_1^\dagger\vqq\vy_1+\vy_2^\dagger\vqq\vy_2\right)\biggr]\;.
\end{equation}
To evaluate the latter, we set the first derivatives to zero and recast 
the term $\vy_2^\dagger\vmm^{-1}\delta\vbb\,\vmm^{-1}\vy_1$, whose 
explicit expression is enclosed inside the curly bracket in the 
right-hand side of Eq.(\ref{eq:1nd1nd}), into the following compact form
\begin{equation}
\label{eq:1ndsimple}
\vy_2^\dagger\vmm^{-1}\delta\vbb\,\vmm^{-1}\vy_1=f_2(r)\,\zeta_1^{(0)}
+f_1(r)\,\zeta_2^{(0)}+g_0(r)\,\zeta_1^{(0)}\zeta_2^{(0)}
+g_1(r)\sum_{s=\pm 1}\zeta_1^{(s)}\czeta_2^{(s)}
+g_2(r)\sum_{s=\pm 2}\zeta_1^{(s)}\czeta_2^{(s)}\;,
\end{equation}
where $f_1$, $f_2$, $g_0$, $g_1$ and $g_2$ are functions of $r$ and,
possibly, also  $\nu_i$ and $u_i$ (the exact expressions can be read
off from Eq.\ref{eq:1nd1nd}). Upon squaring Eq.(\ref{eq:1ndsimple}),
taking the average over the principal axis frames and substituting the
relations (\ref{eq:ortho1}) and (\ref{eq:ortho2}), we arrive at
\begin{align}
\left\la\left(\vy_2^\dagger\vmm^{-1}\delta\vbb\,\vmm^{-1}\vy_1\right)^2
\right\ra &=
\frac{3}{10}\left[f_1^2\tr\!\left(\tilde{\zeta}_2^2\right)+f_2^2
\tr\!\left(\tilde{\zeta}_1^2\right)\right]+\left(\frac{3}{10}\right)^2
\left(g_0^2+2 g_1^2+2 g_2^2\right)\tr\!\left(\tilde{\zeta}_1^2\right)
\tr\!\left(\tilde{\zeta}_2^2\right) \\ 
&= \frac{15}{2\sigma_2^2}\biggl[\tr\!\left(\tilde{\zeta}_1^2\right)
\Bigl( b_{\nu 2}\,\xi_2^{(1)}+b_{\zeta 2}\,\xi_2^{(2)}\Bigr)^2 +
\tr\!\left(\tilde{\zeta}_2^2\right) \Bigl( b_{\nu
1}\,\xi_2^{(1)}+b_{\zeta 1}\,\xi_2^{(2)}\Bigr)^2\biggr] +
h(r)\,\tr\!\left(\tilde{\zeta}_1^2\right)
\tr\!\left(\tilde{\zeta}_2^2\right) \nonumber\;, \\ 
h(r) &= \left(\frac{45}{14\sigma_2^4}\right)
\left[5\left(\xi_2^{(2)}\right)^2+9\left(\xi_4^{(2)}\right)^2\right]\;.
\end{align}
The variables $b_{\nu i}$ and $b_{\zeta i}$ are defined as
\begin{equation}
\label{eq:bvariables}
b_{\nu i}= \frac{1}{\sigma_0}\left(\frac{\nu_i-\gamma_1
u_i}{1-\gamma_1^2}\right),~~~ b_{\zeta i}=
\frac{1}{\sigma_2}\left(\frac{u_i-\gamma_1
\nu_i}{1-\gamma_1^2}\right)\;.
\end{equation}
They characterize the large scale bias of density peaks of
significance $\nu_i$ and curvature $u_i$.  As we will see shortly,
product of these two variables generate bias parameters beyond first
order.  Likewise, the angular average of the scalar-valued function
$(\vy_1^\dagger\vqq\vy_1+\vy_2^\dagger\vqq\vy_2)$ can eventually be
expressed as
\begin{equation}
\left\la\vy_1^\dagger\vqq\vy_1+\vy_2^\dagger\vqq\vy_2\right\ra = 
\frac{5}{\sigma_2^2}\Bigl(b_{\nu 1}\,\xi_2^{(1)}+b_{\zeta 1}\, 
\xi_2^{(2)}\Bigr)^2 
+\frac{3}{\sigma_1^2}\Bigl(b_{\nu 1}\,\xi_1^{(1/2)}+b_{\zeta 1}\, 
\xi_1^{(3/2)}\Bigr)^2 
+ q(r)\,\tr\!\left(\tilde{\zeta}_1^2\right) + 1\leftrightarrow 2\;,
\end{equation}
where
\begin{equation}
q(r) = \frac{15\gamma_1^2}{2(1-\gamma_1^2)} 
\Bigl[\frac{1}{\sigma_0^2}\bigl(\xi_2^{(1)}\bigr)^2-2\frac{\gamma_1^2}
{\sigma_1^2}\xi_2^{(1)}\xi_2^{(2)}+\frac{1}{\sigma_2^2}\bigl(\xi_2^{(2)}
\bigr)^2\Bigr]
+\frac{15}{7\sigma_2^4}\Bigl[5\bigl(\xi_2^{(2)}\bigr)^2+9\bigl(\xi_4^{(2)}
\bigr)^2\Bigr]
+\frac{9}{2\sigma_1^2\sigma_2^2}
\Bigl[3\bigl(\xi_3^{(3/2)}\bigr)^2+2\bigl(\xi_1^{(3/2)}\bigr)^2\Bigr]
\end{equation}
is a function of the separation $r$ solely.

\subsection{Second order approximation to the peak correlation function}

At this point, we follow \cite{1986ApJ...304...15B} and transform the
eigenvalues of $\zeta(\vx_1)$ and $\zeta(\vx_2)$ to the new set of
variables $\{u_i,v_i,w_i,i=1,2\}$. Here, $v_i$ and $w_i$ are shape
parameters that characterize the asymmetry of the density profile in
the neighborhood of density maxima. After some algebra, the 2-point
correlation of density peaks at second order in $\xi_{j\neq 0}^{(n)}$
can be written as
\begin{align}
\label{eq:2xpk}
1+\xpk(r) &=\frac{1}{\bnpk^2}\,\frac{5^5 3^4}{(2\pi)^6}\,
\frac{\left(1-\Upsilon_1^2\right)^{-3/2}\left(1-\Upsilon_2^2\right)^{-5/2}}
{R_\star^6\sqrt{\left(1-\gamma_1^2\right)\Delta_{\rm P}}}
\int\!\!\prod_{i=1,2}\left\{du_i dv_i dw_i\,
F(u_i,v_i,w_i)\, e^{-\frac{5}{2}\frac{(3 v_i^2+w_i^2)}{1-\Upsilon_2^2}}
\right\} e^{-\Phi}\,{\cal I}_\beta(\tilde{\Lambda}_1,\tilde{\Lambda}_2)
\\
&\quad + \frac{1}{2\bnpk^2}\,\frac{5^5 3^4}{(2\pi)^6}\,R_\star^{-6}
\left(1-\gamma_1^2\right)^{-1}\int\!\!\prod_{i=1,2}
\left\{du_i dv_i dw_i\,F(u_i,v_i,w_i)\, e^{-\Phi_{0i}}\right\}
\biggl[\frac{2}{9}\left(3v_1^2+w_1^2\right)\left(3v_2^2+w_2^2\right) 
h(r) \nonumber \\
&\qquad + \frac{5}{\sigma_2^2}\left(3 v_1^2+w_1^2-1\right)
\Bigl(b_{\nu 2}\xi_2^{(1)}+b_{\zeta 2}\xi_2^{(2)}\Bigr)^2
-\frac{3}{\sigma_1^2}\Bigl(b_{\nu 2}\xi_1^{(1/2)}+b_{\zeta 2}
\xi_1^{(3/2)}\Bigr)^2-\frac{2}{3}\left(3v_1^2+w_1^2\right)q(r)
+ 1\leftrightarrow 2 \biggr] \nonumber \;.
\end{align}
In what follows, we will restrict ourselves to the cross-correlation
$\xpk(r)\equiv\xpk(\nu_1,\nu_2,R_S,r)$ of peaks of height $\nu_1$ and 
$\nu_2$. Therefore, we must integrate over the peak curvatures $u_i$.
The peak constraint implies that the integration at fixed $u_i\geq 0$ 
is restricted to the interior of the triangle bounded by
$(v_i,w_i)=(0,0)$,  $(u_i/4,-u_i/4)$ and $(u_i/2,u_i/2)$. Moreover,
$\Phi_{0i}$ is the quadratic form that appears in the 1-point
probability density $\wi{P}_1(\vy_i)$,
\begin{equation}
2\Phi_{0i}=\nu_i^2+\frac{\left(\gamma_1\nu_i-u_i\right)^2}{1-\gamma_1^2}
+5\left(3 v_i^2+w_i^2\right)\;,
\end{equation}
$F(u_i,v_i,w_i)$ is the weight function defined as 
\citep{1986ApJ...304...15B}
\begin{equation}
F(u_i,v_i,w_i)\equiv \left(u_i-2 w_i\right)
\left[\left(u_i+w_i\right)^2-9 v_i^2\right]
v_i\left(v_i^2-w_i^2\right)\;,
\end{equation}
and ${\cal I}_\beta(\tilde{\Lambda}_1,\tilde{\Lambda}_2)$, with 
$\beta(\Upsilon_2)\equiv (15/2)\Upsilon_2/(1-\Upsilon_2^2)$, is the integral
\begin{equation}
{\cal I}_\beta(\tilde{\Lambda}_1,\tilde{\Lambda}_2)=
\int_{\rm SO(3)}\!\!\!\!\!\!d\vrr\,\exp\left[\beta\,
\tr\!\left(\tilde{\Lambda}_1\vrr\tilde{\Lambda}_2\vrr^\top\right)
\right]\;.
\label{eq:intso3}
\end{equation}
Here, the integration domain is $0\leq\varphi\leq 2\pi$,
$0\leq\vartheta\leq\pi$, $0\leq\psi < 2\pi$ and  $d\vrr\equiv
(1/8\pi^2)d\!\cos\vartheta d\varphi d\psi$ is the normalized Haar
measure ($\int\!d\vrr=1$) on the group SO(3).  There is no analytic,
closed-form solution to the integral  ${\cal
I}_\beta(\tilde{\Lambda}_1,\tilde{\Lambda}_2)$, although it can still
be expressed as a hyper-geometric series in the argument
$\beta\Lambda_1$ and $\Lambda_2$ \citep[see, e.g.,][and references
therein]{2008PhRvD..78b3527D}. We emphasize that equation
(\ref{eq:2xpk})  is better than an approximation based on a second
order Taylor expansion of the probability density $P_2(\vy_1,\vy_2;r)$
since it retains the isotropic part at all orders.

For convenience, we write the peak correlation up to second order as
follows:
\begin{equation}
\xpk(r)=\xpk^{(1)}\!(r)+\xpk^{(2)}\!(r)\equiv 
\xpk^{(1)}\!(r)+\sum_{i=1}^3\xpk^{(2i)}\!(r)
\end{equation}
where $\xpk^{(1)}$ is the first order piece, eq.(\ref{eq:xpk}), and
$\xpk^{(2i)}$ are distinct second order contributions depending on i)
correlation functions only ii)  the peak height $\nu$ and iii) linear
and 2nd order bias parameters. We will now detail each of these
contributions.

$\bullet$ The first terms, $\xpk^{(21)}$, follows from expanding the
determinant of the covariant matrix $\wi{\vcc}(r)$ at the second
order. We have
\begin{equation}
\sqrt{\frac{1-\gamma_1^2}{\Delta_{\rm P}}}-1 \approx 
\frac{\Upsilon_0^2-4\gamma_1^2\Upsilon_0\Upsilon_1+2\gamma_1^2\Upsilon_1^2
+2\gamma_1^4\Upsilon_1^2
+2\gamma_1^2\Upsilon_0\Upsilon_2-4\gamma_1^2\Upsilon_1\Upsilon_2+\Upsilon_2^2}
{2\left(1-\gamma_1^2\right)^2}\;.
\end{equation}
Including the contribution from the trace, Eq.(\ref{eq:2ndtrace}), and
expanding the multiplicative factor
$(1-\Upsilon_1^2)^{-3/2}(1-\Upsilon_2^2)^{-5/2}\approx
1+(3/2)\Upsilon_1^2 +(5/2)\Upsilon_2^2$ at second order yields
\begin{align}
\label{eq:xpk21i}
\xpk^{(21)}\!(r) &=\frac{\Upsilon_0^2-4\gamma_1^2\Upsilon_0\Upsilon_1
+2\gamma_1^2\Upsilon_1^2+2\gamma_1^4\Upsilon_1^2+2\gamma_1^2\Upsilon_0
\Upsilon_2-4\gamma_1^2\Upsilon_1\Upsilon_2+\Upsilon_2^2}
{2\left(1-\gamma_1^2\right)^2}
+\frac{3}{2}\Upsilon_1^2+\frac{5}{2}\Upsilon_2^2 
+\frac{3}{\sigma_1^4}\bigl(\xi_2^{(1)}\bigr)^2 \nonumber \\
&\quad 
+\frac{3}{\sigma_1^2\sigma_2^2}
\Bigl[3\bigl(\xi_3^{(3/2)}\bigr)^2+2\bigl(\xi_1^{(3/2)}\bigr)^2\Bigr]
+\left(\frac{3}{1-\gamma_1^2}\right)\biggl[\frac{1}{\sigma_0^2\sigma_1^2}
\bigl(\xi_1^{(1/2)}\bigr)^2-2\frac{\gamma_1^2}{\sigma_1^4}\xi_1^{(1/2)}
\xi_1^{(3/2)}+\frac{1}{\sigma_1^2\sigma_2^2}\bigl(\xi_1^{(3/2)}\bigr)^2
\biggr] \nonumber \\
&\quad +\frac{5}{7\sigma_2^4}
\Bigl[5\bigl(\xi_2^{(2)}\bigr)^2+9\bigl(\xi_4^{(2)}\bigr)^2\Bigr]
+\left(\frac{5}{1-\gamma_1^2}\right)\biggl[\frac{1}{\sigma_0^2\sigma_2^2}
\bigl(\xi_2^{(1)}\bigr)^2-2\frac{\gamma_1^2}{\sigma_1^2\sigma_2^2}
\xi_2^{(1)}\xi_2^{(2)}+\frac{1}{\sigma_2^4}\bigl(\xi_2^{(2)}\bigr)^2\biggr]\;.
\end{align}
Notice that $\xpk^{(21)}$ does not depend upon the peak height, though
it depends on the filtering scale $R_S$ at which the peaks are
identified.

$\bullet$ The second contribution, $\xpk^{(22)}$, contains all the terms
for which the $\nu$-dependence  cannot be expressed as a polynomial in
the linear and 2nd order bias parameters (to be defined shortly). For
subsequent use, we introduce the auxiliary function
\begin{align}
\label{eq:fua}
f(u,\alpha) &\equiv \frac{3^2 5^{5/2}}{\sqrt{2\pi}}
\left\{\int_0^{u/4}\!\!d v \int_{-v}^{+v}\!\!d w
+\int_{u/4}^{u/2}\!\!d v \int_{3v-w}^v\!\!d w\right\} F(u,v,w)\, 
e^{-\frac{5\alpha}{2}\left(3 v^2+w^2\right)} \\
&= \frac{1}{\alpha^4}\left\{\frac{e^{-5\alpha u^2/2}}{\sqrt{10\pi}}
\left(-\frac{16}{5}+\alpha u^2\right)+\frac{e^{-5\alpha u^2/8}}
{\sqrt{10\pi}}\left(\frac{16}{5}+\frac{31}{2}\alpha u^2\right)
+\frac{\sqrt{\alpha}}{2}\left(\alpha u^3-3u\right)
\left[{\rm Erf}\left(\sqrt{\frac{5\alpha}{2}}\frac{u}{2}\right)+
{\rm Erf}\left(\sqrt{\frac{5\alpha}{2}}u\right)\right]\right\} 
\nonumber
\end{align}
and its integral over the $n$th power of the peak curvature $u$ times
the  $u$-dependent part of the one-point probability distribution,
\begin{equation}
G_n^{(\alpha)}(\gamma,w)=\int_0^\infty\!\!d x\, x^n f(x,\alpha) 
\frac{e^{-(x-w)^2/2(1-\gamma_1^2)}}{\sqrt{2\pi\left(1-\gamma_1^2\right)}}\;.
\label{eq:Gn}
\end{equation}
These functions are very similar, albeit more general than those
defined in Eqs (A15) and (A19) of \cite{1986ApJ...304...15B}. With the
above, moments of the peak curvature can now conveniently be written as
$\ov{u^n}(\nu)=G_n^{(1)}(\gamma_1,\gamma_1\nu)/
G_0^{(1)}(\gamma_1,\gamma_1\nu)$.

Next, we collect all second order terms that features the product of
binomials $(3 v_1^2+w_1^2)^{n_1}(3 v_2^2+w^2)^{n_2}$ with $n_1+n_2\leq
2$. For instance, expanding the integrand of ${\cal I}_\beta$ about
$\beta=0$ gives
\begin{equation}
{\cal I}_\beta(\Lambda_1,\Lambda_2)\approx 
1+\frac{2\beta^2}{45}\left(3 v_1^2+w_1^2\right)
\left(3v_2^2+w_2^2\right)
=1+\frac{5\Upsilon_2^2}{2}\left(3 v_1^2+w_1^2\right)
\left(3v_2^2+w_2^2\right)\;.
\end{equation}
Similarly, 
\begin{equation}
e^{-\frac{5}{2}\frac{\left(3 v_i^2+w_i^2\right)}{1-\Upsilon_2^2}}\approx 
e^{-\frac{5}{2}\left(3v_i^2+w_i^2\right)}
\left[1-\frac{5}{2}\Upsilon_2^2\left(3v_i^2+w_i^2\right)\right]\;.
\end{equation}
On inserting these expansions into the integral over the asymmetry
parameters and using eq.(\ref{eq:fua}), we find
\begin{multline}
\label{eq:xpk22c1}
\frac{5^5 3^4}{2\pi}\int\!\!\prod_{i=1,2}\left\{dv_i
dw_i\,F(u_i,v_i,w_i) \left[1-\frac{5}{2}\Upsilon_2^2\left(3
v_i^2+w_i^2\right)\right]e^{-\frac{5}{2}\left(3v_i^2+w_i^2\right)}
\right\} \left[1+\frac{5}{2}\Upsilon_2^2\left(3
v_1^2+w_1^2\right)\left(3 v_2^2+w_2^2\right)\right] \\ - f(u_1,1)
f(u_2,1) = \Upsilon_2^2\left[f(u_1,1)\partial_\alpha
f(u_2,1)+\partial_\alpha f(u_1,1) f(u_2,1)+\frac{2}{5}\partial_\alpha
f(u_1,1)\partial_\alpha f(u_2,1)\right] \;,
\end{multline}
where $\partial_\alpha f(u_i,1)\equiv \partial_\alpha
f(u_i,\alpha)|_{\alpha=1}$. We are subtracting the zeroth order
contribution from the left-hand side of (\ref{eq:xpk22c1}), which
would  otherwise give unity upon an integration over the  peak
curvature. There are two additional terms proportional to
$(3v_i^2+w_i^2)q(r)$ and a term like $(3v_1^2+w_1^2)
(3v_2^2+w_2^2)h(r)$.  For these terms, integrating out the asymmetry
parameters yields
\begin{multline} 
\label{eq:xpk22c2}
\frac{5^5 3^4}{4\pi}\int\!\!\prod_{i=1,2}\left\{dv_i
dw_i\,F(u_i,v_i,w_i)e^{-\frac{5}{2}\left(3v_i^2+w_i^2\right)}
\right\}\left\{\left[-\frac{2}{3}\left(3v_1^2+w_1^2\right)
-\frac{2}{3}\left(3 v_2^2+w_2^2\right)\right]q(r)
+\frac{4}{9}\left(3v_1^2+w_1^2\right)
\left(3v_2^2+w_2^2\right)h(r)\right\} \\ 
= \frac{2}{15}q(r)\biggl[f(u_1,1)\partial_\alpha f(u_2,1)
+\partial_\alpha f(u_1,1)f(u_1,1)\biggr]+2\left(\frac{2}{15}\right)^2 
h(r)\partial_\alpha f(u_1,1)\partial_\alpha f(u_2,1)\;.
\end{multline}
Adding Eqs (\ref{eq:xpk22c1}) and (\ref{eq:xpk22c2}), we can
eventually express the second order contribution $\xpk^{(22)}(r)$ to
the peak correlation function as
\begin{align}
\xpk^{(22)}\!(r) &= \frac{\left(1-\gamma_1^2\right)^{-1}}
{(2\pi)^5\bnpk^2 R_1^6} 
\int_0^\infty\!\!d u_1\int_0^\infty\!\!d u_2\,\Biggl\{\Upsilon_2^2
\biggl[f(u_1,1)\partial_\alpha f(u_2,1)+\partial_\alpha f(u_1,1)f(u_2,1)
+\frac{2}{5}\partial_\alpha f(u_1,1)\partial_\alpha f(u_2,1)\biggr]
\nonumber \\
&\quad + \frac{2}{15} q(r) \biggl[f(u_1,1)\partial_\alpha 
f(u_2,1) +\partial_\alpha f(u_1,1)f(u_1,1)\biggr]
+\frac{8}{45}h(r)\partial_\alpha  f(u_1,1)\partial_\alpha f(u_2,1)\Biggr\}
\nonumber \\
&\qquad \times \exp\left[-\frac{u_1^2-2\gamma_1 u_1\nu_1+\nu_1^2}
{2\left(1-\gamma_1^2\right)}-\frac{u_2^2-2\gamma_1 u_2\nu_2+\nu_2^2}
{2\left(1-\gamma_1^2\right)}\right] \nonumber \\ 
&= \left(\Upsilon_2^2+\frac{2}{15}q(r)\right)  
\biggl(\partial_\alpha\ln G_0^{(\alpha)}(\gamma_1,\gamma_1\nu_1)
+\partial_\alpha\ln G_0^{(\alpha)}(\gamma_1,\gamma_1\nu_2)\biggr)
\biggr\rvert_{\alpha=1}+\frac{2}{5}\left(\Upsilon_2^2+\frac{4}{45}h(r)
\right) \nonumber \\
&\qquad\times\partial_\alpha\ln G_0^{(\alpha)}(\gamma_1,\gamma_1\nu_1)
\Bigr\lvert_{\alpha=1}\partial_\alpha\ln 
G_0^{(\alpha)}(\gamma_1,\gamma_1\nu_2)\Bigr\rvert_{\alpha=1}\;.
\end{align}
The last equality follows from the well-known relation Eq.~(\ref{eq:bnpk})
for the  differential density of peaks of height $\nu$.
The logarithmic derivative of $G_0^{(\alpha)}$ with respect to $\alpha$ 
must be evaluated numerically. Nevertheless, it is worth noticing that
$G_0^{(\alpha)}(\gamma_1,\omega)$ and  $\partial_\alpha
G_0^{(\alpha)}(\gamma_1,\omega)$ are sharply peaked around their
maximum. For large values of $\omega$, the former asymptotes to
$G_0^{(\alpha)}\approx \alpha^{-5/2}\omega^3$. Hence, this implies
$\partial_\alpha\ln G_0^{(\alpha)}(\gamma_1,\omega)|_{\alpha=1} 
\approx -5/2$ in the limit $\omega\gg 1$.

\begin{figure*}
\center
\resizebox{0.45\textwidth}{!}{\includegraphics{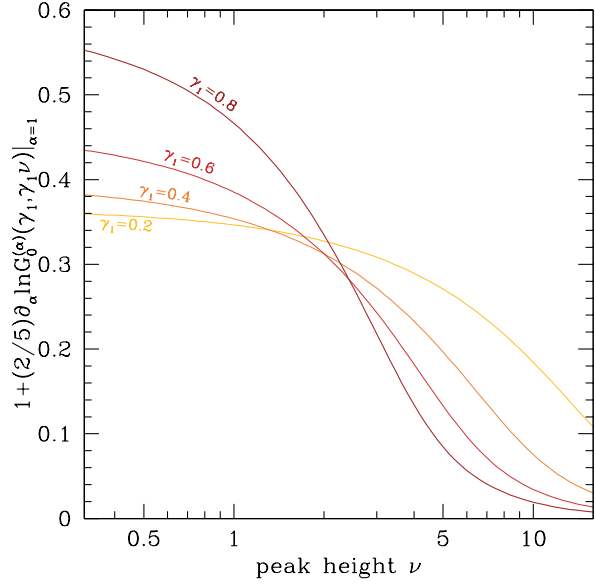}}
\caption{The function $1+(2/5)\partial_\alpha\ln
G_0^{(\alpha)}(\gamma_1,\gamma_1\nu)|_{\alpha=1}$ for several values
of the correlation strength $\gamma_1=0.2$, 0.4, 0.6 and 0.8. This
function vanishes in the limit $\nu\to\infty$ since the logarithmic
derivative of  $G_0^{(\alpha)}$ tends towards $-5/2$.}
\label{fig:lng0}
\end{figure*}

$\bullet$ The last contribution, $\xpk^{(23)}$, is the sum of two parts: a
term which arises from the exponential $e^{-\Phi}$ and, thus, involves
the angle average correlations $\xi_0^{(n)}$; and a second part which
involves the correlations $\xi_{j\neq 0}^{(n)}$.  Upon expanding
$e^{-\Phi}$ at the second order and integrating over the shape
parameters $w_i$, $v_i$ and  the peak curvature $u_i$, we obtain
(after much tedious algebra)
\begin{align}
\label{eq:xpk23i}
\xpk^{(23)}\!(r) &= 
\frac{1}{2}\biggl\{\bb_{\nu\nu 1}\bb_{\nu\nu 2}\bigl(\xi_0^{(0)}\bigr)^2
+2\Bigl(\bb_{\nu\nu 1}\bb_{\nu\zeta 2}+\bb_{\nu\zeta 1}\bb_{\nu\nu 2}\Bigr)
\xi_0^{(0)}\xi_0^{(1)}+\Bigl(\bb_{\nu\nu 1}\bb_{\zeta\zeta 2}
+\bb_{\zeta\zeta 1}\bb_{\nu\nu 2}\Bigr)\bigl(\xi_0^{(1)}\bigr)^2 
+2\bb_{\nu\zeta 1}\bb_{\nu\zeta 2} \\ 
& \qquad\times \Bigl[\bigl(\xi_0^{(1)}\bigr)^2
+\xi_0^{(0)}\xi_0^{(2)}\Bigr]+2\Bigl(\bb_{\zeta\zeta 1}\bb_{\nu\zeta 2}
+\bb_{\nu\zeta 1}\bb_{\zeta\zeta 2}\Bigr)\xi_0^{(1)}\xi_0^{(2)}
+\bb_{\zeta\zeta 1}\bb_{\zeta\zeta 2}\bigl(\xi_0^{(2)}\bigr)^2\biggr\} 
\nonumber \\
& \quad -\frac{1}{2\left(1-\gamma_1^2\right)}\Biggl\{2\Bigl(\bb_{\nu\zeta 1}
+\bb_{\nu\zeta 2}\Bigr)\biggl[\frac{1}{\sigma_0^2}\xi_0^{(0)}\xi_0^{(1)}
-\frac{\gamma_1^2}{\sigma_1^2}\Bigl(\xi_0^{(0)}\xi_0^{(2)}
+\bigl(\xi_0^{(1)}\bigr)^2\Bigr)+\xi_0^{(1)}\xi_0^{(2)}/\sigma_2^2\biggr]
+\Bigl(\bb_{\nu\nu 1}+\bb_{\nu\nu 2}\Bigr) \nonumber \\
& \qquad \times \biggl[\frac{1}{\sigma_0^2}\bigl(\xi_0^{(0)}\bigr)^2
-2\frac{\gamma_1^2}{\sigma_1^2}\xi_0^{(0)}\xi_0^{(1)}+\frac{1}{\sigma_2^2}
\bigl(\xi_0^{(1)}\bigr)^2\biggr]+\Bigl(\bb_{\zeta\zeta 1}+\bb_{\zeta\zeta 2}
\Bigr)\Biggl[\frac{1}{\sigma_0^2}\bigl(\xi_0^{(1)}\bigr)^2-2\frac{\gamma_1^2}
{\sigma_1^2}\xi_0^{(1)}\xi_0^{(2)}+\frac{1}{\sigma_2^2}\bigl(\xi_0^{(2)}
\bigr)^2\Biggr]\Biggr\} \nonumber \\
&\quad -\Biggl\{
\frac{5}{2\sigma_2^2}
\Bigl[\bb_{\nu\nu 1}\bigl(\xi_2^{(1)}\bigr)^2+2\,\bb_{\nu\zeta 1}\,
\xi_2^{(1)}\xi_2^{(2)}+\bb_{\zeta\zeta 1}\bigl(\xi_2^{(2)}\bigr)^2\Bigr]^2
\Biggl(1+\frac{2}{5}\partial_\alpha\ln G_0^{(\alpha)}(\gamma_1,\gamma_1\nu_1)
\Bigr\rvert_{\alpha=1}\Biggr) \nonumber \\
&\qquad +\frac{3}{2\sigma_1^2}
\Bigl[\bb_{\nu\nu 1}\bigl(\xi_1^{(1/2)}\bigr)^2+2\,\bb_{\nu\zeta 1}\,
\xi_1^{(1/2)}\xi_1^{(3/2)}+\bb_{\zeta\zeta 1}
\bigl(\xi_1^{(3/2)}\bigr)^2\Bigr]^2
+ 1\leftrightarrow 2 \Biggr\} \nonumber \;,
\end{align}
where the second order bias parameters $\bb_{\nu\nu
i}=\bb_{\nu\nu}(\nu_i,\gamma_1)$, $\bb_{\nu\zeta
i}=\bb_{\nu\zeta}(\nu_i,\gamma_1)$ and  $\bb_{\zeta\zeta
i}=\bb_{\zeta\zeta}(\nu_i,\gamma_1)$ are constructed from products of
the variables $b_{\nu}$ and $b_\zeta$ defined in Eq.(\ref{eq:bvariables}),
\begin{align}
\bb_{\nu\nu}(\nu_i,\gamma_1) &\equiv \ov{b_{\nu i}^2} =
\frac{\nu_i^2-2\gamma_1\nu_i\bar{u}(\nu_i)+\gamma_1^2\ov{u^2}(\nu_i)}
{\sigma_0^2\left(1-\gamma_1^2\right)^2} \\ 
\bb_{\nu\zeta}(\nu_i,\gamma_1) &\equiv \ov{b_{\nu i}b_{\zeta i}} =
\frac{\left(1+\gamma_1^2\right)\nu_i\bar{u}(\nu_i)-\gamma_1\bigl[\nu_i^2
+\ov{u^2}(\nu_i)\bigr]}{\sigma_0\sigma_2\left(1-\gamma_1^2\right)^2} \\
\bb_{\zeta\zeta}(\nu_i,\gamma_1) &\equiv  \ov{b_{\zeta i}^2} =
\frac{\ov{u^2}(\nu_i)-2\gamma_1\nu_i\bar{u}(\nu_i)+\gamma_1^2\nu_i^2}
{\sigma_2^2\left(1-\gamma_1^2\right)^2}\;.
\label{eq:2ndbias}
\end{align}
Here, the overline designates the average over the peak curvature.
Note that the last  line in the right-hand side of
eq.(\ref{eq:xpk23i}) is the contribution from the correlations
$\xi_{j\neq 0}^{(n)}$. 

\subsection{A compact expression}

The second order contribution
$\xpk^{(2)}=\xpk^{(21)}+\xpk^{(22)}+\xpk^{(23)}$ may be written down
in compact form with the aid of the second order peak bias operator
$\bias{IIi}$ defined through the Fourier space relation
\begin{equation}
\bias{IIi}(q_1,q_2)\equiv 
\ov{b_{\rm spki}(q_1)b_{\rm spki}(q_2)}
-\left(1-\gamma_1^2\right)^{-1}\biggl[\frac{1}{\sigma_0^2}
+\frac{(q_1 q_2)^2}{\sigma_2^2}
-\frac{\gamma_1^2}{\sigma_1^2}\bigl(q_1^2+q_2^2\bigr)\biggr]\;,
\end{equation}
where $b_{\rm spki}(q)\equiv b_{\nu i}+b_{\zeta i}q^2$ and $q_1$ and 
$q_2$ are wavemodes. By definition, its action on the functions 
$\xi_{\ell_1}^{(n_1)}(r)$ and $\xi_{\ell_2}^{(n_2)}(r)$ is
\begin{equation}
\bigl(\xi_{\ell_1}^{(n_1)}\bias{IIi}\xi_{\ell_2}^{(n_2)}\bigr)(r)
\equiv\frac{1}{4\pi^4}\int_0^\infty\!\! dq_1 \int_0^\infty\!\!dq_2\,
q_1^{2(n_1+1)}q_2^{2(n_2+1)}\bias{IIi}(q_1,q_2)P_{\delta_S}(q_1)
P_{\delta_S}(q_2)j_{\ell_1}(q_1 r) j_{\ell_2}(q_2 r)\;.
\end{equation} 
The second order terms can be rearranged so as to recast the 2-point
correlation of peaks of height $\nu_1$ and $\nu_2$ into the more
compact form
\begin{align}
\label{eq:digestible}
\lefteqn{\xpk(\nu_1,\nu_2,R_S,r) = 
\bigl(\bias{I1}\bias{I2}\xi_0^{(0)}\bigr)
+\frac{1}{2}\bigl(\xi_0^{(0)}\bias{II1}\bias{II2}
\xi_0^{(0)}\bigr)-\frac{3}{2\sigma_1^2}
\bigl(\xi_1^{(1/2)}\bias{II1}\xi_1^{(1/2)}\bigr)
-\frac{3}{2\sigma_1^2}
\bigl(\xi_1^{(1/2)}\bias{II2}\xi_1^{(1/2)}\bigr)} \\
&\quad -\frac{5}{2\sigma_2^2}\bigl(\xi_2^{(1)}\bias{II1}
\xi_2^{(1)}\bigr)\biggl(1+\frac{2}{5}\partial_\alpha
\ln G_0^{(\alpha)}\!(\gamma_1,\gamma_1\nu_2)
\Bigl\rvert_{\alpha=1}\biggr) 
-\frac{5}{2\sigma_2^2}\bigl(\xi_2^{(1)}\bias{II2}
\xi_2^{(1)}\bigr)\biggl(1+\frac{2}{5}\partial_\alpha
\ln G_0^{(\alpha)}\!(\gamma_1,\gamma_1\nu_1)
\Bigl\rvert_{\alpha=1}\biggr) 
\nonumber \\
&\quad + \frac{5}{2\sigma_2^4}\Bigl[\bigl(\xi_0^{(2)}\bigr)^2+\frac{10}{7}
\bigl(\xi_2^{(2)}\bigr)^2+\frac{18}{7}\bigl(\xi_4^{(2)}\bigr)^2\Bigr]
\biggl(1+\frac{2}{5}\partial_\alpha\ln G_0^{(\alpha)}\!(\gamma_1,\gamma_1\nu_1)
\Bigl\rvert_{\alpha=1}\biggr)
\biggl(1+\frac{2}{5}\partial_\alpha\ln G_0^{(\alpha)}\!(\gamma_1,\gamma_1\nu_2)
\Bigl\rvert_{\alpha=1}\biggr) \nonumber \\ 
&\quad +\frac{3}{\sigma_1^2\sigma_2^2}\Bigl[3\bigl(\xi_3^{(3/2)}\bigr)^2
+2\bigl(\xi_1^{(3/2)}\bigr)^2\Bigr]+\frac{3}{2\sigma_1^4}
\Bigl[\bigl(\xi_0^{(1)}\bigr)^2+2\bigl(\xi_2^{(1)}\bigr)^2\Bigr]
\nonumber\;.
\end{align}
Notice that the first term of $\xpk^{(21)}$ combines with the first
curly bracket of $\xpk^{(23)}$ to give the second term in the
right-hand side of Eq.(\ref{eq:digestible}), whereas the last term in
Eq.(\ref{eq:digestible}) is the sum of  $(3/2)\Upsilon_1^2$ and
$(3/\sigma_1^4)(\xi_2^{(1)})^2$ in $\xpk^{(21)}$.  We recover
Eq.(\ref{eq:xpkeasy}) in the particular case $\nu_1=\nu_2=\nu$.
For sake of illustration, the function $1+(2/5)\partial_\alpha\ln
G_0^{(\alpha)}(\gamma_1,\gamma_1\nu)$ is shown in Fig.~\ref{fig:lng0}
for several values of $\gamma_1$. Note that it decreases monotonically
and vanishes in the limit $\nu\to\infty$.

In the local bias model, the $N$-order bias parameters are related to
the $N$th-order derivative of the mass function $n(\nu)$ through a 
peak-background split argument (see \S\ref{sub:pksplit} for details). 
Setting $\nu_1=\nu_2=\nu$ for simplicity and collecting the second order 
terms proportional to $(\xi_0^{(0)})^2$ that are present in $\xpk^{(21)}$ 
and $\xpk^{(23)}$, we find their sum is
\begin{equation}
\frac{1}{2}\bb_{\nu\nu}^2\left(\xi_0^{(0)}\right)^2-\frac{\bb_{\nu\nu}}
{\sigma_0^2\left(1-\gamma_1^2\right)}\left(\xi_0^{(0)}\right)^2
+\frac{1}{2}\left(\frac{\xi_0^{(0)}}{\sigma_0^2\left(1-\gamma_1^2\right)}
\right)^2\equiv \frac{1}{2} b_{II}^2\left(\xi_0^{(0)}\right)^2\;.
\end{equation}
Here, $b_{II}$ is the second order peak-background split bias.
Even though we do not calculate the peak correlation at the third
order, it is straightforward to compute the coefficient multiplying
$(\xi_0^{(0)})^3$. This term arises from $e^{-\Phi}$ at order ${\cal
O}(\xi^3)$ in the correlation functions, and $e^{-\Phi}$ at  ${\cal
O}(\xi)$ times $\sqrt{(1-\gamma_1^2/\Delta_{\rm P}}$  at ${\cal
O}(\xi^2)$. Adding these two contributions yields
\begin{equation}
\left[\frac{1}{6}\bb_{\nu\nu\nu}^2-\frac{\bv\bb_{\nu\nu\nu}}
{\sigma_0^2\left(1-\gamma_1^2\right)}+\frac{\bv^2}{\sigma_0^4
\left(1-\gamma_1^2\right)^2}\right]\bigl(\xi_0^{(0)}\bigr)^3
+\frac{1}{2}\frac{\bv^2}{\sigma_0^4\left(1-\gamma_1\right)^2}
\bigl(\xi_0^{(0)}\bigr)^3 =
\frac{1}{6} \left[\bb_{\nu\nu\nu}-\frac{3\,\bv}{\sigma_0^2
\left(1-\gamma_1^2\right)}\right]^2 \bigl(\xi_0^{(0)}\bigr)^3
\equiv \frac{1}{2} b_{III}^2\left(\xi_0^{(0)}\right)^3\;,
\end{equation}
where $b_{III}$ is the third order peak-background split bias and 
the coefficient $\bb_{\nu\nu\nu}\equiv \ov{b_\nu^3}$ is defined in 
Eq.(\ref{eq:bvvv}). This demonstrates that the peak-background split 
derivation of the scale-independent peak bias factors holds at least 
up to third order. We speculate that this remains true at higher order.

\section{Nonlinear evolution of the peak correlation function}
\label{app:xpkevol}

In this Appendix, we provide technical details regarding the nonlinear
evolution of the correlation of initial density peaks under
gravitational instabilities. Here again, we will consider two different 
populations of density peaks of height $\nu_1$ and $\nu_2$ identified 
on the same filtering scale $R_S$. 

\subsection{The joint velocity distribution}

Computing the redshift evolution of the peak correlation function
requires knowledge of the conditional probability density
$P_2(\vups_1,\vups_2|\vy_1,\vy_2;\vr,z_i)$ which, in light of the SVT
decomposition described in the previous Section, is
\begin{equation}
\label{eq:p2w}
P_2(\vups_1,\vups_2|\vy_1,\vy_2;\vr',z_i)=\prod_{m=0,\pm 1} 
P_2\left(\ups_1^{(m)},\ov{\ups}_2^{(m)}|\vy_1^{(m)},
\ov{\vy}_2^{(m)};\vr',z_i\right)\;.
\end{equation}
Here and henceforth, we omit writing the dependence on the
initial redshift $z_i$ for brevity. Furthermore, the velocities 
$\vups$ are in unit of $aHF\sigma_{-1}$ (so there are dimensionless).
Clearly, the line-of-sight
velocity $\ups^{(0)}=\vups\cdot\rvh'$ is a spin-0 component that couples
only to the  other spin-0 variables $\vy_i^{(0)}\equiv
\vy^{(0)}(\vx_i)=(\eta_i^{(0)},\nu_i,u_i,\zeta_i^{(0)})$, while the
components $\ups^{(\pm 1)}=\vups\cdot\ve_{\pm}$ of the transverse vector
correlate only with $\vy_1^{(\pm 1)}\equiv\vy^{(\pm
1)}(\vx_1)=(\eta_1^{(\pm 1)},\zeta_1^{(\pm 1)})$ and the corresponding
conjugates at $\vx_2$.  Using Schur's identities, we can write the
conditional covariance matrix $\Sigma^{(m)}$ and  mean value 
$\Xi^{(m)}$ for the velocity components as
\begin{align}
\Sigma^{(m)} &= \vww^{(m)}-\vvv^{(m)\top}\left(\vcc^{(m)}\right)^{-1}
\vvv^{(m)} \\
\Xi^{(m)} &= \vvv^{(m)\top}\!\!\left(\vcc^{(m)}\right)^{-1}\!\vy^{(m)}\;.
\end{align}
Here, $W^{(m)}$ is the covariance of the spin-$m$ velocity components
$\ups^{(m)}$, $V^{(m)}$ is the cross-covariance between $\ups^{(m)}$ and 
the vector $\vy^{(m)}=(\vy_1^{(m)},\ov{\vy}_2^{(m)})$, and
$\vcc^{(m)}=\la\vy^{(m)}\vy^{(m)\dagger}\ra$ is the covariance matrix of 
$\vy^{(m)}$. Like
the 20-dimensional matrix $\vcc(r)$, $\vcc^{(m)}$ can also be
partitioned into four block matrices, with the auto-covariance
$\vmm^{(m)}$ along the diagonal and the  cross-covariance $\vbb^{(m)}$
and its transpose in the bottom left and top right corners,
respectively. The matrices $\vmm^{(m)}$ and $\vbb^{(m)}$ are defined
in Eqs. (\ref{eq:covm}) and (\ref{eq:covb}). To calculate the
mode-coupling power, it is quite convenient to work with the Fourier
transform of $\Sigma^{(m)}$ and $\Xi^{(m)}$ rather than their real
space counterparts (this allows us to circumvent the addition of three
angular momenta). To this purpose, we will Fourier transform the entries 
of $\vww^{(m)}$, $\vbb^{(m)}$ and $\vcc^{(m)}$ so that 
$\vxx^{(m)}(\vr)=(1/8\pi^3)\int\!d^3\vq\,\vxx^{(m)}\!(\vq)\,
e^{i\vq\cdot\vr}$. For the spin-0 variables, we obtain
\begin{equation}
\vbb^{(0)}\!(\vq) = \left(\begin{array}{cccc} 
\frac{q^2}{\sigma_1^2}\left(\rvh\cdot\qvh\right)^2 &
\frac{iq}{\sigma_0\sigma_1}\left(\rvh\cdot\qvh\right) &
\frac{iq^3}{\sigma_1\sigma_2}\left(\rvh\cdot\qvh\right) &
-\frac{i q^3}{2\sigma_1\sigma_2}
\left[3\left(\rvh\cdot\qvh\right)^3-\left(\rvh\cdot\qvh\right)\right] \\
-\frac{iq}{\sigma_0\sigma_1}\left(\rvh\cdot\qvh\right) & 
1/\sigma_0^2 & q^2/\sigma_0\sigma_2 &
-\frac{q^2}{2\sigma_0\sigma_2}\left[3\left(\rvh\cdot\qvh\right)^2-1\right] \\
-\frac{iq^3}{\sigma_1\sigma_2}\left(\rvh\cdot\qvh\right) &
q^2/\sigma_0\sigma_2 & q^4/\sigma_2^2 &
-\frac{q^4}{2\sigma_2^2}\left[3\left(\rvh\cdot\qvh\right)^2-1\right] \\
\frac{i q^3}{2\sigma_1\sigma_2}
\left[3\left(\rvh\cdot\qvh\right)^3-\left(\rvh\cdot\qvh\right)\right] &
-\frac{q^2}{2\sigma_0\sigma_2}\left[3\left(\rvh\cdot\qvh\right)^2-1\right] &
-\frac{q^4}{2\sigma_2^2}\left[3\left(\rvh\cdot\qvh\right)^2-1\right] &
\frac{q^4}{4\sigma_2^2}\left[3\left(\rvh\cdot\qvh\right)^2-1\right]^2
\end{array}\right)
\end{equation}
and 
\begin{equation}
\vvv^{(0)\top}\!(\vq) =
\left(\begin{array}{cccc} \gamma_0/3 & 0_{1\times 3} & 
\frac{1}{\sigma_{-1}\sigma_1}\left(\rvh\cdot\qvh\right)^2 & 
-{\bf{\cal V}}^\top \\
\frac{1}{\sigma_{-1}\sigma_1}\left(\rvh\cdot\qvh\right)^2 & 
{\bf{\cal V}}^\top & 
\gamma_0/3 & 0_{1\times 3}
\end{array}\right),\qquad
\vww^{(0)}\!(\vq) = \left(\begin{array}{cc} 
1/3 & \frac{q^{-2}}{\sigma_{-1}^2}\left(\rvh\cdot\qvh\right)^2 \\
\frac{q^{-2}}{\sigma_{-1}^2}\left(\rvh\cdot\qvh\right)^2 & 1/3
\end{array}\right)\;,
\end{equation}
where $0_{1\times 3}\equiv (0,0,0)$ and
\begin{equation}
{\bf{\cal V}}^\top=\frac{1}{\sigma_{-1}}
\left(\frac{iq^{-1}}{\sigma_0}\left(\rvh\cdot\qvh\right),\frac{iq}
{\sigma_2}\left(\rvh\cdot\qvh\right),-\frac{i q}{2\sigma_2}\left[3
\left(\rvh\cdot\qvh\right)^3-\left(\rvh\cdot\qvh\right)\right]\right)\;.
\end{equation}
For the spin-1 variables, the cross-covariances are
\begin{equation}
\vbb^{(\pm 1)}\!(\vq)=\left(\begin{array}{cc}
\frac{q^2}{\sigma_1^2}
\left(\ve_{\pm}\cdot\qvh\right)\left(\ov{\ve}_{\pm}\cdot\qvh\right) &
-\frac{\sqrt{3}iq^3}{\sigma_1\sigma_2}\left(\rvh\cdot\qvh\right)
\left(\ve_{\pm}\cdot\qvh\right)\left(\ov{\ve}_{\pm}\cdot\qvh\right) \\
\frac{\sqrt{3}iq^3}{\sigma_1\sigma_2}\left(\rvh\cdot\qvh\right)
\left(\ve_{\pm}\cdot\qvh\right)\left(\ov{\ve}_{\pm}\cdot\qvh\right) &
\frac{3 q^4}{\sigma_2^2}\left(\rvh\cdot\qvh\right)^2
\left(\ve_{\pm}\cdot\qvh\right)\left(\ov{\ve}_{\pm}\cdot\qvh\right)
\end{array}\right)
\end{equation}
whereas
\begin{gather}
\vvv^{(\pm 1)\top}\!(\vq)=\left(\begin{array}{cccc}
\gamma_0/3 & 0 & \frac{1}{\sigma_{-1}\sigma_1}
\left(\ve_{\pm}\cdot\qvh\right)\left(\ov{\ve}_{\pm}\cdot\qvh\right) &
\frac{\sqrt{3}i q}{\sigma_{-1}\sigma_2}\left(\rvh\cdot\qvh\right)
\left(\ve_{\pm}\cdot\qvh\right)\left(\ov{\ve}_{\pm}\cdot\qvh\right) \\
\frac{1}{\sigma_{-1}\sigma_1}\left(\ve_{\pm}\cdot\qvh\right)
\left(\ov{\ve}_{\pm}\cdot\qvh\right) &
-\frac{\sqrt{3}i q}{\sigma_{-1}\sigma_2}\left(\rvh\cdot\qvh\right)
\left(\ve_{\pm}\cdot\qvh\right)\left(\ov{\ve}_{\pm}\cdot\qvh\right) &
\gamma_0/3 & 0
\end{array}\right) \\
\vww^{(\pm 1)}\!(\vq)=\left(\begin{array}{cc} 
1/3 & \frac{q^{-2}}{\sigma_{-1}^2}\left(\ve_{\pm}\cdot\qvh\right)
\left(\ov{\ve}_{\pm}\cdot\qvh\right) \\
\frac{q^{-2}}{\sigma_{-1}^2}\left(\ve_{\pm}\cdot\qvh\right)
\left(\ov{\ve}_{\pm}\cdot\qvh\right) & 1/3
\end{array}\right)\;.
\end{gather}
We have omitted a factor of $P_\delta(q)$ in  all matrix elements for
shorthand purposes. Furthermore, we have not substituted
$\rvh\cdot\qvh$ by $\qh^{(0)}$ and $\ve_{\pm}\cdot\qvh$ by  $\qh^{(\pm
1)}$ to avoid heavy notation for their complex conjugates. We also note
that  entries involving a odd power of $\rvh$ change sign under the
space  reflection $\rvh\to -\rvh$.

With the SVT decomposition introduced above, the integral over the
peak velocities can be written
\begin{equation}
\label{eq:vint}
\int\!\!d^N\vy\,P_2(\vups_1,\vups_2|\vy_1,\vy_2;\vr')\,
e^{i\sigma_v\vk\cdot\Delta\vups_{12}} 
= \exp\left(-\frac{1}{2}\vj^\dagger\Sigma\vj+i\vj^\dagger\Xi\right)
\end{equation}
where $\Sigma={\rm diag}(\Sigma^{(0)},\Sigma^{(+1)},\Sigma^{(-1)})$
and $\Xi=(\Xi^{(0)},\Xi^{(+1)},\Xi^{(-1)})$ are the covariance and
mean of the multivariate Gaussian; and $\vj$, $\vj^\dagger$ are the
6-dimensional vector
$\sigma_v(k^{(0)},-k^{(0)},k^{(+1)},-k^{(+1)},k^{(-1)},-k^{(-1)})$
and its conjugate transpose, respectively. We will now Taylor expand 
the right-hand side of Eq.(\ref{eq:vint}) around the zeroth order 
contribution to $\Sigma^{(m)}$ and $\Xi^{(m)}$.

\subsection{Zeroth order: diffusion damping due to random velocities}

At the zeroth order, the covariance matrices of the spin-0 and spin-1 
conditional velocity distributions reduce to 
\begin{equation}
\label{eq:pvza0nd}
\Sigma^{(0)}=\Sigma^{(\pm 1)}\approx 
\frac{1}{3}\left(1-\gamma_0^2\right)\,\vii_2\;,
\end{equation}
whereas there is no net transverse velocity $\Xi$ at this order.
Inserting this result in Eq.(\ref{eq:vint}) yields
\begin{equation}
\label{eq:ZAdamp0nd}
\int\!\!d^N\vy\,P_2(\vv_1,\vv_2|\vy_1,\vy_2;\vr')\,
e^{i\sigma_v\vk\cdot\Delta\vv_{12}} \approx 
e^{-\frac{1}{3}k^2\sigma_v^2\left(1-\gamma_0^2\right)}=
e^{-\frac{1}{3}k^2\sigma_{\rm vpk}^2}\;,
\end{equation}
where $\sigma_{\rm vpk}^2\!(z)=\sigma_v^2\!(z)(1-\gamma_0^2)$ is the
3-dimensional velocity dispersion of peaks identified on the filtering
scale $R_S$ (Eq.~(\ref{eq:sigmavpk}).
Hence, at the zeroth order, the evolved correlation of density peaks
is simply obtained through a convolution of the initial correlation
$\xpk(R_S,\nu,r)$ with the diffusion kernel $\exp[-(1/3)k^2\sigma_{\rm
vpk}^2\!(z)]$. 

\subsection{First order: linear growth due to coherent motions}

At the first order, the Fourier transform of the covariance matrices 
$\Sigma^{(m)}(\vq)$ receive a contribution $\delta\Sigma^{(m)}(\vq)$ 
given by
\begin{align}
\delta\Sigma^{(0)}(\vq) &=\frac{1}{\sigma_{-1}^2}
\frac{\left(\rvh\cdot\qvh\right)^2}{q^2}\, \bvpk^2(q) P_{\delta_S}(q) 
\left(\begin{array}{cc} 0 & 1 \\ 1 & 0 \end{array}\right) \\
\delta\Sigma^{(\pm 1)}(\vq) &= \frac{1}{\sigma_{-1}^2}
\frac{\left(\ve_{\pm}\cdot\qvh\right)\left(\ov{\ve}_{\pm}\cdot\qvh\right)}
{q^2}\, \bvpk^2(q) P_{\delta_S}(q)
\left(\begin{array}{cc} 0 & 1 \\ 1 & 0 \end{array}\right)
\nonumber \;.
\end{align}
The velocity bias factor $\bvpk(q)$, Eq.~(\ref{eq:vpkbiask}), is the 
same for the two peak populations because it depends only on the 
filtering scale $R_S$. In addition, there is a non-zero mean velocity 
with line-of-sight components
\begin{align}
\label{eq:Xi0_1st}
\delta\Xi_1^{(0)}(\vq) &= -\frac{i}{\sigma_{-1}}
\frac{\left(\rvh\cdot\qvh\right)}{q}\bvpk(q) b_{\rm spk2}(q) 
P_{\delta_S}(q)+\frac{5i}{2\sigma_{-1}\sigma_2}\,
q\left[3\left(\rvh\cdot\qvh\right)^3-\left(\rvh\cdot\qvh\right)\right]
\bvpk(q) \zeta_2^{(0)} P_{\delta_S}(q) \\
\delta\Xi_2^{(0)}(\vq) &= \frac{i}{\sigma_{-1}}
\frac{\left(\rvh\cdot\qvh\right)}{q}
\bvpk(q) b_{\rm spk1}(q) P_{\delta_S}(q)
-\frac{5i}{2\sigma_{-1}\sigma_2}\,
q\left[3\left(\rvh\cdot\qvh\right)^3-\left(\rvh\cdot\qvh\right)\right]
\bvpk(q)\zeta_1^{(0)} P_{\delta_S}(q) \nonumber\;,
\end{align}
and transverse components
\begin{align}
\delta\Xi_1^{(\pm 1)}(\vq) &= \frac{5\sqrt{3}i}{\sigma_{-1}\sigma_2}
q\left(\rvh\cdot\qvh\right)
\left(\ve_{\pm}\cdot\qvh\right)\left(\ov{\ve}_{\pm}\cdot\qvh\right)
\bvpk(q) \zeta_2^{(\pm 1)} P_{\delta_S}(q) \\
\delta\Xi_2^{(\pm 1)}(\vq) &= -\frac{5\sqrt{3}i}{\sigma_{-1}\sigma_2}
q\left(\rvh\cdot\qvh\right)
\left(\ve_{\pm}\cdot\qvh\right)\left(\ov{\ve}_{\pm}\cdot\qvh\right)
\bvpk(q) \zeta_1^{(\pm 1)} P_{\delta_S}(q) \nonumber\;.
\end{align}
Here, $b_{\rm spki}(q)=b_{\nu i}+b_{\zeta i}q^2$ is the linear bias 
for peaks of height $\nu_i$ and curvature $u_i$. Even if the peaks
had all the same height, we would have to distinguish between the linear 
bias of the two populations because the peak curvatures would not
necessarily be the same. We also remark that the mean transverse velocity 
components $\Xi^{(\pm 1)}$ vanish upon averaging  over the orientations 
of the principal axis frames.
With the aid of these results, we now expand the right-hand side of
Eq.(\ref{eq:vint}) up to first order and impose the peak constraint 
to arrive at
\begin{multline}
\frac{1}{\bnpk^2}\int\!\!d^6\zeta_1d^6\zeta_2\,
\npk(\vx_1')\npk(\vx_2')P_2(\vy_1,\vy_2;\vr')
\exp\left(-\frac{1}{2}\vj^\dagger\Sigma\vj+i\vj^\dagger\Xi\right) \\
\approx e^{-\frac{1}{3}k^2\sigma_{\rm vpk}^2}
\Biggl\{1+\bb_{\nu 1}\bb_{\nu 2}\xi_0^{(0)}
+\Bigl(\bb_{\nu 1}\bb_{\zeta 2}+\bb_{\nu 2}\bb_{\zeta 1}\Bigr)
\xi_0^{(1)}+\bb_{\zeta 1}\bb_{\zeta 2}\xi_0^{(2)} 
\\ 
+\prod_{i=1,2}\left[\frac{1}{\bnpk}\int\!\!d^6\zeta_i\,\npk(\vx_i')\,
\wi{P}_1(\vy_i)\right] 
\Bigl(-\frac{1}{2}\vj^\dagger\delta\Sigma\vj+i\vj^\dagger\delta\Xi\Bigr)
\Biggr\} \\
=e^{-\frac{1}{3}k^2\sigma_{\rm vpk}^2}
\Biggl\{1+\int\!\!\frac{d^3\vq}{(2\pi)^3}\biggl[\bias{I1}(q)\bias{I2}(q)
+\left(\frac{\sigma_v}{\sigma_{-1}}\right)\frac{k}{q}
(\rvh'\cdot\qvh)(\rvh'\cdot\kvh)\Bigl[\bias{I1}(q)+\bias{I2}(q)\Bigr]
\bias{vpk}(q) \\ 
+\left(\frac{\sigma_v}{\sigma_{-1}}\right)^2
\frac{k^2}{q^2}\Bigl[(\rvh'\cdot\qvh)^2(\rvh'\cdot\kvh)^2
+\sum_{a=\pm}(\ve_a'\cdot\qvh)(\ov{\ve}_a'\cdot\qvh)
(\ve_a'\cdot\kvh)(\ov{\ve}_a'\cdot\kvh)\Bigr]
\bias{vpk}^2(q)\biggr] P_{\delta_S}(q)\,e^{i\vq\cdot\vr'}\Biggr\}
\end{multline}
upon an integration over the peak asymmetry parameters $v_i$, $w_i$
and the peak curvatures $u_i$ (see Appendix \S\ref{app:xpk2L}). In the 
second term of this equation, $\wi{P}_1(\vy_i)$ is a 1-point probability 
density and, in the third, $\rvh'\cdot\ve_{\pm}'=\rvh'\cdot\ov{\ve}_{\pm}'=0$ 
where $\rvh'=\vr'/r'$.
It is important to bear in mind that the spectral moments $\sigma_n$
are evaluated at initial redshift $z_i$,
e.g. $\sigma_{-1}=\sigma_{-1}(z_i)$. Hence, the ratio
$\sigma_v/\sigma_{-1}$ equals $D(z)/D(z_i)$. To perform the integral
over $\vr'$, we utilize the following relations
\begin{gather}
\int\!\!d^3\vr\,\rh_i \rh_j\, e^{i(\vq-\vk)\cdot\vr}
=(2\pi)^3\delta^{(3)}\!(\vq-\vk)\kh_i\kh_j,~~~
\int\!\!d^3\vr\,e_{\pm i}\ov{e}_{\pm j}\, e^{i(\vq-\vk)\cdot\vr}
=(2\pi)^3\delta^{(3)}\!(\vq-\vk)
\mathfrak{e}_{\pm i}\ov{\mathfrak{e}}_{\pm j}\;,
\end{gather}
where $\mathfrak{e}_+$ and $\mathfrak{e}_-$ denote the unit vectors
orthogonal to the wavevector $\kvh$,
i.e. $\kvh\cdot\mathfrak{e}_\pm=0$. As a consequence,  all the terms
involving $\ve_{\pm}'\cdot\vk$ or $\ov{\ve}_{\pm}'\cdot\vk$ vanish. This
cancellation reflects the fact that, on average, only streaming motions 
along the separation vector $\vr'$ of a peak pair can affect the peak
correlation $\xpk$. On employing the identity 
$(\rvh'\cdot\qvh)^2+\sum_{a=\pm}(\ve_a'\cdot\qvh)(\ov{\ve}_a'\cdot\qvh)=1$ 
(which follows from Eq.\ref{eq:spin1dot}), we thus obtain
\begin{multline}
\frac{1}{\bnpk^2}\int\!\!d^3\vr' e^{-i\vk\cdot\vr'}
\int\!\!d^6\zeta_1d^6\zeta_2\,\npk(\vx_1')\npk(\vx_2')
P_2(\vy_1,\vy_2;\vr')
\exp\left(-\frac{1}{2}\vj^\dagger\Sigma\vj+i\vj^\dagger\Xi\right) \\
\approx e^{-\frac{1}{3}k^2\sigma_{\rm vpk}^2}
\Biggl\{\bias{I1}(q)\bias{I2}(q)+\left(\frac{D(z)}{D(z_i)}\right)
\Bigl[\bias{I1}(q)+\bias{I2}(q)\Bigr]\bias{vpk}(q)
+\left(\frac{D(z)}{D(z_i)}\right)^2\bias{vpk}^2(q)\Biggr\}
P_{\delta_S}(q)\;. 
\end{multline}
After some further manipulation, this yields the desired result 
Eq.(\ref{eq:xpkevol}) where it is assumed $\nu_1=\nu_2=\nu$.

\subsection{Second order: Lagrangian and gravity mode-coupling}

The calculation of the 2nd order term in the Taylor expansion of 
Eq.~(\ref{eq:vint}), which is the lowest order contribution to the 
mode-coupling power, is long and fastidious. For clarity, we can 
decompose this second order mode-coupling power into three distinct 
pieces which we will compute successively~: i) the second order 
contribution to the initial peak correlation $\xpk(\nu_1,\nu_2,R_S,r)$ 
convolved with the diffusion kernel, ii) $\xpk^{(1)}(\nu_1,\nu_2,R_S,r)$ 
times the first order term in the expansion of  
$\exp(-\frac{1}{2}\vj^\dagger\Sigma\vj+i\vj^\dagger\Xi)$ and iii)
the second order term in the expansion of
$\exp(-\frac{1}{2}\vj^\dagger\Sigma\vj+i\vj^\dagger\Xi)$.

$\bullet$ The first piece reflects the fact that the initial, second order 
contribution is progressively smeared out by the peak motions. It is
trivially
\begin{equation}
\int\!\!\frac{d^3\vk}{(2\pi)^3}\,
e^{-\frac{1}{3}k^2\sigma_{\rm vpk}^2}\ppk^{(2)}(\nu_1,\nu_2,R_S,k)\, 
e^{i\vk\cdot\vr}\;,
\end{equation}
where $\ppk^{(2)}(\nu_1,\nu_2,R_S,k)$ is the Fourier transform of the 
second order correlation of initial density peaks, 
$\xpk^{(2)}(\nu_1,\nu_2,R_S,r)$.

$\bullet$ Taking the exponential damping factor out of the integral, the 
second part of the mode-coupling  can be written as
\begin{equation}
\label{eq:modec2}
e^{-\frac{1}{3}k^2\sigma_{\rm vpk}^2}
\prod_{i=1,2}\left[\frac{1}{\bnpk}\int\!\!d^6\zeta_i\,\npk(\vx_i')
\wi{P}_1(\vy_i)\right]\left(\vy_1^\dagger\vmm^{-1}\vbb\vmm^{-1}
\vy_2\right)\left(-\frac{1}{2}\vj^\dagger\delta\Sigma\vj
+i\vj^\dagger\delta\Xi\right)\;,
\end{equation}
where the Fourier transform of $\vy_2^\dagger\vmm^{-1}\vbb\vmm^{-1}\vy_1$
is given by
\begin{multline}
\int\!\!d^3\vr\,\left(\vy_2^\dagger\vmm^{-1}\vbb\vmm^{-1}\vy_1\right)
e^{-i\vq\cdot\vr}= 
\Biggl\{b_{\rm spk1}(q) b_{\rm spk2}(q)-\frac{5}{2\sigma_2}q^2
\left[3(\rvh\cdot\qvh)^2-1\right]\left(b_{\rm spk1}(q)\zeta_2^{(0)}
+b_{\rm spk2}(q)\zeta_1^{(0)}\right) \\ +\frac{25}{4\sigma_2^2}q^4
\left[3(\rvh\cdot\qvh)^2-1\right]^2\zeta_1^{(0)}\zeta_2^{(0)}
+\frac{75}{\sigma_2^2}q^4(\rvh\cdot\qvh)^2 \\
\times \sum (\ve_{\pm}\cdot\qvh)(\ov{\ve}_{\pm}\cdot\qvh)\zeta_1^{(\pm 1)}
\czeta_2^{(\pm 1)}+\frac{3}{2\sigma_2^2}q^4\sum (\ve_{\pm}\cdot\qvh)^2
(\ov{\ve}_{\pm}\cdot\qvh)^2\zeta_1^{(\pm 2)}\czeta_2^{(\pm 2)}\Biggr\}
P_{\delta_S}(q)\;.
\end{multline}
As we can see from Eq.(\ref{eq:Xi0_1st}), there are also terms linear
in $\zeta_1^{(0)}$ and $\zeta_2^{(0)}$ in the first order mean
velocity $\delta\Xi$. On angle averaging over the variables which 
define the orientation of the principal axes, product of the form
$\zeta_i^{(0)}\zeta_i^{(0)}$ reduce to $(3/10)\,\tr\!(\tilde{\zeta}_i^2)$ 
as already shown in Appendix \S\ref{app:xpk2L}. Hence, the Fourier 
transform  of $\vy_2^\dagger\vmm^{-1}\vbb\vmm^{-1}\vy_1$ and 
$i\vj^\dagger\delta\Xi$ becomes
\begin{multline}
\left\la\left(\vy_2^\dagger\vmm^{-1}\vbb\vmm^{-1}\vy_1\right)
\left(i\vj^\dagger\delta\Xi\right)\right\ra =
\Biggl\{\left(\frac{\sigma_v}{\sigma_{-1}}\right) (\rvh\cdot\vk)
\frac{(\rvh\cdot\qvh_2)}{q_2}b_{\rm spk1}(q_1)b_{\rm spk2}(q_1)
\Bigl[b_{\rm spk1}(q_2)+b_{\rm spk2}(q_2)\Bigr]\bvpk(q_2) \\
+\frac{15}{8\sigma_2^2}\left(\frac{\sigma_v}{\sigma_{-1}}\right) 
q_1^2 q_2 (\rvh\cdot\vk)\left[3(\rvh\cdot\qvh_1)^2-1\right]
\left[3(\rvh\cdot\qvh_2)^3-(\rvh\cdot\qvh_2)\right] \\
\times \Bigl[b_{\rm spk1}(q_1)\tr\!\left(\tilde{\zeta}_2^2\right)
+b_{\rm spk2}(q_1)\tr\!\left(\tilde{\zeta}_1^2\right)\Bigr] 
\bvpk(q_2)\Biggr\} P_{\delta_S}(q_1)P_{\delta_S}(q_2)\;.
\end{multline}
Upon substituting this result in Eq.(\ref{eq:modec2}) and integrating
over the other variables, the second piece ii) can eventually be 
expressed as
\begin{multline}
\label{eq:piece2}
e^{-\frac{1}{3}k^2\sigma_{\rm vpk}^2}
\int\!\!\frac{d^3\vq_1}{(2\pi)^3}\int\!\!\frac{d^3\vq_2}{(2\pi)^3}
\Biggl\{\left(\frac{\sigma_v}{\sigma_{-1}}\right)^2\frac{k^2}{q_2^2}
\left[(\rvh'\cdot\qvh_2)^2(\rvh'\cdot\kvh)^2
+\sum_{a=\pm}(\ve_a'\cdot\qvh_2)(\ov{\ve}_a'\cdot\qvh_2)
(\ve_a'\cdot\kvh)(\ov{\ve}_a'\cdot\kvh)\right]
\bias{I1}(q_1)\bias{I2}(q_1)\bias{vpk}^2(q_2) \\
+\left(\frac{\sigma_v}{\sigma_{-1}}\right)(\rvh'\cdot\vk)
\frac{(\rvh'\cdot\qvh_2)}{q_2}\Bigl[\ov{b_{\rm spk1}(q_1)b_{\rm spk1}(q_2)}\,
\bias{I2}(q_1)+\ov{b_{\rm spk2}(q_1)b_{\rm spk2}(q_2)}\,
\bias{I1}(q_1)\Bigr] \bias{vpk}(q_2) \\
-\frac{1}{2\sigma_2^2}\left(\frac{\sigma_v}{\sigma_{-1}}\right)
q_1^2 q_2(\rvh'\cdot\vk) \left[3(\rvh'\cdot\qvh_1)^2-1\right]
\left[3(\rvh'\cdot\qvh_2)^3-(\rvh'\cdot\qvh_2)\right] \\
\times \biggl(\partial_\alpha
\ln G_0^{(\alpha)}\!(\gamma_1,\gamma_1\nu_2)\bias{I1}(q_1)
+\partial_\alpha\ln G_0^{(\alpha)}\!(\gamma_1,\gamma_1\nu_2)
\bias{I2}(q_1)\biggr)\biggr\lvert_{\alpha=1}
\bias{vpk}(q_2)\Biggr\}P_{\delta_S}(q_1) P_{\delta_S}(q_2) 
e^{i(\vq_1+\vq_2)\cdot\vr'}\;,
\end{multline}
where we have replaced $\tr(\tilde{\zeta}_i^2)$ by $(2/3)(3
v_i^2+w_i^2)$ before proceeding with the integration over the
asymmetry parameters (see Appendix \ref{app:xpk2L}).
Eq. (\ref{eq:2ndbias}) can be employed to rewrite the average
$\ov{b_{\rm pk i}(q_1)b_{\rm pk i}(q_2)}$ in terms of the second order
peak bias factors $\bias{20}$, $\bias{11}$ and $\bias{02}$. 

$\bullet$ The third part of the mode-coupling power is the most difficult
to compute. It can be written as
\begin{equation}
\label{eq:modec3}
e^{-\frac{1}{3}k^2\sigma_{\rm vpk}^2}
\prod_{i=1,2}\left[\frac{1}{\bnpk}\int\!\!d^6\zeta_i\,\npk(\vx_i')
\wi{P}_1(\vy_i)\right]\left\{\frac{1}{2}
\left(-\frac{1}{2}\vj^\dagger\delta\Sigma\vj+i\vj^\dagger\delta\Xi\right)^2
-\frac{1}{2}\vj^\dagger\delta^2\Sigma\vj+i\vj^\dagger\delta^2\Xi\right\}\;.
\end{equation}
Let us first concentrate on the integral over the terms quadratic in
$\delta\Sigma$ and $\delta\Xi$. We must proceed carefully with the
calculation of  $\vj^\dagger\delta\Xi$ because, once again, there are
terms of the form $\zeta_i^{(\pm s)}\zeta_i^{(\mp s)}$ which do not
vanish upon averaging over the angular variables. A straightforward
computation yields
\begin{align}
\lefteqn{\left\la\left(\vj^\dagger\delta\Xi\right)^2\right\ra =
\Biggl\{
-\left(\frac{\sigma_v}{\sigma_{-1}}\right)^2(\rvh\cdot\vk)^2
\frac{(\rvh\cdot\qvh_1)}{q_1}\frac{(\rvh\cdot\qvh_2)}{q_2}
\Bigl[b_{\rm spk1}(q_1)+b_{\rm spk2}(q_1)\Bigr]
\Bigl[b_{\rm spk1}(q_2)+b_{\rm spk2}(q_2)\Bigr]\bvpk(q_1)
\bvpk(q_2)} \nonumber \\
&\quad -\frac{15}{8\sigma_2^2}\left(\frac{\sigma_v}{\sigma_{-1}}\right)^2
q_1 q_2 (\rvh\cdot\vk)^2 \left[3(\rvh\cdot\qvh_1)^3-(\rvh\cdot\qvh_1)\right]
\left[3(\rvh\cdot\qvh_2)^3-(\rvh\cdot\qvh_2)\right]
\Bigl[\tr\!\left(\tilde{\zeta}_1^2\right)
+\tr\!\left(\tilde{\zeta}_2^2\right)\Bigr]\bvpk(q_1)\bvpk(q_2)
\nonumber \\
&\quad +\frac{45}{2\sigma_2^2}\left(\frac{\sigma_v}{\sigma_{-1}}\right)^2
(\rvh\cdot\qvh_1)(\rvh\cdot\qvh_2)(\ov{\ve}_+\cdot\vk)(\ov{\ve}_-\cdot\vk)
\Biggl[\sum_{a=\pm}(\ve_a\cdot\qvh_1)(\ov{\ve}_a\cdot\qvh_1)
(\ve_{-a}\cdot\qvh_2)(\ov{\ve}_{-a}\cdot\qvh_2)\Biggr] \nonumber \\
&\qquad \times \Bigl[\tr\!\left(\tilde{\zeta}_1^2\right)
+\tr\!\left(\tilde{\zeta}_2^2\right)\Bigr]\bvpk(q_1)\bvpk(q_2)\Biggr\}
P_{\delta_S}(q_1) P_{\delta_S}(q_2)\;.
\end{align}
Adding the contribution proportional to $(\vj^\dagger\delta\Sigma\vj)^2$
and $(\vj^\dagger\delta\Sigma\vj)(\vj^\dagger\delta\Xi)$ and integrating
out the asymmetry parameters, we finally obtain
\begin{multline}
\label{eq:piece3a}
\frac{1}{2}\, e^{-\frac{1}{3}k^2\sigma_{\rm vpk}^2}
\int\!\!\frac{d^3\vq_1}{(2\pi)^3}\int\!\!\frac{d^3\vq_2}{(2\pi)^3}
\Biggl\{\left(\frac{\sigma_v}{\sigma_{-1}}\right)^4 \frac{k^4}{q_1^2 q_2^2} 
\left[(\rvh'\cdot\qvh_1)^2(\rvh'\cdot\kvh)^2
+\sum_{a=\pm}(\ve_a'\cdot\qvh_1)(\ov{\ve}_a'\cdot\qvh_1)
(\ve_a'\cdot\kvh)(\ov{\ve}_a'\cdot\kvh)\right] \\
\times \,\left[(\rvh'\cdot\qvh_2)^2(\rvh'\cdot\kvh)^2
+\sum_{a=\pm}(\ve_a'\cdot\qvh_2)(\ov{\ve}_a'\cdot\qvh_2)
(\ve_a'\cdot\kvh)(\ov{\ve}_a'\cdot\kvh)\right]
\bias{vpk}^2(q_1)\bias{vpk}^2(q_2)
+2\left(\frac{\sigma_v}{\sigma_{-1}}\right)^3 \frac{k^3}{q_1^2}
(\rvh'\cdot\kvh)\frac{(\rvh'\cdot\qvh_2)}{q_2} \\ 
\times \, \left[(\rvh'\cdot\qvh_1)^2(\rvh'\cdot\kvh)^2
+\sum_{a=\pm}(\ve_a'\cdot\qvh_1)(\ov{\ve}_a'\cdot\qvh_1)
(\ve_a'\cdot\kvh)(\ov{\ve}_a'\cdot\kvh)\right]
\Bigl[\bias{I1}(q_2)+\bias{I2}(q_2)\Bigr]
\bias{vpk}^2(q_1)\bias{vpk}(q_2)
+\left(\frac{\sigma_v}{\sigma_{-1}}\right)^2(\rvh'\cdot\vk)^2
\frac{(\rvh'\cdot\qvh_1)}{q_1}\frac{(\rvh'\cdot\qvh_2)}{q_2} \\
\times\,\Bigl[\ov{b_{\rm spk1}(q_1)b_{\rm spk1}(q_2)}
+\bias{I1}(q_1)\bias{I2}(q_2)+\bias{I1}(q_2)\bias{I2}(q_1)
+\ov{b_{\rm spk2}(q_1)b_{\rm spk2}(q_2)}\Bigr]
\bias{vpk}(q_1)\bias{vpk}(q_2)
-\frac{1}{\sigma_2^2}\left(\frac{\sigma_v}{\sigma_{-1}}\right)^2
(\rvh'\cdot\qvh_1)(\rvh'\cdot\qvh_2) \\
\times \,\Biggl(q_1 q_2(\rvh'\cdot\vk)^2
\left[3(\rvh'\cdot\qvh_1)^2-1\right]\left[3(\rvh'\cdot\qvh_2)^2-1\right]
-12(\ov{\ve}_+'\cdot\vk)(\ov{\ve}_-'\cdot\vk)
\Bigl[\sum_{a=\pm}(\ve_a'\cdot\qvh_1)(\ov{\ve}_a'\cdot\qvh_1)
(\ve_{-a}'\cdot\qvh_2)(\ov{\ve}_{-a}'\cdot\qvh_2)\Bigr]\Biggr) \\
\times \, \biggl(\partial_\alpha
\ln G_0^{(\alpha)}\!(\gamma_1,\gamma_1\nu_1)+\partial_\alpha
\ln G_0^{(\alpha)}\!(\gamma_1,\gamma_1\nu_2)\biggr)
\biggr\lvert_{\alpha=1}\bias{vpk}(q_1)\bias{vpk}(q_2)\Biggr\}
P_{\delta_S}(q_1) P_{\delta_S}(q_2) e^{i(\vq_1+\vq_2)\cdot\vr'}\;,
\end{multline}
where, for the sake of completeness, we have included all the terms
involving $(\ov{\ve}_\pm'\cdot\kvh)$ even though they will vanish when we
carry out the integral over $\vr'$.  Next, we consider the integral
over the second order terms $\delta^2\Sigma$ and $\delta^2\Xi$. The
Fourier transform of the second order contribution to the covariance
matrices $\Sigma^{(m)}$ are
\begin{align}
\delta^2\Sigma^{(0)}(\vq_1,\vq_2) &= 
\frac{1}{\sigma_{-1}^2}\Biggl\{-\frac{3}{\sigma_1^2}
(\rvh\cdot\qvh_1)(\rvh\cdot\qvh_2)+\frac{5}{4\sigma_2^2}q_1 q_2
\left[3(\rvh\cdot\qvh_1)^2-1\right]\left[3(\rvh\cdot\qvh_2)^2-1\right] \\
&\qquad +\left(1-\gamma_1^2\right)^{-1}\left[\frac{1}{\sigma_0^2}
(q_1 q_2)^{-1}+\frac{1}{\sigma_2^2}q_1 q_2-\frac{2\gamma_1}
{\sigma_0\sigma_2}q_1^{-1} q_2\right]\Biggr\}(\rvh\cdot\qvh_1)
(\rvh\cdot\qvh_2) \nonumber \\ 
&\qquad\times\bvpk(q_1)\bvpk(q_2) P_{\delta_S}(q_1)P_{\delta_S}(q_2)\,\vii_2 
\nonumber \\
\delta^2\Sigma^{(\pm 1)}(\vq_1,\vq_2) &= \frac{1}{\sigma_{-1}^2}
\left[-\frac{3}{\sigma_1^2}+\frac{15}{\sigma_2^2}q_1 q_2(\rvh\cdot\qvh_1)
(\rvh\cdot\qvh_2)\right](\ve_{\pm}\cdot\qvh_1)(\ov{\ve}_{\pm}\cdot\qvh_1)
(\ve_{\pm}\cdot\qvh_2)(\ov{\ve}_{\pm}\cdot\qvh_2) \nonumber \\
&\qquad\times\bvpk(q_1)\bvpk(q_2)P_{\delta_S}(q_1) P_{\delta_S}(q_2)\,\vii_2 
\nonumber\;,
\end{align}
whereas, for the line-of-sight components of the mean velocity, we arrive at
\begin{align}
\delta^2\Xi_1^{(0)}(\vq_1,\vq_2) &= -\frac{i}{\sigma_{-1}}
\Biggl\{\left(1-\gamma_1^2\right)^{-1}\Biggl(\frac{1}{\sigma_0^2}q_1^{-1}
+\frac{1}{\sigma_2^2}q_1 q_2^2 -\frac{\gamma_1}{\sigma_0\sigma_2} 
q_1^{-1} q_2^2-\frac{\gamma_1}{\sigma_0\sigma_2}q_1\Biggr)
+\frac{3}{\sigma_1^2}q_2(\rvh\cdot\qvh_1)(\rvh\cdot\qvh_2) \\ 
& \quad -\frac{5}{4\sigma_2^2}q_1 q_2^2\left[3(\rvh\cdot\qvh_1)^2-1\right]
\left[3(\rvh\cdot\qvh_2)^2-1\right]\Biggr\}(\rvh\cdot\qvh_1)
b_{\rm spk1}(q_2)\bvpk(q_1)P_{\delta_S}(q_1) P_{\delta_S}(q_2) 
\nonumber \\
\delta^2\Xi_2^{(0)}(\vq_1,\vq_2) &= \frac{i}{\sigma_{-1}}
\Biggl\{\left(1-\gamma_1^2\right)^{-1}\Biggl(\frac{1}{\sigma_0^2}q_1^{-1}
+\frac{1}{\sigma_2^2}q_1 q_2^2 -\frac{\gamma_1}{\sigma_0\sigma_2} 
q_1^{-1} q_2^2-\frac{\gamma_1}{\sigma_0\sigma_2}q_1\Biggr)
+\frac{3}{\sigma_1^2}q_2(\rvh\cdot\qvh_1)(\rvh\cdot\qvh_2) \nonumber \\ 
& \quad -\frac{5}{4\sigma_2^2}q_1 q_2^2\left[3(\rvh\cdot\qvh_1)^2-1\right]
\left[3(\rvh\cdot\qvh_2)^2-1\right]\Biggr\}(\rvh\cdot\qvh_1)
b_{\rm spk2}(q_2)\bvpk(q_1)P_{\delta_S}(q_1) P_{\delta_S}(q_2) 
\nonumber\;.
\end{align}
We have ignored all terms linear in $\zeta_i^{(0)}$ because these will
cancel  out when we integrate over the angular variables. Furthermore,
the second order contribution  to the transverse velocity component,
$\delta^2\Xi_i^{(\pm 1)}$, can be ignored  because it vanishes upon
averaging over the orientations of the principal axis frames.  On
integrating over the asymmetry parameters and the peak curvature, the 
third piece iii) can eventually be  cast into the form
\begin{multline}
\label{eq:piece3b}
e^{-\frac{1}{3}k^2\sigma_{\rm vpk}^2}
\int\!\!\frac{d^3\vq_1}{(2\pi)^3}\int\!\!\frac{d^3\vq_2}{(2\pi)^3}
\left\{\left(\frac{\sigma_v}{\sigma_{-1}}\right)^2
\Biggl[\frac{3}{\sigma_1^2}(\rvh'\cdot\qvh_1)(\rvh'\cdot\qvh_2)
-\frac{5}{4\sigma_2^2}q_1 q_2
\left[3(\rvh'\cdot\qvh_1)^2-1\right]\left[3(\rvh'\cdot\qvh_2)^2-1\right]
\right. \\ \left.
-\left(1-\gamma_1^2\right)^{-1}(q_1 q_2)^{-1}\Biggl(\frac{1}{\sigma_0^2}
+\frac{(q_1 q_2)^2}{\sigma_2^2}-2\frac{\gamma_1^2}{\sigma_1^2}q_2^2
\Biggr)\Biggr](\rvh'\cdot\qvh_1)(\rvh'\cdot\qvh_2)(\rvh'\cdot\vk)^2
\bias{vpk}(q_1)\bias{vpk}(q_2) 
\right. \\ \left.
+\left(\frac{\sigma_v}{\sigma_{-1}}\right)^2\Biggl(\frac{3}{\sigma_1^2}
-\frac{15}{\sigma_2^2}q_1 q_2(\rvh'\cdot\qvh_1)(\rvh'\cdot\qvh_2)\Biggr)
\Biggl(\sum_{a=\pm}(\ve_a'\cdot\qvh_1)(\ov{\ve}_a'\cdot\qvh_1)
(\ve_a'\cdot\qvh_2)(\ov{\ve}_a'\cdot\qvh_2)(\ve_a'\cdot\vk)
(\ov{\ve}_a'\cdot\vk)\Biggr)\bias{vpk}(q_1)\bias{vpk}(q_2) 
\right. \\ \left.
+\left(\frac{\sigma_v}{\sigma_{-1}}\right)\Biggl[\frac{3}{\sigma_1^2}
q_2(\rvh'\cdot\qvh_1)(\rvh'\cdot\qvh_2)-\frac{5}{4\sigma_2^2}q_1 q_2^2
\left[3(\rvh'\cdot\qvh_1)^2-1\right]\left[3(\rvh'\cdot\qvh_2)^2-1\right]
-\left(1-\gamma_1^2\right)^{-1}
\right. \\ \left. 
\times\, q_1^{-1}\Biggl(\frac{1}{\sigma_0^2}+\frac{(q_1 q_2)^2}{\sigma_2^2}
-\frac{\gamma_1^2}{\sigma_1^2} (q_1^2+q_2^2)\Biggr)
\Biggr](\rvh'\cdot\qvh_1)(\rvh'\cdot\vk)^2
\Bigl[\bias{I1}(q_2)+\bias{I2}(q_2)\Bigr]\bias{vpk}(q_1)\right\}
P_{\delta_S}(q_1)P_{\delta_S}(q_2) e^{i(\vq_1+\vq_2)\cdot\vr'}
\end{multline}
To derive the mode-coupling power $\pmc(k)$, we must now add the
contributions  Eqs (\ref{eq:piece2}), (\ref{eq:piece3a}) and
(\ref{eq:piece3b}) and  perform the integration over $\vr'$. At this
point, it is convenient to express the results in terms of quantities
at the collapse redshift $z_0$ rather than  the initial redshift
$z_i\gg 1$. This change of fiducial redshift is readily achieved by
making the replacement $z_i\to z_0$.

Again, all the terms involving the multiplicative factors
$(\ve_{\pm}'\cdot\vk)$ or $(\ov{\ve}_{\pm}'\cdot\vk)$ cancel out and, as
for the correlation of initial density peaks, the mode-coupling power
can be drastically simplified upon substituting the expression  of the
second order peak bias $\bias{II}(q_1,q_2,z_0)$,
Eq.(\ref{eq:2ndbias}). Adding   Eqs (\ref{eq:piece2}),
(\ref{eq:piece3a}) and  (\ref{eq:piece3b}), the mode-coupling power
can be written
\begin{align}
\label{eq:pmc1}
\lefteqn{\pmc(\nu,R_S,k,z)= 
e^{-\frac{1}{3}k^2\sigma_{\rm vpk}^2(z)}\ppk^{(2)}(R_S,\nu,k)}
\nonumber \\
&\quad+\frac{e^{-\frac{1}{3}k^2\sigma_{\rm vpk}^2(z)}}{2(2\pi)^3}
\int\!\!d^3\vq_1\!\int\!\!d^3\vq_2\,
\Biggl\{\left(\frac{D(z)}{D(z_0)}\right) {\cal F}_1(\vq_1)\Bigl[
\bias{II1}(q_1,q_2)\bias{I2}(q_2)+\bias{II2}(q_1,q_2)\bias{I1}(q_2)
\Bigr] \nonumber \\ 
&\quad +\left(\frac{D(z)}{D(z_0)}\right) {\cal F}_1(\vq_2)\Bigl[
\bias{II1}(q_1,q_2)\bias{I2}(q_1)+\bias{II2}(q_1,q_2)\bias{I1}(q_1)
\Bigr]+\left(\frac{D(z)}{D(z_0)}\right)^2\bigl[{\cal F}_1(\vq_2)\bigr]^2
\bias{I1}(q_1)\bias{I2}(q_1) \nonumber \\
&\quad +\left(\frac{D(z)}{D(z_0)}\right)^2
\bigl[{\cal F}_1(\vq_1)\bigr]^2\bias{I1}(q_2)\bias{I2}(q_2)
+\left(\frac{D(z)}{D(z_0)}\right)^2 {\cal F}_1(\vq_1){\cal F}_1(\vq_2)
\Bigl[\bias{II1}(q_1,q_2)+\bias{II2}(q_1,q_2)\Bigr] \nonumber \\
&\quad +\left(\frac{D(z)}{D(z_0)}\right)^2 {\cal F}_1(\vq_1){\cal F}_1(\vq_2)
\Bigl[\bias{I1}(q_1)\bias{I2}(q_2)+\bias{I1}(q_2)\bias{I2}(q_1)\Bigr]
+\left(\frac{D(z)}{D(z_0)}\right)^3\bigl[{\cal F}_1(\vq_1)\bigr]^2
{\cal F}_1(\vq_2)\Bigl[\bias{I1}(q_2)+\bias{I2}(q_2)\Bigr] \nonumber \\
&\quad +\left(\frac{D(z)}{D(z_0)}\right)^3{\cal F}_1(\vq_1)
\bigl[{\cal F}_1(\vq_2)\bigr]^2\Bigl[\bias{I1}(q_1)+\bias{I2}(q_1)\Bigr]
+\left(\frac{D(z)}{D(z_0)}\right)^4\bigl[{\cal F}_1(\vq_1)\bigr]^2
\bigl[{\cal F}_1(\vq_2)\bigr]^2 -\frac{1}{2\sigma_2^2}
\left(\frac{D(z)}{D(z_0)}\right) q_1^2 q_2^2 \nonumber \\
&\quad \times {\cal F}_1(\vq_2) \Bigl[3(\kvh\cdot\qvh_1)^2-1\Bigr]
\Bigl[3(\kvh\cdot\qvh_2)^2-1\Bigr]\biggl(\partial_\alpha
\ln G_0^{(\alpha)}\!(\gamma_1,\gamma_1\nu_2)\bias{I1}(q_1)
+\partial_\alpha\ln G_0^{(\alpha)}\!(\gamma_1,\gamma_1\nu_1)
\bias{I2}(q_1)\biggr)\biggr\lvert_{\alpha=1} \nonumber \\
&\quad -\frac{1}{2\sigma_2^2}\left(\frac{D(z)}{D(z_0)}\right)^2
{\cal F}_1(\vq_1){\cal F}_1(\vq_2)\Bigl[3(\kvh\cdot\qvh_1)^2-1\Bigr]
\Bigl[3(\kvh\cdot\qvh_2)^2-1\Bigr]\biggl(\partial_\alpha
\ln G_0^{(\alpha)}\!(\gamma_1,\gamma_1\nu_1)+\partial_\alpha
\ln G_0^{(\alpha)}\!(\gamma_1,\gamma_1\nu_2)\biggr)
\biggr\lvert_{\alpha=1} \nonumber \\
&\quad +\left(\frac{D(z)}{D(z_0)}\right)^2 q_1 q_2{\cal F}_1(\vq_1)
{\cal F}_1(\vq_2)\biggl(\frac{3}{\sigma_1^2}(\kvh\cdot\qvh_1)
(\kvh\cdot\qvh_2)-\frac{5}{4\sigma_2^2}q_1 q_2 
\Bigl[3(\kvh\cdot\qvh_1)^2-1\Bigr]\Bigl[3(\kvh\cdot\qvh_2)^2-1\Bigr]
\biggr) \nonumber \\
&\quad +\left(\frac{D(z)}{D(z_0)}\right) q_1 q_2 {\cal F}_1(\vq_1)
\biggl(\frac{3}{\sigma_1^2}(\kvh\cdot\qvh_1)(\kvh\cdot\qvh_2)
-\frac{5}{4\sigma_2^2}q_1 q_2\Bigl[3(\kvh\cdot\qvh_1)^2-1\Bigr]
\Bigl[3(\kvh\cdot\qvh_2)^2-1\Bigr]\biggr)
\Bigl[\bias{I1}(q_2)+\bias{I2}(q_2)\Bigr]\Biggr\} \nonumber \\
&\qquad \times P_{\delta_S}(q_1) P_{\delta_S}(q_2)\, 
\delta^{(3)}(\vk-\vq_1-\vq_2) \;,
\end{align}
where, unless otherwise specified, all quantities are evaluated at
redshift $z_0$.  In analogy with standard PT, we have defined the
kernels ${\cal F}_n$ as
\begin{equation}
{\cal F}_n(\vq_1,\cdots\vq_n)\equiv \frac{1}{n!}\,
\frac{(\vk\cdot\qvh_1)}{q_1}\cdots
\frac{(\vk\cdot\qvh_n)}{q_n}\,
\bias{vpk}(q_1)\cdots\bias{vpk}(q_n)\;.
\end{equation}
We have also introduced the symmetric function 
$\bias{II i}^{\rm E}(\vq_1,\vq_2,z)$,
\begin{equation}
\bias{II i}^{\rm E}(q_1,q_2,z)\equiv {\cal F}_2(\vq_1,\vq_2)
+\frac{1}{2}\left(\frac{D(z_0)}{D(z)}\right)
\Bigl[{\cal F}_1(\vq_1)\bias{I i}(q_2,z_0)
+{\cal F}_1(\vq_2)\bias{I i}(q_1,z_0)\Bigr]
+\frac{1}{2}\left(\frac{D(z_0)}{D(z)}\right)^2\bias{II i}(q_1,q_2,z_0)\;,
\end{equation}
which represents the evolved (Eulerian), second order bias of initial
density maxima in the Zel'dovich approximation. The action of
$\bias{II}^{\rm E}(q_1,q_2,z)$ on fields and correlation
functions is  identical to that of $\bias{II}(q_1,q_2,z_0)$
(see \S\ref{sub:xipk2nd}).  With the aid of these auxiliary functions,
we can rearrange the Fourier transform of
$(1/2)\xi_0^{(0)}\bias{II 1}\bias{II 2}\xi_0^{(0)}$
times the exponential damping with the first nine terms in the curly
bracket of Eq.(\ref{eq:pmc1}) into the compact expression
\begin{equation}
\frac{2}{(2\pi)^3}\left(\frac{D(z)}{D(z_0)}\right)^4
e^{-\frac{1}{3}k^2\sigma_{\rm vpk}^2(z)}\int\!\!d^3\vq_1\!
\int\!\!d^3\vq_2\,\bias{II 1}^{\rm E}(\vq_1,\vq_2,z)
\bias{II 2}^{\rm E}(\vq_1,\vq_2,z) P_{\delta_S}(q_1,z_0) 
P_{\delta_S}(q_2,z_0)\,\delta^{(3)}(\vk-\vq_1-\vq_2)\;.
\end{equation}
Observing that $-i q(\rvh\cdot\qvh)$ and
$-(1/2)q^2[3(\rvh\cdot\qvh)^2-1]$ are the Fourier transform of
$\xi_1^{(1/2)}$ and $\xi_2^{(1)}$,  the four last terms in the curly
bracket of Eq.(\ref{eq:pmc1}) can be combined in a similar way with
some of the terms present in the initial  peak correlation
$\xpk(R_S,\nu,r)$ and eventually arrive at Eq.(\ref{eq:pmc}). Upon 
changing to the variables $x\equiv q_1/k$ and $\mu=\kvh\cdot\qvh_1$, 
the mode-coupling power spectrum can be explicitly written as
\begin{align}
\label{eq:pmc2}
\lefteqn{\pmc(\nu,R_S,k,z) = \frac{k^3}{(2\pi)^2}
e^{-\frac{1}{3}k^2\sigma_{\rm vpk}^2(z)} 
\Biggl\{2\left(\frac{D(z)}{D(z_0)}\right)^4\int_0^\infty\!\!dx\, x^2
P_\delta(kx,z_0)\int_{-1}^{+1}\!\!d\mu\,\Bigl[\bias{II}^{\rm E}
(k,x,\mu,z)\Bigr]^2} \nonumber \\
&\qquad \times P_\delta(k\sqrt{1+x^2-2x\mu},z_0) \nonumber \\
&\quad + \frac{6k^2}{\sigma_1^2(z_0)}\left(\frac{D(z)}{D(z_0)}\right)^2
\int_0^\infty\!\! dx\, x^3 P_\delta(kx,z_0) \int_{-1}^{+1}\!\!d\mu\,
\mu\left(1-x\mu\right)\bias{II}^{\rm E}(k,x,\mu,z) 
P_\delta(k\sqrt{1+x^2-2 x\mu},z_0) \nonumber \\ 
&\quad -\frac{5k^4}{2\sigma_2^2(z_0)}\left(\frac{D(z)}{D(z_0)}\right)^3
\biggl(1+\frac{2}{5}\partial_\alpha\ln G_0^{(\alpha)}\!(\gamma_1,\gamma_1\nu)
\Bigl\rvert_{\alpha=1}\biggr)\int_0^\infty\!\!dx\,x^4 P_\delta(kx,z_0)
\int_{-1}^{+1}\!\!d\mu\,\left(3\mu^2-1\right) \nonumber \\
&\qquad \times\Bigl[2-x^2\left(1-3\mu^2\right)-4 x\mu\Bigr]\,
\bias{II}^{\rm E}(k,x,\mu,z)\,P_\delta(k\sqrt{1+x^2-2x\mu},z_0) 
\nonumber \\
&\quad +\frac{25 k^8}{64\sigma_2^4(z_0)}
\biggl(1+\frac{2}{5}\partial_\alpha\ln G_0^{(\alpha)}\!(\gamma_1,\gamma_1\nu)
\Bigl\rvert_{\alpha=1}\biggr)^2
\int_0^\infty\!\!dx\,x^6 P_\delta(kx,z_0)\int_{-1}^{+1}\!\!d\mu\,
\Bigl[\left(11-30\mu^2+27\mu^4\right) \nonumber \\
&\qquad \times \left(1+x^2-2x\mu\right)^2
-6\left(5-42\mu^2+45\mu^4\right)\left(1+x^2-2x\mu\right)
\left(1-x\mu\right)^2+9\left(3-30\mu^2+35\mu^4\right)\left(1-x\mu\right)^4
\Bigr] \nonumber \\ 
&\qquad \times P_\delta(k\sqrt{1+x^2-2x\mu},z_0) \nonumber \\
&\quad + \frac{27k^4}{8\sigma_1^4(z_0)}\int_0^\infty\!\!dx\,x^4\,P_\delta(kx)
\int_{-1}^{+1}\!\!d\mu\,\Bigl[\left(3\mu^2-1\right)\left(1-x\mu\right)^2
+\left(1-\mu^2\right)\left(1+x^2-2x\mu\right)\Bigr] \nonumber \\
&\qquad \times P_\delta(k\sqrt{1+x^2-2x\mu},z_0) \nonumber \\
&\quad -\frac{15k^5}{4\sigma_1^2(z_0)\sigma_2^2(z_0)}\int_0^\infty\!\!dx\,
x^5\, P_\delta(kx)\int_{-1}^{+1}\!\!d\mu\,\mu\left(1-x\mu\right)\biggl[15
\mu^2\left(1-x\mu\right)^2-9\Bigl[\mu^2\left(1+x^2-2x\mu\right)
\nonumber \\
&\qquad +\left(1-x\mu\right)^2\Bigr]+7\left(1+x^2-2x\mu\right)\biggr] 
P_\delta(k\sqrt{1+x^2-2x\mu},z_0) \Biggr\}\;,
\end{align}
where the second order Eulerian peak bias is
\begin{align}
\bias{II}^{\rm E}(k,x,\mu,z) &= \frac{1}{2}
\Biggl\{\frac{\mu\left(1-x\mu\right)}{x\left(1+x^2-2x\mu\right)}
\biggl(1-\frac{\sigma_0^2}{\sigma_1^2}k^2 x^2\biggr)
\biggr[1-\frac{\sigma_0^2}{\sigma_1^2}k^2\left(1+x^2-2x\mu\right)
\biggr]+\left(\frac{D(z_0)}{D(z)}\right)\frac{\mu}{x}
\biggl(1-\frac{\sigma_0^2}{\sigma_1^2}k^2 x^2\biggr) \nonumber \\
&\qquad \times \Bigl[\bias{10}+\bias{01} k^2\bigl(1+x^2-2x\mu\bigr)\Bigr]
+\left(\frac{D(z_0)}{D(z)}\right)\frac{\left(1-x\mu\right)}
{\left(1+x^2-2x\mu\right)}\biggl[1-\frac{\sigma_0^2}{\sigma_1^2}k^2
\left(1+x^2-2x\mu\right)\biggr]\bigl(\bias{10}+\bias{01}k^2 x^2\bigr)
\nonumber \\
&\qquad +\left(\frac{D(z_0)}{D(z)}\right)^2\Bigl[\bias{20}+\bias{11}k^2 
\left(1+2x^2-2x\mu\right)+\bias{02}k^4 x^2\left(1+x^2-2x\mu\right)\Bigr]
\Biggr\}\;.
\end{align}
To calculate the mode-coupling in configuration space, we simply
Fourier transform $\pmc(\nu,R_S,k,z_0)$.

\bibliographystyle{prsty}
\bibliography{pk2LPT}

\end{document}

%% file: defs.tex
\def\lsim{~\rlap{$<$}{\lower 1.0ex\hbox{$\sim$}}}
\def\bsim{~\rlap{$>$}{\lower 1.0ex\hbox{$\sim$}}}

\def\hkmsmpc{\ {\rm km\,s^{-1}\,{\it h}Mpc^{-1}}}

\def\hmpc{\ {\rm {\it h}^{-1}Mpc}}

\def\hmsun{\ {\rm M_\odot/{\it h}}}
\def\hhmpc{\ {\rm h^{2}Mpc^{-3}}}
\def\hhhmpc{\ {\rm h^{3}Mpc^{-3}}}

\def\hmmpc{\ {\rm {\it h}Mpc^{-1}}}
\def\hgpc{\ {\rm {\it h}^{-1}Gpc}}
\def\hhhgpc{\ {\rm h^{3}Gpc^{-3}}}

\def\la{\langle}
\def\ra{\rangle}

\def\ln{{\rm ln}}
\def\tr{{\rm tr}}
\def\det{{\rm det}}

\def\mathbi#1{\textbf{\em #1}}

\def\ov{\overline}
\def\bb{\bar{b}}

\def\dsc{\delta_{\rm sc}}

\def\rb{\bar{\rho}_m}

\def\veta{{\boldsymbol\eta}}
\def\vzeta{{\boldsymbol\zeta}}
\def\vups{{\boldsymbol\upsilon}}
\def\czeta{\smash[b]{\ov{\zeta}}}
\def\evh{\mathrm{\hat{\bf{e}}}}
\def\kvh{\mathrm{\hat{\bf{k}}}}

\def\qvh{\mathrm{\hat{\bf{q}}}}
\def\rvh{\mathrm{\hat{\bf{r}}}}
\def\xvh{\mathrm{\hat{\bf{x}}}}
\def\yvh{\mathrm{\hat{\bf{y}}}}
\def\zvh{\mathrm{\hat{\bf{z}}}}
\def\kh{\mathit{\hat{k}}}
\def\qh{\mathit{\hat{q}}}

\def\rh{\mathit{\hat{r}}}
\def\ve{\mathrm{\bf e}}
\def\vk{\mathrm{\bf k}}
\def\vq{\mathrm{\bf q}}
\def\vr{\mathrm{\bf r}}

\def\vu{\mathrm{\bf u}}
\def\vv{\mathrm{\bf v}}
\def\vw{\mathrm{\bf w}}
\def\vx{\mathrm{\bf x}}
\def\vy{\mathrm{\bf y}}
\def\vj{\mathrm{\bf J}}
\def\vvs{\mathrm{\bf S}}
\def\vbb{\mathrm{B}}
\def\vcc{\mathrm{C}}
\def\vii{\mathrm{I}}
\def\vmm{\mathrm{M}}
\def\vpp{\mathrm{P}}
\def\vqq{\mathrm{Q}}
\def\vrr{\mathrm{R}}
\def\vvv{\mathrm{V}}
\def\vww{\mathrm{W}}
\def\vxx{\mathrm{X}}

\def\ups{\upsilon}

\def\grad{\mathbi{$\nabla$}}

\def\bvpk{b_{\rm vpk}}
\def\bspk{b_{\rm spk}}
\def\bv{\bar{b}_\nu}
\def\bz{\bar{b}_\zeta}
\def\bias#1{\mathfrak{\tilde{b}}_{\rm{#1}}}
\def\npk{n_{\rm pk}}
\def\bnpk{\bar{n}_{\rm pk}}
\def\dpk{\delta_{\rm pk}}
\def\vxpk{\mathrm{\bf x}_{\rm pk}}
\def\vvpk{\mathrm{\bf v}_{\rm pk}}
\def\xpk{\xi_{\rm pk}}
\def\xpd{\xi_{{\rm pk},\delta}}

\def\xim{\xi_{\rm m}}
\def\ppk{P_{\rm pk}}
\def\ppd{P_{{\rm pk},\delta}}
\def\pmc{P_{\rm MC}}
\def\xmc{\xi_{\rm MC}}
\def\wi#1{\widehat{#1}}

\def\apjl{Astrophys.\ J.\ Lett.}

\def\aap{Astron.\ Astrophys.}

\def\apjs{Astrophys.\ J. Supp.}

%% file: peaks.v2.bbl
\begin{thebibliography}{100}

\bibitem{1990Natur.348..705E}
G. {Efstathiou}, W.~J. {Sutherland}, and S.~J. {Maddox}, \nat {\bf 348},  705
  (1990).

\bibitem{1995MNRAS.276L..59B}
W.~E. {Ballinger}, A.~F. {Heavens}, and A.~N. {Taylor}, \mnras {\bf 276},  L59
  (1995).

\bibitem{1999MNRAS.305..527T}
H. {Tadros}, W.~E. {Ballinger}, A.~N. {Taylor}, A.~F. {Heavens}, G.
  {Efstathiou}, W. {Saunders}, C.~S. {Frenk}, O. {Keeble}, R. {McMahon}, S.~J.
  {Maddox}, S. {Oliver}, M. {Rowan-Robinson}, W.~J. {Sutherland}, and S.~D.~M.
  {White}, \mnras {\bf 305},  527  (1999).

\bibitem{2001MNRAS.327.1297P}
W.~J. {Percival}, C.~M. {Baugh}, J. {Bland-Hawthorn}, T. {Bridges}, R.
  {Cannon}, S. {Cole}, M. {Colless}, C. {Collins}, W. {Couch}, G. {Dalton}, R.
  {De Propris}, S.~P. {Driver}, G. {Efstathiou}, R.~S. {Ellis}, C.~S. {Frenk},
  K. {Glazebrook}, C. {Jackson}, O. {Lahav}, I. {Lewis}, S. {Lumsden}, S.
  {Maddox}, S. {Moody}, P. {Norberg}, J.~A. {Peacock}, B.~A. {Peterson}, W.
  {Sutherland}, and K. {Taylor}, \mnras {\bf 327},  1297  (2001).

\bibitem{2004ApJ...606..702T}
M. {Tegmark}, M.~R. {Blanton}, M.~A. {Strauss}, F. {Hoyle}, D. {Schlegel}, R.
  {Scoccimarro}, M.~S. {Vogeley}, D.~H. {Weinberg}, I. {Zehavi}, A. {Berlind},
  T. {Budavari}, A. {Connolly}, D.~J. {Eisenstein}, D. {Finkbeiner}, J.~A.
  {Frieman}, J.~E. {Gunn}, A.~J.~S. {Hamilton}, L. {Hui}, B. {Jain}, D.
  {Johnston}, S. {Kent}, H. {Lin}, R. {Nakajima}, R.~C. {Nichol}, J.~P.
  {Ostriker}, A. {Pope}, R. {Scranton}, U. {Seljak}, R.~K. {Sheth}, A.
  {Stebbins}, A.~S. {Szalay}, I. {Szapudi}, L. {Verde}, Y. {Xu}, J. {Annis},
  N.~A. {Bahcall}, J. {Brinkmann}, S. {Burles}, F.~J. {Castander}, I. {Csabai},
  J. {Loveday}, M. {Doi}, M. {Fukugita}, J.~R. {Gott}, III, G. {Hennessy},
  D.~W. {Hogg}, {\v Z}. {Ivezi{\'c}}, G.~R. {Knapp}, D.~Q. {Lamb}, B.~C. {Lee},
  R.~H. {Lupton}, T.~A. {McKay}, P. {Kunszt}, J.~A. {Munn}, L. {O'Connell}, J.
  {Peoples}, J.~R. {Pier}, M. {Richmond}, C. {Rockosi}, D.~P. {Schneider}, C.
  {Stoughton}, D.~L. {Tucker}, D.~E. {Vanden Berk}, B. {Yanny}, and D.~G.
  {York}, \apj {\bf 606},  702  (2004).

\bibitem{2005MNRAS.362..505C}
S. {Cole}, W.~J. {Percival}, J.~A. {Peacock}, P. {Norberg}, C.~M. {Baugh},
  C.~S. {Frenk}, I. {Baldry}, J. {Bland-Hawthorn}, T. {Bridges}, R. {Cannon},
  M. {Colless}, C. {Collins}, W. {Couch}, N.~J.~G. {Cross}, G. {Dalton}, V.~R.
  {Eke}, R. {De Propris}, S.~P. {Driver}, G. {Efstathiou}, R.~S. {Ellis}, K.
  {Glazebrook}, C. {Jackson}, A. {Jenkins}, O. {Lahav}, I. {Lewis}, S.
  {Lumsden}, S. {Maddox}, D. {Madgwick}, B.~A. {Peterson}, W. {Sutherland}, and
  K. {Taylor}, \mnras {\bf 362},  505  (2005).

\bibitem{2005ApJ...633..560E}
D.~J. {Eisenstein}, I. {Zehavi}, D.~W. {Hogg}, R. {Scoccimarro}, M.~R.
  {Blanton}, R.~C. {Nichol}, R. {Scranton}, H. {Seo}, M. {Tegmark}, Z. {Zheng},
  S.~F. {Anderson}, J. {Annis}, N. {Bahcall}, J. {Brinkmann}, S. {Burles},
  F.~J. {Castander}, A. {Connolly}, I. {Csabai}, M. {Doi}, M. {Fukugita}, J.~A.
  {Frieman}, K. {Glazebrook}, J.~E. {Gunn}, J.~S. {Hendry}, G. {Hennessy}, Z.
  {Ivezi{\'c}}, S. {Kent}, G.~R. {Knapp}, H. {Lin}, Y. {Loh}, R.~H. {Lupton},
  B. {Margon}, T.~A. {McKay}, A. {Meiksin}, J.~A. {Munn}, A. {Pope}, M.~W.
  {Richmond}, D. {Schlegel}, D.~P. {Schneider}, K. {Shimasaku}, C. {Stoughton},
  M.~A. {Strauss}, M. {SubbaRao}, A.~S. {Szalay}, I. {Szapudi}, D.~L. {Tucker},
  B. {Yanny}, and D.~G. {York}, \apj {\bf 633},  560  (2005).

\bibitem{2006PhRvD..74l3507T}
M. {Tegmark}, D.~J. {Eisenstein}, M.~A. {Strauss}, D.~H. {Weinberg}, M.~R.
  {Blanton}, J.~A. {Frieman}, M. {Fukugita}, J.~E. {Gunn}, A.~J.~S. {Hamilton},
  G.~R. {Knapp}, R.~C. {Nichol}, J.~P. {Ostriker}, N. {Padmanabhan}, W.~J.
  {Percival}, D.~J. {Schlegel}, D.~P. {Schneider}, R. {Scoccimarro}, U.
  {Seljak}, H. {Seo}, M. {Swanson}, A.~S. {Szalay}, M.~S. {Vogeley}, J. {Yoo},
  I. {Zehavi}, K. {Abazajian}, S.~F. {Anderson}, J. {Annis}, N.~A. {Bahcall},
  B. {Bassett}, A. {Berlind}, J. {Brinkmann}, T. {Budavari}, F. {Castander}, A.
  {Connolly}, I. {Csabai}, M. {Doi}, D.~P. {Finkbeiner}, B. {Gillespie}, K.
  {Glazebrook}, G.~S. {Hennessy}, D.~W. {Hogg}, {\v Z}. {Ivezi{\'c}}, B.
  {Jain}, D. {Johnston}, S. {Kent}, D.~Q. {Lamb}, B.~C. {Lee}, H. {Lin}, J.
  {Loveday}, R.~H. {Lupton}, J.~A. {Munn}, K. {Pan}, C. {Park}, J. {Peoples},
  J.~R. {Pier}, A. {Pope}, M. {Richmond}, C. {Rockosi}, R. {Scranton}, R.~K.
  {Sheth}, A. {Stebbins}, C. {Stoughton}, I. {Szapudi}, D.~L. {Tucker}, D.~E.
  {vanden Berk}, B. {Yanny}, and D.~G. {York}, \prd {\bf 74},  123507  (2006).

\bibitem{2006A&A...459..375H}
G. {H{\"u}tsi}, \aap {\bf 459},  375  (2006).

\bibitem{2007MNRAS.381.1053P}
W.~J. {Percival}, S. {Cole}, D.~J. {Eisenstein}, R.~C. {Nichol}, J.~A.
  {Peacock}, A.~C. {Pope}, and A.~S. {Szalay}, \mnras {\bf 381},  1053  (2007).

\bibitem{2009MNRAS.400.1643S}
A.~G. {S{\'a}nchez}, M. {Crocce}, A. {Cabr{\'e}}, C.~M. {Baugh}, and E.
  {Gazta{\~n}aga}, \mnras {\bf 400},  1643  (2009).

\bibitem{2009MNRAS.393.1183C}
A. {Cabr{\'e}} and E. {Gazta{\~n}aga}, \mnras {\bf 393},  1183  (2009).

\bibitem{2009ApJ...696L..93M}
V.~J. {Mart{\'{\i}}nez}, P. {Arnalte-Mur}, E. {Saar}, P. {de la Cruz}, M.~J.
  {Pons-Border{\'{\i}}a}, S. {Paredes}, A. {Fern{\'a}ndez-Soto}, and E.
  {Tempel}, \apjl {\bf 696},  L93  (2009).

\bibitem{2010MNRAS.404...60R}
B.~A. {Reid}, W.~J. {Percival}, D.~J. {Eisenstein}, L. {Verde}, D.~N.
  {Spergel}, R.~A. {Skibba}, N.~A. {Bahcall}, T. {Budavari}, J.~A. {Frieman},
  M. {Fukugita}, J.~R. {Gott}, J.~E. {Gunn}, {\v Z}. {Ivezi{\'c}}, G.~R.
  {Knapp}, R.~G. {Kron}, R.~H. {Lupton}, T.~A. {McKay}, A. {Meiksin}, R.~C.
  {Nichol}, A.~C. {Pope}, D.~J. {Schlegel}, D.~P. {Schneider}, C. {Stoughton},
  M.~A. {Strauss}, A.~S. {Szalay}, M. {Tegmark}, M.~S. {Vogeley}, D.~H.
  {Weinberg}, D.~G. {York}, and I. {Zehavi}, \mnras {\bf 404},  60  (2010).

\bibitem{1996ApJ...471...30H}
W. {Hu} and M. {White}, \apj {\bf 471},  30  (1996).

\bibitem{1998ApJ...504L..57E}
D.~J. {Eisenstein}, W. {Hu}, and M. {Tegmark}, \apjl {\bf 504},  L57+  (1998).

\bibitem{2001ApJ...557L...7C}
A. {Cooray}, W. {Hu}, D. {Huterer}, and M. {Joffre}, \apjl {\bf 557},  L7
  (2001).

\bibitem{2003PhRvD..68f3004H}
W. {Hu} and Z. {Haiman}, \prd {\bf 68},  063004  (2003).

\bibitem{2003ApJ...594..665B}
C. {Blake} and K. {Glazebrook}, \apj {\bf 594},  665  (2003).

\bibitem{2003PhRvD..68h3504L}
E.~V. {Linder}, \prd {\bf 68},  083504  (2003).

\bibitem{2004ApJ...615..573M}
T. {Matsubara}, \apj {\bf 615},  573  (2004).

\bibitem{2005MNRAS.357..429A}
L. {Amendola}, C. {Quercellini}, and E. {Giallongo}, \mnras {\bf 357},  429
  (2005).

\bibitem{2005MNRAS.363.1329B}
C. {Blake} and S. {Bridle}, \mnras {\bf 363},  1329  (2005).

\bibitem{2005ApJ...631....1G}
K. {Glazebrook} and C. {Blake}, \apj {\bf 631},  1  (2005).

\bibitem{2006MNRAS.366..884D}
D. {Dolney}, B. {Jain}, and M. {Takada}, \mnras {\bf 366},  884  (2006).

\bibitem{2006ApJ...644..663Z}
H. {Zhan} and L. {Knox}, \apj {\bf 644},  663  (2006).

\bibitem{2006MNRAS.365..255B}
C. {Blake}, D. {Parkinson}, B. {Bassett}, K. {Glazebrook}, M. {Kunz}, and R.~C.
  {Nichol}, \mnras {\bf 365},  255  (2006).

\bibitem{2008PhRvD..77l3540P}
N. {Padmanabhan} and M. {White}, \prd {\bf 77},  123540  (2008).

\bibitem{2009MNRAS.399.1663G}
E. {Gazta{\~n}aga}, A. {Cabr{\'e}}, and L. {Hui}, \mnras {\bf 399},  1663
  (2009).

\bibitem{2009ApJ...693.1404S}
M. {Shoji}, D. {Jeong}, and E. {Komatsu}, \apj {\bf 693},  1404  (2009).

\bibitem{2009ApJ...698..967Y}
J. {Yoo}, D.~H. {Weinberg}, J.~L. {Tinker}, Z. {Zheng}, and M.~S. {Warren},
  \apj {\bf 698},  967  (2009).

\bibitem{2010ApJ...710.1444K}
E.~A. {Kazin}, M.~R. {Blanton}, R. {Scoccimarro}, C.~K. {McBride}, A.~A.
  {Berlind}, N.~A. {Bahcall}, J. {Brinkmann}, P. {Czarapata}, J.~A. {Frieman},
  S.~M. {Kent}, D.~P. {Schneider}, and A.~S. {Szalay}, \apj {\bf 710},  1444
  (2010).

\bibitem{1996MNRAS.282..347M}
H.~J. {Mo} and S.~D.~M. {White}, \mnras {\bf 282},  347  (1996).

\bibitem{1999MNRAS.304..767S}
R.~K. {Sheth} and G. {Lemson}, \mnras {\bf 304},  767  (1999).

\bibitem{1984ApJ...284L...9K}
N. {Kaiser}, \apjl {\bf 284},  L9  (1984).

\bibitem{1986ApJ...304...15B}
J.~M. {Bardeen}, J.~R. {Bond}, N. {Kaiser}, and A.~S. {Szalay}, \apj {\bf 304},
   15  (1986).

\bibitem{1989MNRAS.237.1127C}
S. {Cole} and N. {Kaiser}, \mnras {\bf 237},  1127  (1989).

\bibitem{1999ApJ...525..543M}
T. {Matsubara}, \apj {\bf 525},  543  (1999).

\bibitem{1993ApJ...413..447F}
J.~N. {Fry} and E. {Gaztanaga}, \apj {\bf 413},  447  (1993).

\bibitem{1988ApJ...333...21S}
A.~S. {Szalay}, \apj {\bf 333},  21  (1988).

\bibitem{2009JCAP...08..020M}
P. {McDonald} and A. {Roy}, \jcap {\bf 8},  20  (2009).

\bibitem{1997MNRAS.284..189M}
H.~J. {Mo}, Y.~P. {Jing}, and S.~D.~M. {White}, \mnras {\bf 284},  189  (1997).

\bibitem{2001ApJ...546...20S}
R. {Scoccimarro}, R.~K. {Sheth}, L. {Hui}, and B. {Jain}, \apj {\bf 546},  20
  (2001).

\bibitem{2010MNRAS.402..589M}
M. {Manera}, R.~K. {Sheth}, and R. {Scoccimarro}, \mnras {\bf 402},  589
  (2010).

\bibitem{2002PhR...367....1B}
F. {Bernardeau}, S. {Colombi}, E. {Gazta{\~n}aga}, and R. {Scoccimarro},
  \physrep {\bf 367},  1  (2002).

\bibitem{1998MNRAS.301..797H}
A.~F. {Heavens}, S. {Matarrese}, and L. {Verde}, \mnras {\bf 301},  797
  (1998).

\bibitem{1998MNRAS.297..692C}
P. {Catelan}, F. {Lucchin}, S. {Matarrese}, and C. {Porciani}, \mnras {\bf
  297},  692  (1998).

\bibitem{1998MNRAS.298.1097P}
C. {Porciani}, S. {Matarrese}, F. {Lucchin}, and P. {Catelan}, \mnras {\bf
  298},  1097  (1998).

\bibitem{2000ApJ...537...37T}
A. {Taruya}, \apj {\bf 537},  37  (2000).

\bibitem{2000ApJ...530...36B}
A. {Buchalter}, M. {Kamionkowski}, and A.~H. {Jaffe}, \apj {\bf 530},  36
  (2000).

\bibitem{2006PhRvD..74j3512M}
P. {McDonald}, \prd {\bf 74},  103512  (2006).

\bibitem{2007PhRvD..75f3512S}
R.~E. {Smith}, R. {Scoccimarro}, and R.~K. {Sheth}, \prd {\bf 75},  063512
  (2007).

\bibitem{2007PhRvD..76h3004S}
E. {Sefusatti} and E. {Komatsu}, \prd {\bf 76},  083004  (2007).

\bibitem{2007PASJ...59...93N}
T. {Nishimichi}, I. {Kayo}, C. {Hikage}, K. {Yahata}, A. {Taruya}, Y.~P.
  {Jing}, R.~K. {Sheth}, and Y. {Suto}, \pasj {\bf 59},  93  (2007).

\bibitem{2008PhRvD..78h3519M}
T. {Matsubara}, \prd {\bf 78},  083519  (2008).

\bibitem{2009ApJ...691..569J}
D. {Jeong} and E. {Komatsu}, \apj {\bf 691},  569  (2009).

\bibitem{2009PhRvD..80h3528S}
S. {Saito}, M. {Takada}, and A. {Taruya}, \prd {\bf 80},  083528  (2009).

\bibitem{2010arXiv1009.1131V}
P. {Valageas}, ArXiv e-prints  (2010).

\bibitem{1998ApJ...504..601P}
U. {Pen}, \apj {\bf 504},  601  (1998).

\bibitem{1999ApJ...520...24D}
A. {Dekel} and O. {Lahav}, \apj {\bf 520},  24  (1999).

\bibitem{2000ApJ...542..559T}
A. {Taruya} and Y. {Suto}, \apj {\bf 542},  559  (2000).

\bibitem{1988ApJ...327..507F}
C.~S. {Frenk}, S.~D.~M. {White}, M. {Davis}, and G. {Efstathiou}, \apj {\bf
  327},  507  (1988).

\bibitem{1993MNRAS.265..689K}
N. {Katz}, T. {Quinn}, and J.~M. {Gelb}, \mnras {\bf 265},  689  (1993).

\bibitem{2002MNRAS.332..339P}
C. {Porciani}, A. {Dekel}, and Y. {Hoffman}, \mnras {\bf 332},  339  (2002).

\bibitem{1985MNRAS.217..805P}
J.~A. {Peacock} and A.~F. {Heavens}, \mnras {\bf 217},  805  (1985).

\bibitem{1985ApJ...297...16H}
Y. {Hoffman} and J. {Shaham}, \apj {\bf 297},  16  (1985).

\bibitem{1986PhRvL..56.1878O}
S. {Otto}, H.~D. {Politzer}, and M.~B. {Wise}, Physical Review Letters {\bf
  56},  1878  (1986).

\bibitem{1987MNRAS.225..777C}
H.~M.~P. {Couchman}, \mnras {\bf 225},  777  (1987).

\bibitem{1989MNRAS.238..319C}
P. {Coles}, \mnras {\bf 238},  319  (1989).

\bibitem{1989MNRAS.238..293L}
S.~L. {Lumsden}, A.~F. {Heavens}, and J.~A. {Peacock}, \mnras {\bf 238},  293
  (1989).

\bibitem{1990MNRAS.243..133P}
J.~A. {Peacock} and A.~F. {Heavens}, \mnras {\bf 243},  133  (1990).

\bibitem{1995MNRAS.272..447R}
E. {Reg\"os} and A.~S. {Szalay}, \mnras {\bf 272},  447  (1995).

\bibitem{1998ApJ...499..548M}
A. {Manrique}, A. {Raig}, J.~M. {Solanes}, G. {Gonzalez-Casado}, P. {Stein},
  and E. {Salvador-Sole}, \apj {\bf 499},  548  (1998).

\bibitem{2008PhRvD..78j3503D}
V. {Desjacques}, \prd {\bf 78},  103503  (2008).

\bibitem{2010PhRvD..81b3526D}
V. {Desjacques} and R.~K. {Sheth}, \prd {\bf 81},  023526  (2010).

\bibitem{1985ApJ...297..365K}
N. {Kaiser} and M. {Davis}, \apj {\bf 297},  365  (1985).

\bibitem{1988lsmu.book..419B}
J.~R. {Bond},  in {\em Large-Scale Motions in the Universe: A Vatican study
  Week}, edited by {Rubin, V.~C.~\& Coyne, G.~V.} (PUBLISHER, ADDRESS, 1988),
  pp.\ 419--435.

\bibitem{1989ApJ...345....3C}
S. {Colafrancesco}, F. {Lucchin}, and S. {Matarrese}, \apj {\bf 345},  3
  (1989).

\bibitem{1998ApJ...509..494C}
R. {Cen}, \apj {\bf 509},  494  (1998).

\bibitem{2001NYASA.927....1S}
R.~K. {Sheth},  in {\em The Onset of Nonlinearity in Cosmology}, Vol.~927 of
  {\em New York Academy Sciences Annals}, edited by {J.~N.~Fry, J.~R.~Buchler,
  \& H.~Kandrup} (PUBLISHER, ADDRESS, 2001), pp.\ 1--+.

\bibitem{1998ApJ...495..554C}
R.~A.~C. {Croft} and E. {Gaztanaga}, \apj {\bf 495},  554  (1998).

\bibitem{2010MNRAS.401.1989D}
S. {de} and R.~A.~C. {Croft}, \mnras {\bf 401},  1989  (2010).

\bibitem{2008MNRAS.385L..78P}
W.~J. {Percival} and B.~M. {Sch{\"a}fer}, \mnras {\bf 385},  L78  (2008).

\bibitem{2008ApJ...687...12D}
N. {Dalal}, M. {White}, J.~R. {Bond}, and A. {Shirokov}, \apj {\bf 687},  12
  (2008).

\bibitem{1984ApJ...285L...1P}
H.~D. {Politzer} and M.~B. {Wise}, \apjl {\bf 285},  L1  (1984).

\bibitem{2009arXiv0910.5224B}
B.~A. {Bassett} and R. {Hlozek}, ArXiv e-prints  (2009).

\bibitem{2010PhRvD..81f3530G}
T. {Giannantonio} and C. {Porciani}, \prd {\bf 81},  063530  (2010).

\bibitem{1987AcPhH..62..263S}
A.~S. {Szalay} and L.~G. {Jensen}, Acta Physica Hungarica {\bf 62},  263
  (1987).

\bibitem{2010arXiv1001.4538K}
E. {Komatsu}, K.~M. {Smith}, J. {Dunkley}, C.~L. {Bennett}, B. {Gold}, G.
  {Hinshaw}, N. {Jarosik}, D. {Larson}, M.~R. {Nolta}, L. {Page}, D.~N.
  {Spergel}, M. {Halpern}, R.~S. {Hill}, A. {Kogut}, M. {Limon}, S.~S. {Meyer},
  N. {Odegard}, G.~S. {Tucker}, J.~L. {Weiland}, E. {Wollack}, and E.~L.
  {Wright}, ArXiv e-prints  (2010).

\bibitem{1972ApJ...176....1G}
J.~E. {Gunn} and J.~R.~I. {Gott}, \apj {\bf 176},  1  (1972).

\bibitem{1974ApJ...187..425P}
W.~H. {Press} and P. {Schechter}, \apj {\bf 187},  425  (1974).

\bibitem{1970Afz.....6..581D}
A.~G. {Doroshkevich}, Astrofizika {\bf 6},  581  (1970).

\bibitem{1996ApJS..103....1B}
J.~R. {Bond} and S.~T. {Myers}, \apjs {\bf 103},  1  (1996).

\bibitem{2001MNRAS.323....1S}
R.~K. {Sheth}, H.~J. {Mo}, and G. {Tormen}, \mnras {\bf 323},  1  (2001).

\bibitem{2008MNRAS.388..638D}
V. {Desjacques}, \mnras {\bf 388},  638  (2008).

\bibitem{Kac1943}
M. {Kac}, Bull.~Am.~Math.~Soc. {\bf 49},  938  (1943).

\bibitem{Rice1945}
S.~O. {Rice}, Bell~System~Tech.~J. {\bf 25},  46  (1945).

\bibitem{1999MNRAS.308..119S}
R.~K. {Sheth} and G. {Tormen}, \mnras {\bf 308},  119  (1999).

\bibitem{2001PASJ...53..155T}
A. {Taruya}, H. {Magara}, Y.~P. {Jing}, and Y. {Suto}, \pasj {\bf 53},  155
  (2001).

\bibitem{2010arXiv1001.3162T}
J.~L. {Tinker}, B.~E. {Robertson}, A.~V. {Kravtsov}, A. {Klypin}, M.~S.
  {Warren}, G. {Yepes}, and S. {Gottlober}, ArXiv e-prints  (2010).

\bibitem{2006PhRvD..73f3519C}
M. {Crocce} and R. {Scoccimarro}, \prd {\bf 73},  063519  (2006).

\bibitem{1970A&A.....5...84Z}
Y.~B. {Zel'dovich}, \aap {\bf 5},  84  (1970).

\bibitem{1980lssu.book.....P}
P.~J.~E. {Peebles},  in {\em Research supported by the National Science
  Foundation.~Princeton, N.J., Princeton University Press, 1980.~435 p.},
  edited by {Peebles, P.~J.~E.} (PUBLISHER, ADDRESS, 1980).

\bibitem{1996ApJ...472....1B}
S. {Bharadwaj}, \apj {\bf 472},  1  (1996).

\bibitem{2006PhRvD..73f3520C}
M. {Crocce} and R. {Scoccimarro}, \prd {\bf 73},  063520  (2006).

\bibitem{1984ApJ...279..499F}
J.~N. {Fry}, \apj {\bf 279},  499  (1984).

\bibitem{1986ApJ...311....6G}
M.~H. {Goroff}, B. {Grinstein}, S. {Rey}, and M.~B. {Wise}, \apj {\bf 311},  6
  (1986).

\bibitem{1987ApJ...320..448G}
B. {Grinstein} and M.~B. {Wise}, \apj {\bf 320},  448  (1987).

\bibitem{2008PhRvD..77f3530M}
T. {Matsubara}, \prd {\bf 77},  063530  (2008).

\bibitem{Zeldovich1965}
Y.~B. {Zel'dovich}, \AvAA {\bf 3},  241  (1965).

\bibitem{1974A&A....32..391P}
P.~J.~E. {Peebles}, \aap {\bf 32},  391  (1974).

\bibitem{1993MNRAS.262.1065C}
P. {Coles}, \mnras {\bf 262},  1065  (1993).

\bibitem{1998ApJ...504..607S}
R.~J. {Scherrer} and D.~H. {Weinberg}, \apj {\bf 504},  607  (1998).

\bibitem{1996ApJ...461L..65F}
J.~N. {Fry}, \apjl {\bf 461},  L65+  (1996).

\bibitem{1998ApJ...500L..79T}
M. {Tegmark} and P.~J.~E. {Peebles}, \apjl {\bf 500},  L79+  (1998).

\bibitem{2008PhRvD..77d3527H}
L. {Hui} and K.~P. {Parfrey}, \prd {\bf 77},  043527  (2008).

\bibitem{2000MNRAS.318L..39C}
P. {Catelan}, C. {Porciani}, and M. {Kamionkowski}, \mnras {\bf 318},  L39
  (2000).

\bibitem{2002MNRAS.333..730C}
R. {Casas-Miranda}, H.~J. {Mo}, R.~K. {Sheth}, and G. {Boerner}, \mnras {\bf
  333},  730  (2002).

\bibitem{2010PhRvD..82d3515H}
N. {Hamaus}, U. {Seljak}, V. {Desjacques}, R.~E. {Smith}, and T. {Baldauf},
  \prd {\bf 82},  043515  (2010).

\bibitem{2010arXiv1006.2343B}
J. {Beltr{\'a}n Jim{\'e}nez} and R. {Durrer}, ArXiv e-prints  (2010).

\bibitem{2008PhRvD..77b3533C}
M. {Crocce} and R. {Scoccimarro}, \prd {\bf 77},  023533  (2008).

\bibitem{2008PhRvD..77d3525S}
R.~E. {Smith}, R. {Scoccimarro}, and R.~K. {Sheth}, \prd {\bf 77},  043525
  (2008).

\bibitem{2009PhRvD..80f3508P}
N. {Padmanabhan} and M. {White}, \prd {\bf 80},  063508  (2009).

\bibitem{2008MNRAS.383..755A}
R.~E. {Angulo}, C.~M. {Baugh}, C.~S. {Frenk}, and C.~G. {Lacey}, \mnras {\bf
  383},  755  (2008).

\bibitem{2007ApJ...664..660E}
D.~J. {Eisenstein}, H. {Seo}, and M. {White}, \apj {\bf 664},  660  (2007).

\bibitem{2008MNRAS.391..435F}
P. {Fosalba}, E. {Gazta{\~n}aga}, F.~J. {Castander}, and M. {Manera}, \mnras
  {\bf 391},  435  (2008).

\bibitem{2010MNRAS.403.1353C}
M. {Crocce}, P. {Fosalba}, F.~J. {Castander}, and E. {Gazta{\~n}aga}, \mnras
  {\bf 403},  1353  (2010).

\bibitem{1993ApJ...412...64L}
S.~D. {Landy} and A.~S. {Szalay}, \apj {\bf 412},  64  (1993).

\bibitem{2009ApJ...701.1547K}
J. {Kim}, C. {Park}, J.~R. {Gott}, and J. {Dubinski}, \apj {\bf 701},  1547
  (2009).

\bibitem{1990MNRAS.245..522A}
L. {Appel} and B.~J.~T. {Jones}, \mnras {\bf 245},  522  (1990).

\bibitem{1995ApJ...453....6M}
A. {Manrique} and E. {Salvador-Sole}, \apj {\bf 453},  6  (1995).

\bibitem{2010ApJ...720.1650S}
H. {Seo}, J. {Eckel}, D.~J. {Eisenstein}, K. {Mehta}, M. {Metchnik}, N.
  {Padmanabhan}, P. {Pinto}, R. {Takahashi}, M. {White}, and X. {Xu}, \apj {\bf
  720},  1650  (2010).

\bibitem{Lifshiftz1946}
E.~M. {Lifshiftz}, J.~Phys.~USSR {\bf 10},  116  (1946).

\bibitem{1970ApJ...162..815P}
P.~J.~E. {Peebles} and J.~T. {Yu}, \apj {\bf 162},  815  (1970).

\bibitem{1980PhRvD..22.1882B}
J.~M. {Bardeen}, \prd {\bf 22},  1882  (1980).

\bibitem{1984PThPS..78....1K}
H. {Kodama} and M. {Sasaki}, Progress of Theoretical Physics Supplement {\bf
  78},  1  (1984).

\bibitem{1992PhR...215..203M}
V.~F. {Mukhanov}, H.~A. {Feldman}, and R.~H. {Brandenberger}, \physrep {\bf
  215},  203  (1992).

\bibitem{1994FCPh...15..209D}
R. {D\"urrer}, Fundamentals of Cosmic Physics {\bf 15},  209  (1994).

\bibitem{1995ApJ...455....7M}
C. {Ma} and E. {Bertschinger}, \apj {\bf 455},  7  (1995).

\bibitem{2004astro.ph..2060H}
W. {Hu}, ArXiv Astrophysics e-prints  (2004).

\bibitem{1988ApJ...332L...7G}
K. {Gorski}, \apjl {\bf 332},  L7  (1988).

\bibitem{1998PhRvD..57.3199D}
R. {D\"urrer} and M. {Kunz}, \prd {\bf 57},  3199  (1998).

\bibitem{1975ctf..book.....L}
L.~D. {Landau} and E.~M. {Lifshitz},  in {\em Course of theoretical physics -
  Pergamon International Library of Science, Technology, Engineering and Social
  Studies, Oxford: Pergamon Press, 1975, 4th rev.engl.ed.}, edited by E.~M.
  Landau, L. D. \&~Lifshitz (PUBLISHER, ADDRESS, 1975).

\bibitem{2008PhRvD..78b3527D}
V. {Desjacques} and R.~E. {Smith}, \prd {\bf 78},  023527  (2008).

\end{thebibliography}
